\documentclass{article}

\usepackage{booktabs}
\usepackage{natbib}
\usepackage{epubtk}
\usepackage{graphicx}
\usepackage{lscape}
\usepackage{xspace}

\newcommand{\kms}{~km~s\super{-1}\xspace}

\begin{document}

\title{The FIP and Inverse FIP Effects in Solar and Stellar Coronae}

\author{\epubtkAuthorData{J.\ Martin Laming}{%
Code 7684 Naval Research Laboratory\\
Washington DC 20375, USA}{%
laming@nrl.navy.mil}{%
}
}

\date{}
\maketitle

\begin{abstract}
We review our state of knowledge of coronal element abundance anomalies in the Sun
and stars. We concentrate on the first ionization potential (FIP) effect observed
in the solar corona and slow-speed wind, and in the coronae of solar-like dwarf stars,
and the ``inverse FIP'' effect seen in the corona of stars of later spectral type;
specifically M~dwarfs. These effects relate to the enhancement or depletion,
respectively, in coronal abundance with respect to photospheric values of elements
with FIP below about 10~eV. They are interpreted in terms of the
ponderomotive force due to the propagation and/or reflection of
magnetohydrodynamic waves in the chromosphere. This acts on chromospheric ions, but not neutrals, and so can lead to ion-neutral fractionation.

A detailed description of the model applied to closed magnetic loops, and to open field regions
is given, accounting for the observed difference in solar FIP fractionation between the slow and
fast wind. It is shown that such a model can also account for the observed depletion of helium
in the solar wind. The helium depletion is sensitive to the chromospheric altitude where
ion-neutral separation occurs, and
the behavior of the helium abundance in the closed magnetic loop strongly suggests that the waves have a coronal origin. This, and other similar inferences may be expected to have a strong bearing on theories of solar coronal heating.

Chromospheric waves originating from below as acoustic waves mode convert, mainly to fast mode waves, can also give rise to ion-neutral separation. Depending on the geometry of the magnetic field, this can result in FIP or Inverse FIP effects.
We argue that such configurations are more
likely to occur in later-type stars (known to have stronger field in any case), and that this
explains the occurrence of the Inverse FIP effect in M~dwarfs. We conclude with a discussion of
possible directions for future work.

\end{abstract}

\epubtkKeywords{Stars: coronae, Sun: abundances, Sun: chromosphere, Sun: corona, Turbulence, Waves}

\newpage

%==============================================================================

\section{Introduction}
\label{section:introduction}

Working during the early years of solar UV and X-ray spectroscopy, \citet{pottasch63} found
evidence for significantly higher abundances
of Mg, Si, and Fe in the low solar corona than in the photosphere, and concluded, somewhat reluctantly that ``the chemical composition in the solar atmosphere differs from the
photosphere to the corona''. These elements, Mg, Si, and Fe, are all elements with
first ionization potential (FIP) less than 10~eV, now known to be routinely enhanced
in abundance in the corona with respect to photospheric values, a phenomenon that has
become known as the ``FIP effect''. High FIP elements such as O, Ne, and He, have much smaller abundance enhancements, or even abundance depletions in the corona.
Although the possibility of elemental fractionation between the solar photosphere and
corona only really began to be taken seriously in the mid 1980s, with the publication
of influential reviews by \citet{meyer85a,meyer85b}, recognizing the work of
\citet{pottasch63} almost fifty years ago makes the problem of understanding the FIP effect nearly as old
as that of coronal heating. In fact, modern models of the effect to be discussed in
detail below, in which the fractionation is driven by the ponderomotive force of Alfv\'en
waves, make an intimate connection between the abundance anomaly and coronal heating
mechanisms, such that the FIP effect may yield several important insights into the nature of the latter. This is a theme we will develop throughout this review.

The solar FIP effect manifests itself in several modes of observation. Spectroscopy, as
pioneered by \citet{pottasch63} and extensively reviewed at intervals over the last
20 years \citep[e.g.,][]{feldman92,feldman00,feldman03,saba95}, reveals the composition
primarily of the ``closed loop'' corona. \citet{meyer85a,meyer85b}
also considered elemental abundances measured \textit{in situ} in the solar wind, and in
solar energetic particle events, which flow out along open magnetic field lines. Observations
with Ulysses, the first mission to fly over the solar polar regions \citep{wenzel92} revealed FIP
fractionation varying with wind speed \citep{zurbuchen99,vonsteiger00}. Slow speed solar wind had abundances resembling those in the closed loop solar corona, whereas high speed wind from polar coronal holes had a much lower level of FIP fractionation.
With the advent of the Solar and Heliospheric Observatory (SOHO),
a more coherent observational picture began to emerge. Coronal holes, the source of the relatively
unfractionated fast solar wind, were themselves shown to have similar abundances to the wind
emanating from them. The lower latitude closed field corona was also shown to have FIP fractionated
plasma, similar to the slow speed solar wind. \citet{schmelz12} give a modern view of
coronal element abundances derived from these various solar physics sources, coronal spectroscopy, solar wind and solar energetic particles.

Astrophysical EUV and X-ray spectroscopy, made possible by the 1990's launches of the Extreme Ultraviolet Explorer (EUVE), the Advanced Satellite for Cosmology and Astrophysics (ASCA), Chandra and XMM-Newton, allowed coronal abundances in stars to be measured for the first time. Again, nearly two decades after the first such observations, various observations seem to be falling into place.
In this review, we will attempt to synthesize these two strands of observations, solar and stellar, into one complete picture of coronal element abundance anomalies.

Following this introduction, in Section~\ref{sec:photosphere} we briefly review recent developments in solar
photospheric abundances, where improved spectroscopic data and the application of 3D radiation
transfer calculations have ushered in a revised solar composition. Section~\ref{sec:solar_FIP} describes the
various facets of the solar FIP effect, while Section~\ref{sec:stellar_FIP} surveys stellar FIP and Inverse FIP
effects. Section~\ref{sec:early_theory} describes early attempts to
model the solar FIP effect. Section~\ref{sec:ponderomotive} lays out
the model advocated here, where the ponderomotive force due to
Alfv\'en waves propagating through, or reflecting from, the
chromosphere accelerates ions up or down, while leaving neutral atoms unaffected. Section~\ref{sec:results} describes the results of such a model, and discusses its important
implications. Section~\ref{sec:conclusions} concludes with suggestion for future research directions.

\newpage

\section{Solar Photospheric Composition}
\label{sec:photosphere}

\subsection{Review}

Any work on solar coronal abundance anomalies must begin with reviewing the photospheric
composition. Despite its long history, the composition of the photosphere, also taken
as a proxy for ``cosmic abundances'', has undergone significant revisions in recent
years. We take as our starting point the composition review of \citet{anders89}, this
being the standard solar composition in use for the early studies of the FIP effect,
and for many years the default abundance set implemented by spectral fitting software in use
in X-ray astronomy. This was updated by \citet{grevesse98}, who revised downwards by
a small amounts N and O (0.13 and 0.10~dex, respectively) and also Fe which moved from
7.67 (on a logarithmic scale where the abundance of H is 12) to 7.50, in agreement
with the meteoritic value. This last modification stemmed from improvements in
atomic data used to analyze solar spectrum \citep[e.g.,][]{holweger91,biemont91}, and is
largely supported by more modern analyzes \citep{asplund00a,asplund00b,asplund00c,rubio02}.

The next major revision came to the photospheric abundance of O by \citet{allende01}, who
applied a three-dimensional time-dependent hydrodynamical model solar atmosphere to
the observed 6300~\AA\ line. These authors recognized that this line, attributed to
a forbidden line of neutral O, is also blended with Ni I, with the result that the
O abundance of 8.83 in \citet{grevesse98} \citep[8.93 in][]{anders89} was revised
down to $8.69\pm 0.05$. A number of subsequent papers analyzing other allowed and
forbidden lines in O I, and also molecular OH supported this change
\citep{asplund04,asplund05a,melendez04,navarro07,melendez08}, though \citet{ayres08}
and \citet{caffau08} offered more cautious views. Joining, O, C
\citep{allende02,asplund05b,caffau10} and N \citep{caffau09} also underwent downward revisions in their abundances.

In Table~\ref{table:photosphere} we collect the recommended solar photospheric abundances
of various commonly observed elements from \citet{grevesse98}, \citet{asplund09}, \citep[also given in][]{grevesse10} \citet{caffau11},
and \citet{scott14a,scott14b} and \citet{scott14c} for comparison and reference. Not given here are result from \citet{lodders10}, who for the elements of
most interest here appears to quote the average of \citet{asplund09} and \citet{caffau11}. The data sources for each element are
all listed in these reviews. For future reference we remark that aside from the
revisions above, the noble gases will be of interest to us, since their photospheric
abundances are often determined from coronal observations
\citep[e.g.,][]{feldman90,young05}, on the assumption that no
fractionation occurs between the photosphere and the corona for these elements. This
is an assumption we shall scrutinize. Only He is determined independently, from
helioseismology \citep{basu04}, in an analysis that includes the revised solar
metallicity.

\begin{table}[htbp]
\caption{Recommended Solar Photospheric Abundances of Common Elements}
\label{table:photosphere}
\centering
{\small
\begin{tabular}{lrrrr}
\toprule
Element & \citet{grevesse98} & \citet{asplund09} & \citet{caffau11}& \citet{scott14a,scott14b}\\
 & & & & \citet{scott14c}\\
\midrule
H & 12.00             & 12.00           & ~ & ~\\
He & $10.93\pm 0.004$ & $10.93\pm 0.01$ & ~ & ~\\
C  & $8.52\pm 0.06$ & $8.43\pm 0.05$ & $8.50\pm 0.06$ \\
N  & $7.92\pm 0.06$ & $7.83\pm 0.05$ & $7.86\pm 0.12$ \\
O  & $8.83\pm 0.06$ & $8.69\pm 0.05$ & $8.76\pm 0.07$ \\
Ne & $8.08\pm 0.06$ & $7.93\pm 0.10$ & \\
Na & $6.33\pm 0.03$ & $6.24\pm 0.04$ & & $6.21\pm 0.04$\\
Mg & $7.58\pm 0.05$ & $7.60\pm 0.04$ & & $7.59\pm 0.04$\\
Al & $6.47\pm 0.07$ & $6.45\pm 0.03$ & & $6.43\pm 0.04$\\
Si & $7.55\pm 0.05$ & $7.51\pm 0.03$ & & $7.51\pm 0.03$\\
P  & $5.45\pm 0.04$ & $5.41\pm 0.03$ & $5.46\pm 0.04$ & $5.41\pm 0.03$\\
S  & $7.33\pm 0.11$ & $7.12\pm 0.03$ & $7.16\pm 0.05$ & $7.13\pm 0.03$\\
Cl & $5.5\pm 0.3$ & $5.50\pm 0.30$ \\
Ar & $6.40\pm 0.06$ & $6.40\pm 0.13$ & ~ \\
K  & $5.12\pm 0.13$ & $5.03\pm 0.09$ & $5.11\pm 0.09$ & $5.04\pm 0.05$\\
Ca & $6.36\pm 0.02$ & $6.34\pm 0.04$ & ~ & $6.32\pm 0.03$\\
Ti & $5.02\pm 0.06$ & $4.95\pm 0.05$ & & $4.90\pm 0.04$\\
Cr & $5.67\pm 0.03$ & $5.64\pm 0.04$ & & $5.62\pm 0.04$\\
Fe & $7.50\pm 0.05$ & $7.50\pm 0.04$ & $7.52\pm 0.06$ & $7.47\pm 0.04$\\
Ni & $6.25\pm 0.04$ & $6.22\pm 0.04$ & ~ & $6.20\pm 0.04$\\
Kr & $3.31\pm 0.08$ & $3.25\pm 0.06$ & & $3.23\pm 0.06$\\
\bottomrule
\end{tabular}}
\end{table}

\subsection{Helioseismology}

The re-evaluation
of the solar composition above has revised the solar
metallicity down from 0.0170 \citep{grevesse98} to the range 0.0153
\citep{caffau11} to 0.0134 \citep{asplund09}, with \citet{asplund05c} giving a
value as low as 0.0122, largely
driven by improved analysis of lines of C, N, and O. The lower
metallicity decreases the sound speed at the base of the solar
convection zone, yielding now a larger disagreement between observed
and modeled sound speeds \citep[see, e.g., Figure~1 in][]{guzik10}. The
reduced abundances also decrease the depth of the convection zone in
solar models, again worsening agreement between models and
helioseismic inversions \citep[e.g.,][]{basu04}.

One early solution proposed was to increase the abundance of Ne to
compensate \citep{antia05,bahcall05a}, by bringing the solar metallicity
back to its prior value. A survey in nearby active stars
had previously yielded the abundance ratio Ne/O in the range 0.3\,--\,0.4
\citep{drake05}, significantly higher than the solar coronal value and
closer to that suggested.  \citet{drake05} argue
that the relative constancy of this ratio in a sample of over 20 stars
suggests that the ``true'' Ne/O abundance ratio should be around 0.4,
and that the variation in the solar corona must be due to some unknown
fractionation. More recent work suggests than
an increased Ne abundance (of about 0.5\,--\,0.67~dex) is not a complete
fix \citep{lin07}, but a more modest increase of 0.45~dex is still
acceptable \citep[see introduction of][]{guzik10}. Ne, and also
possibly Ar, are the focus of revisions to composition because having
no photospheric absorption lines, their abundances can only be
measured in the solar corona, or in astrophysical sources as proxies
for the solar photosphere. \citet{asplund09} argue for an Ne/O
abundance ratio of $0.175\pm 0.031$ following the measurements of
\citet{young05} in the solar corona. Other authors have suggested that increased opacity
\citep{dalsgaard09,serenelli09}, beyond that implied by the increase due to data
from the Opacity Project replacing older OPAL radiative opacities \citep{bahcall05b} might
ease the problem,
while variations to the solar evolutionary history with extra episodes of
mass loss of accretion have also been considered \citep{guzik10,serenelli11}. \citet{villante13}
provide a recent evaluation, accounting for helioseismic and solar neutrino data, again
favouring the older photospheric abundance set of \citet{grevesse98}. Even more
recently, \citet{shearer14} study the variation of Ne/O measured by Ulysses/SWICS and
ACE/SWICS over the solar cycle between 1998 and 2012. Their results also favour a low
Ne/O abundance ratio in the range 0.10 - 0.15, although with unexpected variation.

\citet{bergemann14} review these and other potential solutions to the ``solar abundance problem''. Other ideas include that of
\citet{lopes13}, who revisit the helioseismology problem in the light of revised photospheric
abundance for the Sun and solar-like stars. Relative to solar analogs without planetary systems,
the Sun appears to be underabundant in metals, more so in refractory elements than in volatiles.
The mass of the ``missing'' elements from the solar convection zone appears to be similar to the
combined mass  of the inner terrestrial planets, Mercury, Venus, Earth and Mars, leaving open the
possibility that the metallicity of the solar interior, specifically the radiation zone, is higher
than that of the convection zone, as appears to be required. \citet{lopes13} explore several such
models, computing the solar neutrino spectra with a view to future observational capabilities.
Their models however appear to need high mass loss from the young Sun, which might be
problematic \citep[cf.][]{wood04,wood06b}. \citet{zhang14} considers the effect of a turbulent
kinetic flux within the solar convection zone. A negative flux, i.e. turbulence propagating from
the convection zone to the radiation zone, goes some way to restoring agreement. A downwards
turbulent kinetic flux requires larger outward radiative and convective energy fluxes, which lead
to a deeper boundary between the convection and radiation zones.

Finally the solar helium abundance is most accurately determined from helioseismology, from the
anomaly produced in the sound speed at the depths in the convection zone where helium ionizes.
\citet{villante13} give the value as $Y=0.2485\pm 0.0035$, apparently the mean of the results derived by \citet{basu04} using GONG or MDI data.

\newpage

\section{The Solar FIP Effect: Overview}
\label{sec:solar_FIP}

As mentioned above, the observation of element abundances in the solar corona and
wind has a history spanning decades, and has been reviewed several times during that
period \citep{meyer85a,meyer85b,feldman92,feldman00,feldman02,feldman03,feldman07}. Here, rather
than provide an exhaustive review, we attempt to summarize and update the status
of solar coronal abundances, and refer readers back to these reviews for extensive
references.

In contrast to photospheric abundances which are generally measured and given relative to H
(i.e absolute abundances), with a few exceptions to be noted below, most coronal abundance measurements and results
are relative, in that one minor ion is compared to another minor ion (often O) with no reference to H. This arises due to the difficulties in observing H. H generally has no observable emission lines from the solar corona in bandpasses in which other elements are observed, and requires different {\it in situ} instruments for detection to those for heavy ions. Both difficulties
lead to cross calibration issues. Exceptions occur in the UV where H emission lines can sometimes
be detected, or in the X-ray region where the thermal bremsstrahlung continuum can be taken as
an indicator of the H abundance. We will highlight these features where appropriate below.

Theoretically, it will also turn out to be easier to discuss relative abundances. Modeling the
action of the ponderomotive force on minor ions in an ambient H atmosphere is much more
straightforward than the multifluid calculation that would be necessary to treat the back reaction of the Alfv\'en waves on the H fluid itself, and to incorporate those effects into the predicted
FIP fractionations. In fact in many cases, it turns out the H and O behave similarly because of their strong charge exchange coupling, due to their similar ionization potentials, and this provides some justification for our approach.

In the quiet solar corona and slow speed solar wind, elements with first
ionization potential (FIP) below about 10~eV are enhanced in abundance by a factor
of about 3, with typical variations in the range 2\,--\,5. This is established by
both remotely sensed (i.e., spectroscopic) and \textit{in situ} measurements and at the
time of writing, despite early controversy, is now considered an established fact.
Variations in the fractionation with solar region exist. Coronal holes, and the
fast solar wind emanating from them are known to have a significantly smaller degree
of FIP fractionation than the quiet corona and slow wind \citep[e.g.,][]{bochsler07a,feldman98b}.
\citet{brooks11} show that FIP fractionated abundances measured in an active region with
the Extreme Ultraviolet Imaging Spectrograph (EIS) on Hinode \citep{culhane07} match those
detected a few days later when the solar wind from this active region would have reached 1~AU.
The Sun viewed ``as a star''
\citep{laming95} shows FIP effect only for temperatures above about $10^6$ K. This is illustrated in Figure~\ref{fig:LDW}, taken from \citet{laming95}, where
the full disk solar emission distribution is determined from the solar spectrum of \citet{malinovsky73}, based on emission lines from low FIP (filled symbols) and high FIP
(open symbols) elements. The two emission measures coincide up to $\log T\simeq 6$, and thereafter
diverge, indicating overabundance of low FIP elements.
This result is corroborated by \citet{young05a,young05b}, \citet{feldman93,feldman98a}, and \citet{young98}, and leads to the conclusion that structures emitting at temperature below
$10^6$ K must be distinct entities from those responsible for the higher temperature
emission. \citet{feldman94} elaborate this argument based on earlier observations, most notably
those in \citet{feldman83,feldman87}. It appears likely that the high temperature
FIP fractionated plasma resides in coronal loops fed by chromospheric evaporation
\citep[the response of chromospheric plasma to heat released in the corona conducted downwards,
see e.g.][]{bray91,reale14}, while the
lower temperate plasma, attributed to ``unresolved fine structures" by \citet{feldman83,feldman87},
can now be identified with the recently discovered Type II spicules \citep{depontieu11,martinez11}, or other dynamic loop structures \citep{hansteen14}
recently observed with the Interface Region Imaging Spectrograph (IRIS).
Individual features distinct from Type II spicules at transition region temperatures ($10^4$K $< T < 10^6$K) may show FIP
fractionation \citep[see examples in e.g.][]{feldman92,feldman00,feldman02,feldman03,feldman07}. We emphasize
again that the result of \citet{laming95} applies to the ``Sun as a star''.

Active regions and flares
can also have different fractionation, often reduced from that observed in the
quiet Sun.
This is most clearly seen in spectra acquired of plasma above a sunspot \citep{feldman90}. \citet{phillips94} also measured a maximum Fe enhancement of a factor of 2
between photosphere and corona, by comparing the photospheric Fe K $\beta$ line excited by
flourescence with the collisionally excited coronal Fe XXV resonance line observed during flares
by the YOHKOH Bragg Crystal Spectrometer \citep{culhane91}. Later work with RHESSI \citep{lin04}
found much stronger enhancements in the abundance of Fe relative to H from line to continuum
measurements \citep{phillips06}. Abundances of other elements in flares relative to H have been measured
by RESIK \citep{sylwester05}. FIP fractionations for K, Ar, Cl, S, Si and Al have been given by
\citet{sylwester08}, with more detailed results given in subsequent papers \citep{sylwester10a,
sylwester10p,sylwester11,sylwester12,sylwester13,sylwester14}, and given in more detail in Table~\ref{table:closedfield} for comparison with model results. \citet{warren14} finds almost no FIP fractionation in 21 flares observed with the EUV variability Experiment (EVE) on
the Solar Dynamics Observatory (SDO), while \citet{delzanna14} find a typical enhancement in
the abundance of Fe compared to O and Ne of 3.2 in 9 flares observed by the Flat Crystal Spectrometer (FCS) on the Solar Maximum Mission (SMM). \citet{brooks12} and \citet{widing08}
find similar enhancements.

Coronal mass ejections (CMEs) observed \textit{in situ} often have strong FIP fractionation as shown
in a survey \citep{reisenfeld07} conducted with data the Advanced Composition Explorer \citep[ACE][]{gloeckler98} and the Genesis mission \citep{burnett03}. \citet{zurbuchen04} and
\citet{smith01} show that frequently these large FIP fractionations are associated with the CME flux rope, possibly implying that the flux rope forms in the corona and does not emerge preformed from
the photosphere (in which case it might be expected to exhibit photospheric abundances). CMEs can also exhibit mass dependent fractionation \citep{wurz00}, most likely indicating the role of
processes other than FIP fractionation in modifying the elemental composition of the solar upper
atmosphere.

We have already mentioned the difference in FIP effect between fast and slow speed streams observed by Ulysses \citep{zurbuchen99,vonsteiger00}. \citet{lepri13} study the evolution of
abundances of fast and slow solar wind observed in the ecliptic by the Advanced Composition Explorer (ACE). While other solar wind parameters such as charge states, magnetic fields and
freeze-in temperatures show marked variation, the degree of FIP fractionation (i.e. element
abundances relative to O) does not. Absolute abundances, measured relative to H, do show some variation, with the lowest abundances of He, C, O, Si and Fe being observed in the slowest wind at solar minimum. Interestingly, their absolute abundances for O support the revised photospheric
abundance of O \citep{asplund09}, with the highest mean O/H abundance ratio in the slow speed wind at solar maximum being 8.68 (in logarithmic notation), with upper and lower limits of 8.42 and 8.94 respectively, from their Table 1. Previous studies of this sort \citep{bochsler07b,vonsteiger10} have favoured the ``older'' O abundance \citep[8.83][]{grevesse98}. \citet{vonsteiger10} assume that the high speed solar wind gives
the most faithful representation of photospheric abundances \citep[their O/H value in slow speed
solar wind agrees with][]{asplund09}. \citet{bochsler07b} plots O/H against He/H in his Figure 3. and shows a striking correlation between them, arguing that inefficient Coulomb drag is the cause of both variations. This might appear to be problematic, in that O is not always depleted relative to H \citep[unless one takes oldest O/H value from][as the correct value]{anders89}, but He is, with the problem becoming worse for the more recent O/H values.

\epubtkImage{LDW_FIP}{%
\begin{figure}[htbp]
\centerline{\includegraphics[scale=0.55]{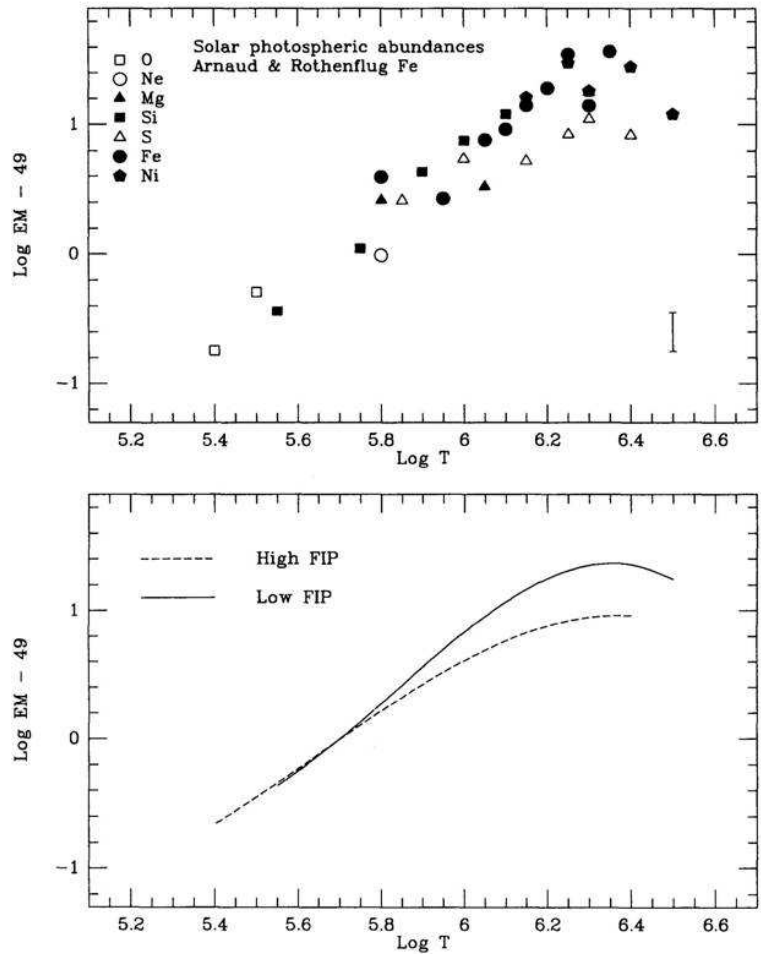}}
\caption{Full disk solar emission measure distribution, with low FIP elements depicted by solid symbols, high FIP elements by open symbols. The transition from photospheric abundances at $\log T < 6.0$ to coronal abundances at higher temperatures, indicating that different types of coronal structures are emitting above and below this temperature. Figure from \citet{laming95}. Reproduced by permission of the AAS.\label{fig:LDW}}
\end{figure}}

\epubtkImage{kasper}{%
\begin{figure}[htbp]
\centerline{\includegraphics[scale=0.4]{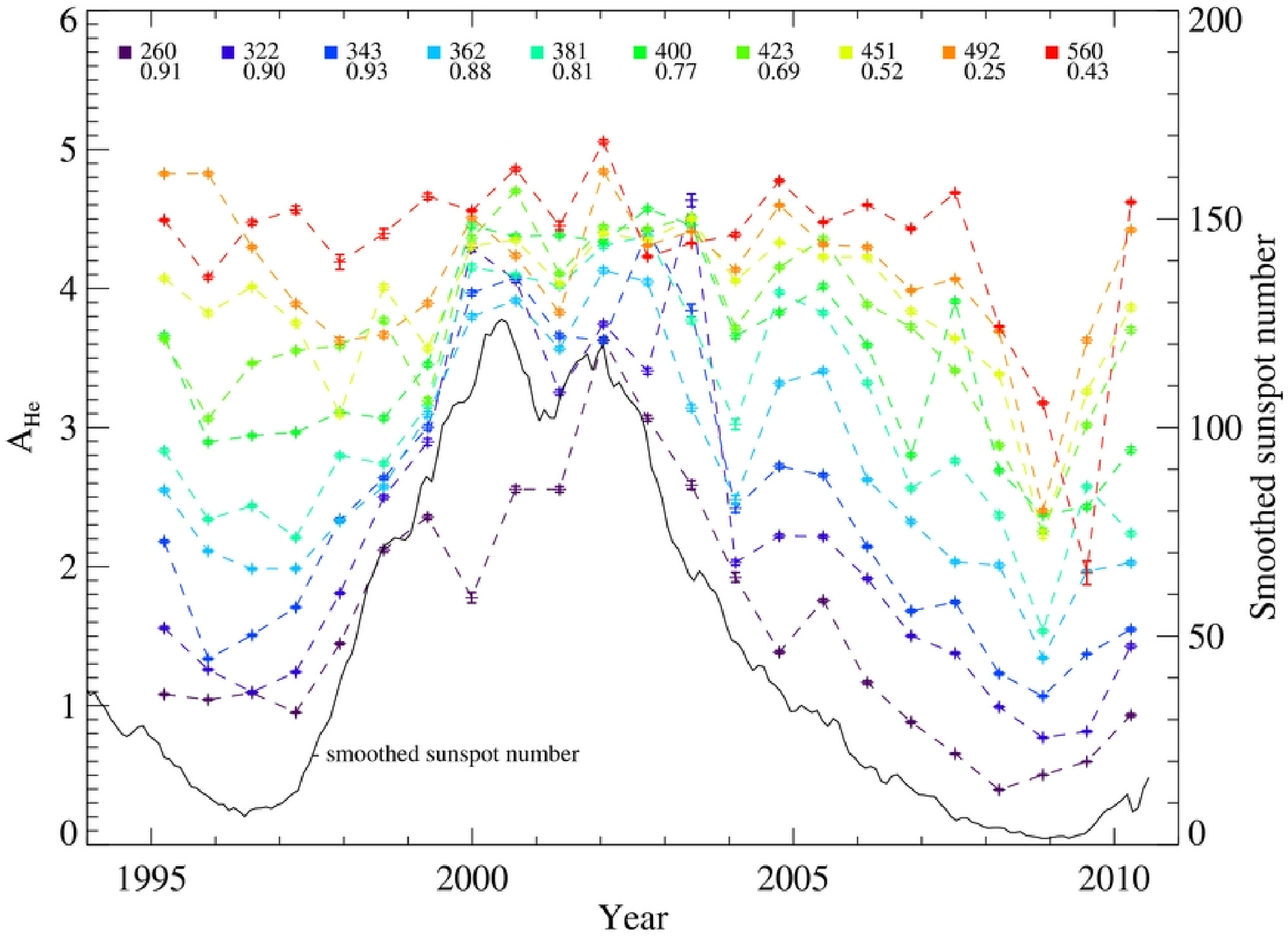}\includegraphics[scale=0.45]{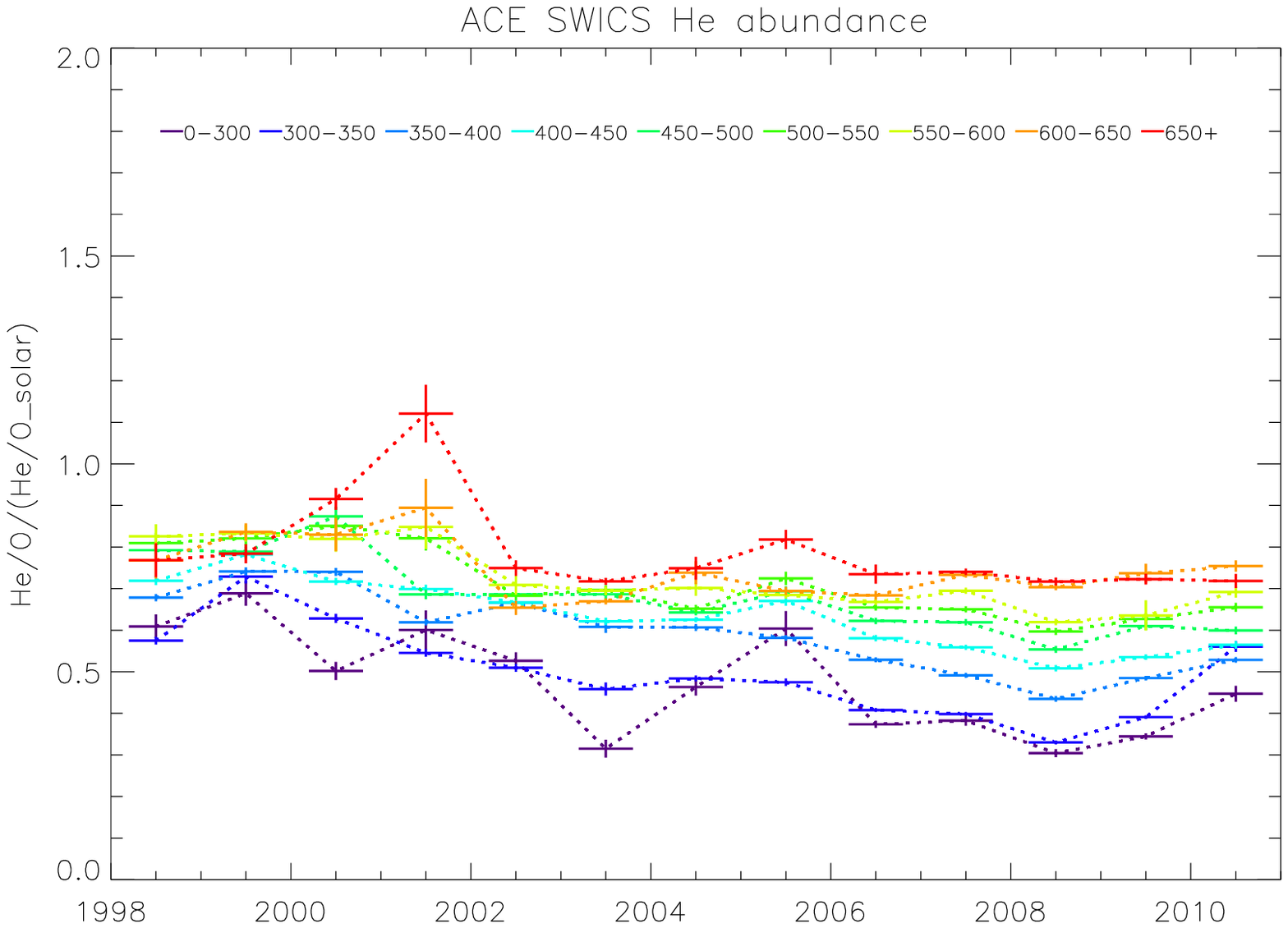}}
\caption{\emph{Left:} Helium abundance relative to hydrogen in the slow speed solar wind over 1.5 solar cycles. The curve colors denoted the wind speed at which at abundance is measured. He is more variable in the slowest slow speed solar wind. The black curve gives the monthly smoothed sunspot number. Figure from \citet{kasper12}. Reproduced by permission of the AAS. \emph{Right:} Helium abundance relative to oxygen in the slow speed solar wind, measured by ACE/SWICS. The same trend of depletion with wind speed as for He/H is seen, but the solar cycle dependence is less pronounced. Figure from \citet{rakowski12}. Reproduced by permission of the AAS. \label{fig:kasper}}
\end{figure}}

\epubtkImage{baker}{%
\begin{figure}[htbp]
%\centerline{\includegraphics[scale=0.8]{baker}}
\centerline{\includegraphics[width=0.5\textwidth]{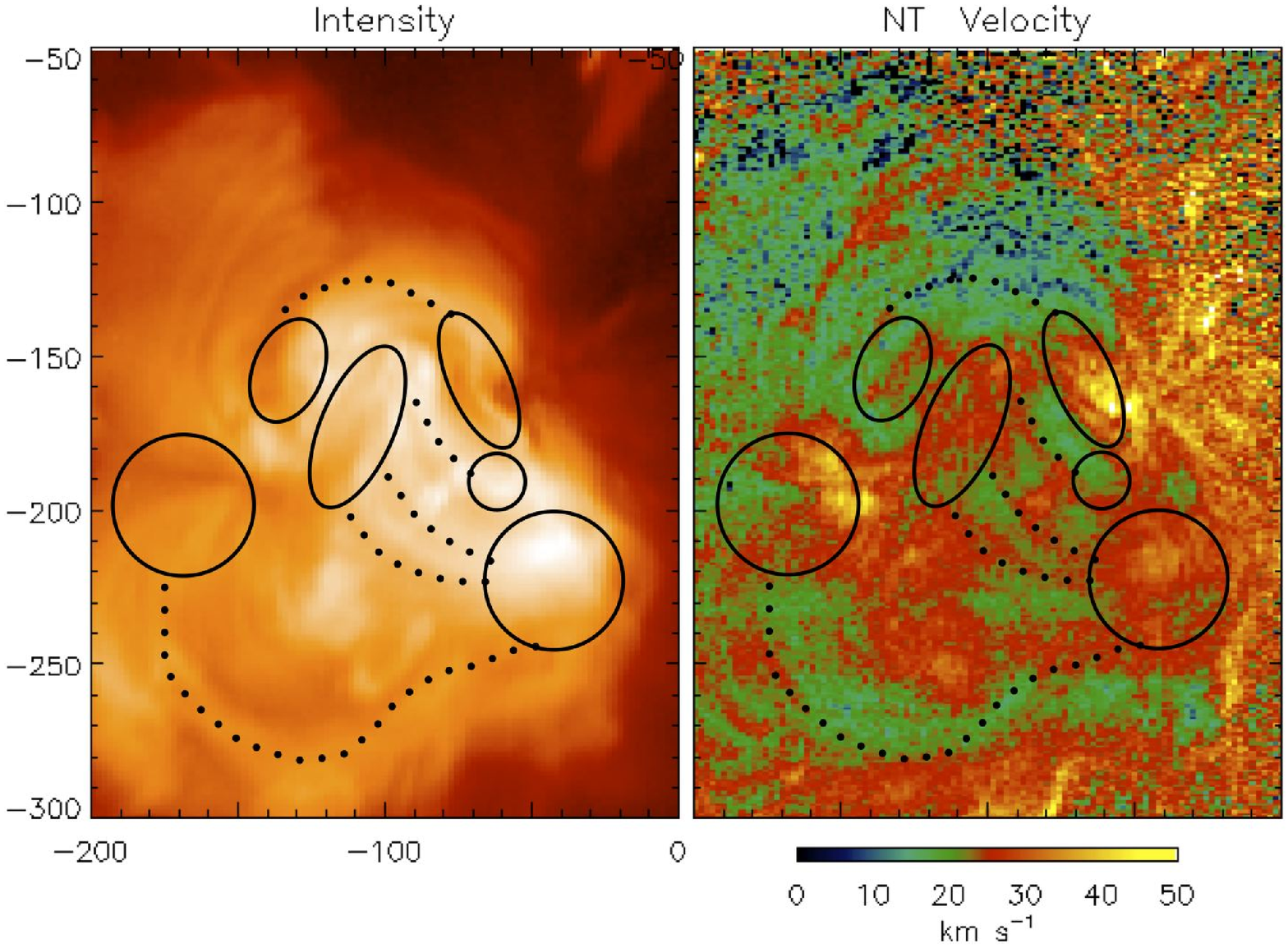}
\includegraphics[width=0.5\textwidth]{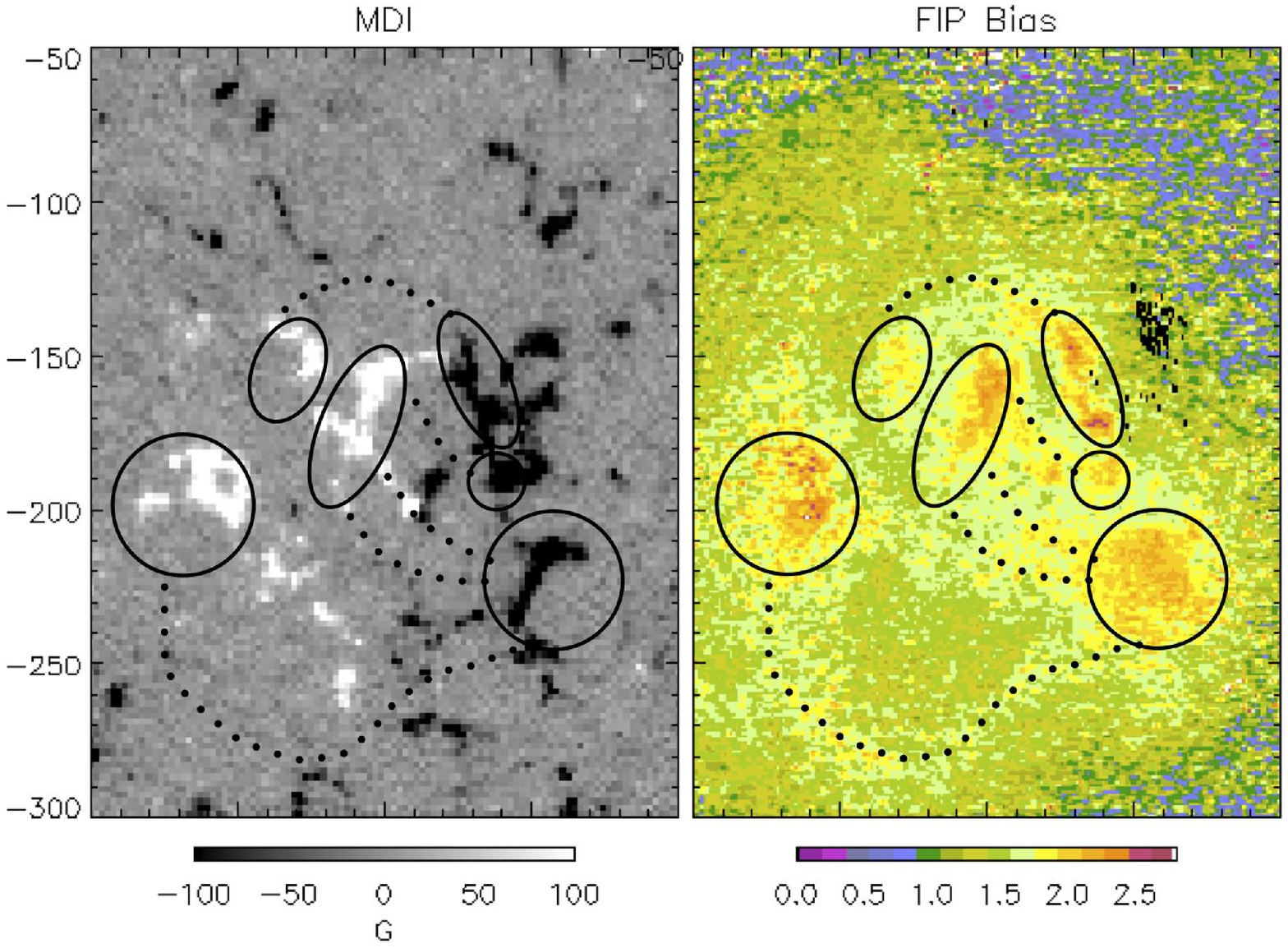}}
\caption{Maps
of intensity, non-thermal velocity, photospheric magnetic field and coronal FIP bias derived for
a sigmoidal anemone active region observed by EIS/Hinode. The overlaid black ellipses show
the footpoints of loops, coincident with strong photospheric magnetic field concentrations, while the dotted black lines show the loop connections between them. The
strongest FIP effect can be seen at the loop footpoints, also coincident with strong nonthermal
velocities. Figure from \citet{baker13}. Reproduced by permission of the AAS.\label{fig:baker}}
\end{figure}}

The solar slow speed wind helium abundance has long been known to be depleted from the photospheric
value, and this depletion is now established to vary with solar slow wind speed and the phase of the solar activity cycle \citep{aellig01,kasper07} in a similar manner to that mentioned above
\citep{lepri13}.
Figure~\ref{fig:kasper} \citep[left panel, from][]{kasper12} shows the He/H abundance ratio measured by the Wind spacecraft
over 1.5 solar cycles, in bins of different solar wind speed. Greater variability is seen in
the slowest wind speed bins, and in the slowest bin, the variability matches the smoothed sunspot
number. The right panel of Figure~\ref{fig:kasper}
show a similar study with ACE by \citet{rakowski12}, where the abundance ratio He/O is plotted. Broadly similar behavior with solar wind speed is seen, but the solar cycle dependence is less pronounced, but still apparent.
The fast solar wind helium abundance is also depleted, but to a lesser degree than the slow wind, and is also less variable. Measurements of the helium abundance in coronal holes \citep{laming03} and
in the quiet solar corona \citep{laming01} generally show similar values to those
measured in the solar wind, indicating that the helium depletion occurs lower down
in the solar atmosphere than the corona. A spectroscopic measurement in a solar flare
\citep{feldman05} showed a much higher abundance of helium, in qualitative agreement
with the by now routine observation \textit{in situ} of enhanced (i.e., less depleted)
helium abundance in CMEs \citep[e.g.,][]{wimmer06}.

\citet{widing01} studied the variation of the FIP effect with time in newly emerged active regions.
They found that the new loops emerged with photospheric abundances, and gradually changed to
coronal abundances, i.e., developed a FIP effect, over the course of a few days. Consequently a new
active region should have a weak FIP effect, while an older one would show strong fractionation. Recent observations by \citet{baker13} corroborate this view. Figure~\ref{fig:baker} shows maps
of intensity, non-thermal velocity, photospheric magnetic field and coronal FIP bias derived for
a sigmoidal anemone active region observed by EIS/Hinode. The overlaid black ellipses show
the footpoints of loops, coincident with strong photospheric magnetic field concentrations, while the dotted black lines show the loop connections between them. The
strongest FIP effect can be seen at the loop footpoints, also coincident with strong nonthermal
velocities. Weaker FIP effect is seen along the loop connections, suggesting that the FIP effect
originates in the chromosphere, and is communicated to the coronal loop by transport processes.
The active region seen here would then be considered relatively new, since only weak FIP effect is seen in its coronal connections.

\newpage

%===================================================================================

\section{Stellar FIP and Inverse FIP Effects}
\label{sec:stellar_FIP}

The measurement of element abundances in stellar coronae became possible for the
first time with the 1992 June 7 launch of the Extreme Ultraviolet Explorer (EUVE)
satellite. The use of grazing incidence gratings allowed strong lines in the EUV
spectra of stars to be resolved, allowing the acquisition of data on stellar corona of
similar quality to the early solar spectra analyzed by e.g., \citet{pottasch63}.
The Advanced Satellite for Cosmology and Astrophysics (ASCA) launched
on 1993 February 20 provided X-ray spectroscopy of stellar coronae with CCD-level
spectral resolution. Individual lines could not be resolved, but the He- and H-like
line complexes of different elements could. Other missions from which a few
results emerged were Ginga (launched on 1987 February 5), ROSAT (launched 1990 June 1)
and BeppoSAX (launched 1996 April 30) which carried proportional counters for
spectroscopy (gas scintillation proportional counters in the cases of Ginga and BeppoSAX)
which allowed measurements of the He-like Fe complex of lines with respect to the
surrounding continuum. More recently Suzaku (launch 2005 July 10) has also provided
CCD X-ray spectra, although the higher resolution calorimeter failed soon after launch.

The review of \citet{feldman00} covered
the status of this new field at a time when the results from EUVE and ASCA were
available, but before the launches of Chandra and XMM-Newton (launched 1999 July 23
and 1999 December 10, respectively). These satellites both carry grating instruments
as well as CCD cameras, and so high spectral resolution X-ray spectra of stellar
coronae are acquired with significantly higher throughput than was the case with
EUVE or ASCA. \citet{testa10} provided a more updated view of the field, encompassing a wider
variety of stellar targets. Here we will recap the discussion of \citet{feldman00}, and
the later reviews of \citet{favata03}, \citet{drake03} and \citet{testa10}, concentrating on stars of varying spectral type but in other respects similar
to the Sun, with a view to the theoretical discussions to follow in Sections~\ref{sec:ponderomotive} and \ref{sec:results}.

\begin{landscape}
\begin{table}[htbp]
\caption[List of FIP Bias Assessments for Main Sequence Stars]{List of FIP Bias Assessments for Main Sequence Stars \citep[updated from][]{wood12}.}
%Reproduced by permission of the AAS.}
\label{table:stellar_FIP}
\centering
{\small
\begin{tabular}{lcccccccc}
\toprule
Star& Alternate& Spectral& Radius& $\log L_X$& $\log F_X$& Ne/Fe& $F_{\mathrm{bias}}$& Ref.\\
 & Name& Type& ($R_{\odot}$)& (ergs s$^{-1}$)& ergs cm$^{-2}$ s$^{-1}$& ~ & ~ & \\
\midrule
$\pi ^3$ Ori& ... & F6V& 2.05& 28.96 & 5.55& 2.00& $-0.55$ & \citet{wood13}\\
$\beta$ Com &HD 114710 & G0 V    & 1.08 & 28.21 & 5.36 & 0.73 & $-0.668$ & \citet{telleschi05} \\
$\pi^1$ UMa &HD 72905  & G1 V    & 0.91 & 28.97 & 6.27 & 1.60 & $-0.645$ & \citet{telleschi05} \\
$\chi^1$ Ori&HD 39587  & G1 V    & 0.98 & 28.99 & 6.22 & 1.91 & $-0.555$ & \citet{telleschi05} \\
Sun         & ...      & G2 V    & 1.00 & 27.35 & 4.57 & 1.02 & $-0.600$ & \citet{feldman00} \\
$\alpha$ Cen A&HD 128620&G2 V    & 1.22 & 27.00 & 4.04 & 1.41 & $-0.410$ & \citet{raassen03a} \\
$\kappa$ Ceti&HD 20630 & G5 V    & 0.98 & 28.79 & 6.02 & 2.51 & $-0.462$ & \citet{telleschi05} \\
$\xi$ Boo A &HD 131156A& G8 V    & 0.83 & 28.86 & 6.24 & 3.80 & $-0.344$ & \citet{wood10} \\
70 Oph A    &HD 165341A& K0 V    & 0.85 & 28.27 & 5.63 & 2.40 & $-0.403$ & \citet{wood06} \\
36 Oph A    &HD 155886 & K1 V    & 0.69 & 28.10 & 5.64 & 4.90 & $-0.250$ & \citet{wood06} \\
36 Oph B    &HD 155885 & K1 V    & 0.59 & 27.96 & 5.63 & 3.80 & $-0.328$ & \citet{wood06} \\
$\alpha$ Cen B&HD 128621&K1 V    & 0.86 & 27.60 & 4.95 & 1.44 & $-0.478$ & \citet{raassen03a} \\
$\epsilon$ Eri&HD 22049& K2 V    & 0.78 & 28.32 & 5.75 & 4.68 & $-0.050$ & \citet{wood06} \\
$\xi$ Boo B &HD 131156B& K4 V    & 0.61 & 27.97 & 5.62 & 7.41 & $-0.185$ & \citet{wood10} \\
70 Oph B    &HD 165341B& K5 V    & 0.66 & 28.09 & 5.67 & 8.71 & $0.138$  & \citet{wood06} \\
GJ 338AB    & ... &M0 V+M0 V& 0.59+0.58 & 27.92 & 5.30 & ...  & $0.305$  & \citet{wood12} \\
EQ Peg A    & GJ 896A  & M3.5 V  & 0.35 & 28.71 & 6.84 &16.12 & $0.450$  & \citet{liefke08} \\
EV Lac      & GJ 873   & M3.5 V  & 0.30 & 28.99 & 7.25 &13.92 & $0.474$  & \citet{liefke08} \\
EQ Peg B    & GJ 896B  & M4.5 V  & 0.25 & 27.89 & 6.31 &14.67 & $0.417$  & \citet{liefke08} \\
AD Leo      & GJ 388   & M4.5 V  & 0.38 & 28.80 & 6.86 &18.25 & $0.536$  & \citet{liefke08} \\
Proxima Cen & GJ 551   & M5.5 V  & 0.15 & 27.22 & 6.08 &11.98 & $0.471$  & \citet{liefke08} \\
\midrule
\multicolumn{9}{l}{Very Active Star Sample ($\log L_X>29$)} \\
\midrule
EK Dra      &HD 129333 & G1.5 V  & 0.89 & 29.93 & 7.25 & 3.71 & $-0.277$ & \citet{telleschi05} \\
AB Dor      &HD 36705  & K0 V    & 0.79 & 30.06 & 7.48 &17.11 & $0.488$  & \citet{gudel01b} \\
AU Mic      &HD 197481 & M1 V    & 0.61 & 29.62 & 7.27 &24.30 & $0.695$  & \citet{liefke08} \\
\bottomrule
\end{tabular}}
%{References: (1) \citet{telleschi05} (2) \citet{feldman00} (3)
%  \citet{raassen03a} (4) \citet{wood10} (5)
%  \citet{wood06} (6) \citet{wood12} (7) \citet{liefke08} (8)
%  \citet{gudel01b} (9) \citet{wood13}.
\end{table}
\end{landscape}

\citet{wood10} conducted a survey of stellar FIP effects restricted to stars with X-ray
luminosities less than $10^{29}$ erg s$^{-1}$, with results shown in Figure~\ref{fig:woodlinsky}
(adapted from Figure~9 of their paper, also including points for $\alpha$Cen B, $\pi^3$ Ori and GJ338). By excluding the most active stars, a clear trend of decreasing stellar FIP effect with later spectral type is uncovered. The Sun is at the bottom
left of Figure~\ref{fig:woodlinsky}, with a logarithmic FIP bias $-0.6$ (expressed as
$\log\left(X/H\right)_{\mathrm{phot}} - \log\left(X/H\right)_{\mathrm{cor}}$, so that this is a coronal low
FIP enhancement of $10^{0.6}\simeq 4$), along with other stars of similar spectral type
\citep[$\pi^1$UMa (G1V) and
$\chi^1$Ori (G0V);][]{gudel02}.
The degree of fractionation diminishes as one moves to later spectral type,
becoming zero at about K5, and an inverse FIP effect is observed in the M stars. The magnetic
fields of a sunspot umbra and penumbra are likely to be similar to those found more ubiquitously
in the atmospheres of later type (i.e. M) stars \citep[see][]{donati09}, so the results
of \citet{feldman90} and \citet{phillips94} appear consistent with these stellar results.
At the extreme left hand side, the result for $\pi^3$ Ori suggests that the FIP effect saturates at a value similar to that found in the Sun, a low FIP enhancement of about 4, and does not continue increasing at spectral types earlier than this. The properties of the various stars are summarized in Table~\ref{table:stellar_FIP}. \citet{wood12} show that the sample of
T Tauri stars from \citet{gudel07} show a similar trend of Ne/Fe as in Figure~\ref{fig:woodlinsky}, but generally at higher FIP bias (i.e., stronger Inverse FIP effect).  Other more active stars (X-ray luminosities larger than $10^{29}$ erg s$^{-1}$) also show generally larger inverse FIP than those in Figure~\ref{fig:woodlinsky}. We do not consider these further because
of complications to the stellar physics introduced by fast rotation and tidal interactions in close binaries that are typical of these types of stars.

\epubtkImage{stellar_FIP}{%
\begin{figure}[htb]
\centerline{\includegraphics[scale=0.55]{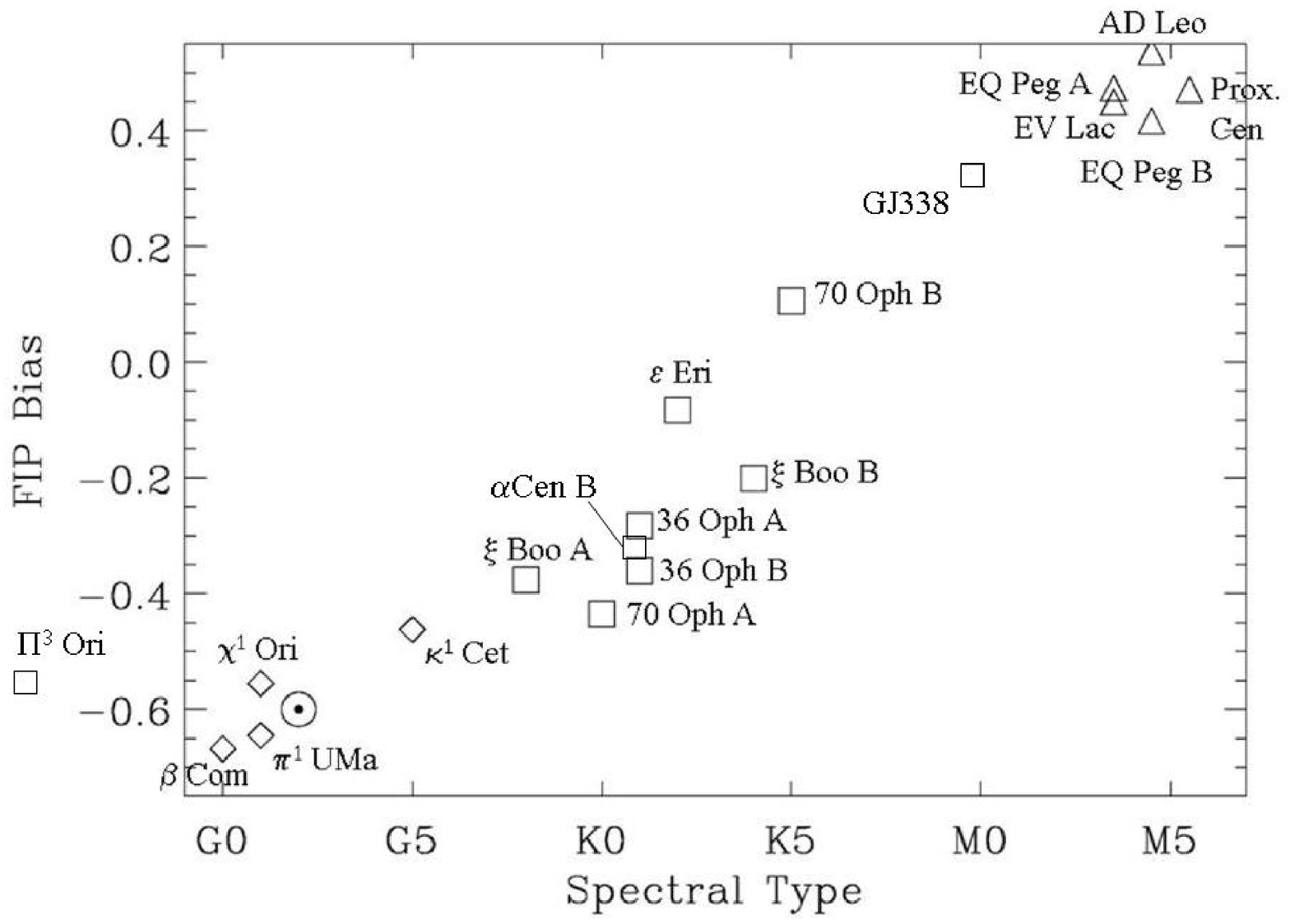}}
\caption{Survey of FIP fractionation ($\log
  \left(X/H\right)_{\mathrm{phot}}-\log\left(X/H\right)_{\mathrm{cor}}$)
  observed in a sample of dwarf stars, updated from \citet{wood10},
  where values below zero indicate a solar-like FIP effect and values
  above zero indicate an inverse FIP effect. Points for GJ338 and $\pi
  ^3$ Ori have been added from \citet{wood12} and \citet{wood13},
  respectively, and $\alpha$Cen B from \citet{drake97}. Diamonds
  indicate measurements from \citet{telleschi05}, triangles from
  \citet{liefke08}, with a solar value from \citet{feldman00}. For all
  GK stars the FIP bias calculations include corrections for stellar
  photospheric abundances from \citet{allende04}, but for the M stars
  there are in general no stellar photospheric measurements available
  so we have to simply assume solar photospheric abundance
  apply. Among these, EV Lac is the strongest case, due to
  chromospheric evaporation of photospheric abundance material
  observed during flares \citep[][and references
    therein]{laming09b}. For the purposes of this figure, we avoid
  extremes of stellar activity, confining our attention to stars with
  $\log L_X\le 29$. Reproduced by permission of the AAS.}
\label{fig:woodlinsky}
\end{figure}}

Anticipating a connection between the coronal abundance anomalies, and the
chromospheric wave field, arising either from waves produced in the corona and propagating down, or as a result of helioseismic or asteroseismic p-modes, we make the following comments about some of the important stars in Figure~\ref{fig:woodlinsky}.

\textbf{$\alpha$ Cen AB:} This inactive G2~V+K1~V binary is not in the
  original sample of \citet{wood10}, but it is of prime interest since its fundamental parameters
  (photospheric abundances, $p$-mode frequencies, mass, radius, surface
  gravity, and effective temperatures) are very well known for both
  stars \citep[e.g.,][]{bedding04,butler04,kjeldsen05,porto08,bruntt10,karoff07,chaplin09}.
  Coronal abundances were first measured by \citet{drake97} from EUVE
  data, yielding a FIP effect of about a factor of two in the
  unresolved binary.  Subsequent observations determined that $\alpha$
  Cen B is the dominant coronal source, which places it in its spot in Figure~1.
  There are many observations by
  both XMM and Chandra \citep[e.g.,][]{raassen03a,liefke06}, and
  model chromospheres of both stars have been developed \citep{ayres76,jordan87,vieytes09}.
  Fe XII 1242 \AA\ linewidths are available in \citet{ayres03}.

\textbf{$\epsilon$ Eri:} \citet{laming96} first measured coronal abundances for this moderately active
  K2~V star with EUVE.  More recent spectra from XMM and Chandra have been
  extensively analyzed as well \citep{sanz04,wood06,ness08}.
  \citet{drake93} give fundamental parameters \citep[see also Table~2 in][]{vieytes09}.
  Only theoretical estimates of $\epsilon$~Eri's $p$-mode spectra are available \citep{gai08}.
  Several sources provide model chromospheres \citep{jordan87,sim05,vieytes09}. \citet{ayres03} give the Fe XII 1242 \AA\ width from which coronal turbulence may be deduced.

\textbf{$\xi$ Boo A:}  This is a G8 V dwarf with coronal abundances first measured
  with EUVE by \citet{laming99} and \citet{drake01b}, and later with Chandra \citep{wood10}.
  Model chromospheres are available from \citet{kelch79} and \citet{jordan87}.

\textbf{70 Oph A:}  This  K0 V dwarf with an X-ray spectrum was analyzed by \citet{wood10}.
  Its fundamental parameters are given by \citet{bruntt10}, including $p$-mode oscillation
  frequencies \citep{carrier06,eggenberger08}. No empirical model chromosphere is available for this star, but there is a theoretical one from \citet{schmitz80}. Further guidance may come
  from the K dwarf model chromospheres of \citet{vieytes09}.
  It is also included in the survey of forbidden line observations by
  \citet{ayres03}.

\textbf{M Dwarfs:}  The cluster of M dwarfs at top right in Figure~1 are
  studied by \citet{liefke08}. A model chromosphere for AD Leo is given by \citet{fuhrmeister05}, and a grid of model chromospheres for M1 dwarfs is given by \citet{houdebine97}. \citet{ayres03} give linewidths for the Fe XII 1242~\AA\ and Fe XXI 1354~\AA\ forbidden lines. EV Lac is most recently studied by \citet{laming09b} using Suzaku observations before and during a flare. It exhibits abundance change during flares; the inverse FIP quiescent corona giving way to a more ``normal'' (i.e., solar photospheric) abundance pattern. This is interpreted as evidence of chromospheric evaporation, where unfractionated plasma is evaporated up into a flaring loop, lending more confidence to the inverse FIP interpretation.

\citet{favata03} and \citet{sanz04} caution that for several stars with apparent metal depletion
in their coronae, determined from the Fe/H ratio, the coronal abundances merely reflect metal poor photospheres. In Figure~\ref{fig:woodlinsky}, AD Leo maybe such a case \citep{jones96}, although it clearly shows
a nonsolar Ne/Fe abundance ratio. However metal
depleted coronae, or inverse FIP effect, do appear to clearly exist in
some cases, e.g.,
II Peg \citep[K2 IV plus an unseen companion;][]{huenemoerder01},
AR Lac \citep[G and K subgiants in a 1.98 day orbit;][]{huenemoerder03} and
AB Dor \citep[K2 IV--V with a 0.515 day spin period;][]{sanz03}.
Further, the variation of element
abundance during stellar flares, in which initially metal depleted
plasma evolves towards the standard composition, interpreted in
terms of the chromospheric evaporation of unfractionated plasma to
the coronal flare site, seems to require the existence of such abundance
anomalies. Such phenomena are observed
in HR 1099 \citep[K1 IV and G5 IV;][]{audard01},
Algol \citep[B8 V and K2 IV;][]{favata99}, and
UX Ari \citep[G5 V and K0 IV;][]{gudel99} where the later type subgiant is
taken to be the main source of coronal emission, as well as
AB Dor \citep{gudel01a},
YY Gem \citep[dMe and dMe with 0.814 day orbit;][]{gudel01b},
II Peg \citep{mewe97}, AT Mic \citep[dM4.5 and dM4.5;][]{raassen03b}, and EV Lac \citep[dMe3.5e;][]{laming09b}. Many of these more active stars show stronger Inverse FIP effects than shown in Figure~\ref{fig:woodlinsky}.

Not included in Figure~\ref{fig:woodlinsky}, but also
of interest for having neither a solar-like FIP effect nor an inverse FIP effect is Procyon. Coronal abundances measured with EUVE, Chandra and XMM \citep{drake95a,drake95b,raassen02,sanz04}. At spectral type F5 does not follow the trend of Figure~1 (it is, however, a subgiant, not a solar-like dwarf star). Extensive asteroseismological observations are available for Procyon \citep[e.g.,][]{mosser08,leccia07}. In fact, its $p$-mode lifetimes are known to be significantly shorter than those of the Sun \citep{bedding10}. Chromospheric models are available for us to use for this well-studied star \citep{ayres74,evans75,brown81}.

Another interesting case is $\tau$ Bootis \citep{maggio11}. It is a F7 V dwarf, with a M2 V companion. Following the assumption of \citet{maggio11} that the F7 V star is responsible for
the X-ray emission, it is also discrepant from the trend in Figure~\ref{fig:woodlinsky}, with
$F_{\mathrm{bias}}=-0.17$ at spectral type F7. $\tau$ Bootis A has already attracted interest because
it hosts a close-in giant planet \citep[$\tau$ Boo b: $P_{\mathrm{rot}}=3.31$ days, $M\sin i=3.9M_J$)][]{butler97}. A star-planet interaction has been noted, in that the planet appears to be
inducing an active region on the star, that leads the planet by $\sim 70^{\circ}$ in longitude
\citep{shkolnik08,walker08}, and this opens the possibility of the planet also affecting the coronal abundances, due to waves induced in the stellar atmosphere. Another possibility of course is that
the XMM observation is contaminated by emission from the M2 companion. Another curiosity is that
the photospheric abundances of $\tau$ Boo A do not agree with the trend reported by \citet{lopes13}, being of order 0.2 dex higher than solar values \citep{maggio11}. $\epsilon$ Eridani also hosts a Jupiter mass planet, but shows no coronal abundance discrepancy from the trend in Figure~\ref{fig:woodlinsky}.

\newpage
%===================================================================================

\section{Early Theoretical Models: Overview}
\label{sec:early_theory}

\subsection{Diffusion Models and Variations}
The earliest attempts to explain that FIP effect invoked
various processes like thermal and ambipolar diffusion (in a stationary atmosphere) or inefficient Coulomb drag (in a moving case). Thermal diffusion arises in a temperature gradient, with minor ions diffusing towards higher temperatures. This happens because collision cross sections
decrease with increasing collision energy, and so the force from the hotter particles from the
direction in which the temperature is rising is lower than than from the colder direction. This
obviously will select ions over neutrals, and accelerate them towards the corona. However such a
process is inherently slow \citep{henoux95,henoux98}, and requires static conditions for periods ranging from tens of hours \citep{hansteen94} to days or weeks \citep{killie07}. This appears
increasingly at odds with the modern view of the solar chromosphere as a dynamic environment,
continually being perturbed by the passage of shocks \citep{carlsson02} or reconnection at
chromospheric layers \citep{isobe08}. It also conflicts with the argument that in the absence of mass supply from below, the solar wind would empty the corona within 1-2 days. Clearly, any fractionation mechanism that takes longer than this to change coronal abundances cannot be right. Ambipolar diffusion refers to the diffusion of neutrals
along an ionization balance gradient, and must also be considered in such models, though by
itself does not appear capable of causing FIP fractionation. \citet{hansteen97} also
demonstrate that chromospheric mixing, in their case between hydrogen and helium, is also
necessary to prevent gravitational settling of helium in the chromosphere and to give a realistic abundance of helium in the solar wind.

In an attempt to speed up the fractionation process by thermal means, various authors have modeled the separation of elements entrained in a flow of neutral hydrogen and protons.
\citet{geiss89} considered such a flow driven across magnetic field lines, either a horizontal flow across vertical field lines, where gravity plays no role in the fractionation, or a vertical
flow driven across horizontal field lines, where it does. FIP fractionations matching observations reasonably well are achieved, but the models themselves must be considered highly idealized and unlikely to represent the real sun. No mechanism is suggested to
produce such a cross-field flow. In a treatment of fractionation in a (more plausible) vertical flow along vertical field lines, \citet{marsch95} solve diffusion equations for ions and
neutrals separately in the background flow. These equations only include photoionization
of neutrals. No recombination of ions and electrons is accounted for (their equations 6 and 9). At the lower boundary the gas is assumed completely
neutral, with density $n\left(s=0\right)=n_0$. The neutral density gradient is specified at the
upper boundary, taken to be where the gas is completely ionized, and is accelerated into the
solar wind. Here $dn/ds\left(s=S\right)=0$ and the ion density also $n^+\left(s=S\right)=0$.
Finally the ion density gradient at the lower boundary $dn^+/ds\left(s=0\right)=0$. The boundary
conditions on $n$ and $n^+$ are justifiable, and do not greatly affect the solutions. Different
choices can lead to the same diffusive fluxes. The choices of boundary conditions for the density
gradients are less clear, and these are crucial to obtaining FIP fractionation in the model.
In fact it is precisely the choice of $dn^+/ds\left(s=0\right)=0$ that specifies the ion flux
at the upper boundary, with value $n_0\sqrt{D/\tau }$ where $D=v_j^2/\nu _{jH}$, the diffusion
coefficient for neutrals of element $j$ in background of neutral H, in terms of its thermal speed
and collision frequency with H, and $\tau$ is photoionization time for neutral $j$ in the chromospheric radiation field. The fact that the ion flux at the top of the chromosphere depends ultimately
on the characteristics of neutrals at the bottom through the diffusion equation, and especially so for the FIP fractionated
low FIP elements that in reality have very small neutral fractions throughout the region of
interest, is indicative of problems. \citet{mckenzie98} and \citet{mckenzie00} point out that in a gas where collisions play the dominant role in coupling
species together, FIP fractionation cannot occur in a one dimensional steady state model. If all
elements enter at the lower boundary with the speed of H, and leave at the upper boundary with
at the proton speed, continuity demands that no fractionation occur. They also conclude that the
boundary condition on the density gradient at the lower boundary is the cause of the
fractionation in this model, and that such fractionation must have taken place {\em below the
lower boundary}. Indeed, equation 71 of \citet{marsch95} giving the upper boundary condition on
the exiting ion fluxes suggests that the FIP fractionation is already embodied in the inputs to the model, rendering the model explanation spurious.

Fractionation in such a class of models is not completely ruled out. If the chromosphere can respond to the enhanced coupling of ions to protons in the upper chromosphere (compared to that of neutrals) by allowing diffusion to supply ions from below at a greater rate, i.e. $dn/ds$ has to vary with species. But this reliance on diffusion places a limit on the speed with which abundance modification may take place.
There are some newer models \citep{pucci10} investigating the changes in abundances with hydrogen flux through a chromospheric magnetic funnel, but this requires significant fine tuning for each element \citep[][consider O, Ne, and Fe]{pucci10}, and appears unlikely to be able to get every element right at the same time,
even if the atmosphere was sufficiently quiescent
\citep[see also][]{byhring11b,byhring11a}. \citet{bo13} make the point that above the chromospheric temperature minimum, the atmosphere is convectively stable, but ignore the fact that waves from the convectively unstable regions may continue to propagate upwards to perturb higher altitudes, as in e.g. \citet{heggland11}.
They consider the effect of gravitational settling
within the chromosphere as a means of fractionation among O, Ne, S, and Fe. The mechanism only appears capable of producing a depletion of high FIP elements (no enhancement of
low FIPs as observed), and then only with a static atmosphere. An episodic upflow to supply the corona, punctuated by periods of stasis to allow the fractionation to occur.
Both \citet{pucci10} and \citet{bo13} use somewhat unrealistic chromospheric models,
in that the degree of hydrogen ionization for a given density is lower than that in \citet{avrett08}, at least for the models which come closest to matching observed abundance anomalies. Although the \citet{avrett08} model is quasi-static designed to match spectroscopic
observations, it matches quite well with average chromospheric structures modeled in
\citet{heggland11}, as does its antecedent \citep[][VALC]{vernazza81} compared with
\citet{carlsson02}, at least in terms of electron density.

The helium depletion explained by such processes \citep{byhring11a} is also at odds with the observation that it (and the other minor ions) flow faster than hydrogen in the solar wind at 1~AU \citep{neugebauer96,kasper08,bourouaine11}. A further stage of solar wind acceleration must set in at a level above that where the He depletion sets in \citep{wang08}. \citet{noci97}, following \citet{geiss70}, make the point that if such processes were the origin of the solar wind He abundance depletion, other heavy ions should also be depleted, which is now known not to be the
case. \citet{bochsler07b} revisits this, comparing He, O, and Ne
for which He appears most affected by gravitational settling, or inefficient Coulomb drag, in the corona. In this case the He abundance should correlate with the proton flux, which is
not observed \citep{wang08}. However all these elements appear also to vary with respect to H, leaving open the possibility that quasi-thermal or diffusive processes play some
role in establishing the absolute coronal abundances.

\subsection{Thermoelectric Driving}
Perhaps the best early model was that of \citet{antiochos94}. Cross B diffusion of chromospheric ions into a flux tube by a thermoelectric force associated with downward heat conducting electrons from a closed loop enhances the loop footpoint abundances of ions, but not neutrals. Given
$\nabla\times {\bf E}=0$ in steady state conditions, the transverse gradient of the longitudinal electric field (due to the plasma resistivity) must be balanced by a longitudinal gradient of the transverse electric field. This transverse electric field points into the flux tube, thus concentrating ions therein. The predicted fractionation pattern is proportional to ion mass$^{1/2}$ (it is independent of mass for the ponderomotive force). In a sense, this is conceptually very similar to the ponderomotive force model, in that it is the loop's response to coronal heating that causes the fractionation,
except here it is mediated by heat conduction rather than Alfv\'en waves. The main problem is that heat must be conducted down to regions of the chromosphere without transverse magnetic structuring (there must be plasma surrounding the flux tube), and this appears unlikely. It is unclear where in the chromosphere horizontal structuring sets in, but a natural place to expect it is likely to be the equipartition layer, where the Alfv\'en speed and sound speed are equal, roughly where the plasma $\beta \simeq 1$.
The stopping distance of 1 keV electrons is of order 1 km at a density of $10^{12}$ cm$^{-3}$, so downwards heat conduction is unlikely to reach this far, and other authors have recently considered energy transport by Alfv\'en waves \citep{fletcher08,haerendel09}. Alfv\'en waves may still fractionate
material in the upper magnetically structured layers of the chromosphere, because ions move vertically along the flux tube rather than horizontally across it. \citet{antiochos94} also does not allow for an ``Inverse FIP Effect''.  \citet{laming09b}
discuss how the ponderomotive force may also inhibit downwards heat conduction, but
large Alfv\'en wave amplitudes are required.

\subsection{Chromospheric Reconnection}
\citet{arge98} consider chromospheric reconnection giving ions a larger density scale height than neutrals. They model this using the Zeus code, iterating between its hydrodynamic and magnetohydrodynamic implementations to treat the neutrals and ions respectively, with these two fluids being coupled by collisions. They only calculate Si and Ne as examples, so it is not possible to evaluate the full fractionation pattern such a process would produce. The ion-neutral coupling rate they use appears to be that due to charge exchange collisions between protons and neutral hydrogen (their equation 6), and this appears to be applied to all elements, with possible modifications to the collision velocities. This is not correct, and the coupling should more properly be described by elastic scattering \cite[cf.][]{malyshkin11}. In fact, looking at these ion-neutral collision rates a mass dependent fractionation appears likely, not the FIP effect, which might be relevant to the observations of \citet{wurz00}.\epubtkFootnote{This is similar to the case in \citet{drake09}, where pick-up ion behavior leads to mass dependent fractionation.} By contrast \citet{laming04} and \citet{laming12} devoted considerable effort to the correct description
of these atomic processes. \citet{arge98} also predict FIP effect everywhere, with no difference between coronal holes and closed loops, unless extra assumptions are made about the presence or absence of chromospheric reconnection in open and closed field, and cannot reproduce
an ``Inverse FIP Effect''.

\subsection{Ion Cyclotron Wave Heating}
\citet{schwadron99} offered the first mechanism of fractionation by wave-particle interactions. Coronal ion cyclotron (IC) waves propagate down and heat chromospheric ions, not neutrals. The FIP effect arises from the preferential heating combined with the coupling of the chromospheric minor
ions to a background flow of H atoms and protons. The formalism for this coupling is elegant, and forms the basis for models invoking the ponderomotive force to be discussed at greater length elsewhere in this paper.

The fractionation pattern depends on the assumed spectrum of IC waves, since a resonant wave-particle interaction is assumed. This is not specified in their paper, except through the
wave interaction rate given as $\nu _{sw}=N_w\Omega _s$ where $\Omega _s$ is the gyrofrequency
of ions of element $s$, and $N_w$ is a constant of order 70. Equating this to the pitch diffusion
coefficient, $D=\left(\pi /4\right)\Omega _s\left(\delta B^2/B^2\right)$, where $\delta B$ is the
magnetic field perturbation due to the wave, and $B$ is the ambient magnetic field, we find
$\delta B/B\sim 10$.\footnote{Bohm diffusion corresponding to $N_w\sim 1$ would give $\delta B/B\sim 1$.} With $B\sim 10$G and a density of order $10^{10}$ cm$^{-3}$, the wave
velocity amplitude is 2000 km s$^{-1}$. Of course these estimates should be treated with caution,
because they are derived by applying quasi-linear theory well beyond its regime of validity, but
the essential point that the wave energy requirements of this model are implausible remains.
If the coronal ion cyclotron wave derived form a turbulent cascade, then even higher wave amplitudes should be found at lower frequencies, assuming for example a Kolmogorov spectrum. Loop
resonant frequencies are lower than ion cyclotron frequencies by several orders of magnitude (typically 6-8), leading to a huge increase in the intensity of these waves if a -5/3 spectrum is
assumed.
Further, waves in the ion cyclotron frequency range rapidly damp by charge exchange reactions
\citep[e.g.][]{kulsrud69}, with rate 1 - 100 s$^{-1}$, depending on the neutral fraction. Thus waves traveling at the chromospheric Alfv\'en speed are damped after traveling a few km, increasing the energy requirements still further.

\subsection{Summary}
To summarize, while some of the models described above might come close to describing the solar
FIP effect, they do so for rather contrived magnetic field geometries or for various other
extreme assumptions. In all cases except those of \citet{antiochos94} and \citet{schwadron99}, whose mechanisms both derive from the coronal response to energy deposition, no natural account of the difference between closed and open field is given. Further, none of the models appear capable of explaining the Inverse FIP Effect.

Given the relative success of \citet{antiochos94} and \citet{schwadron99}, it seems reasonable
to pursue the byproducts of coronal heating as the key to the fractionation.
Given the high frequency waves do not work due their energy requirements, and that lower frequency
Alfv\'en waves carry far more of the wave energy, the next question to ask is how such waves
could interact with chromospheric ions. Clearly a resonant interaction is not possible, but
such waves can interact nonresonantly through the ponderomotive force. This is discussed more
fully in the following sections.

\newpage

%===================================================================================

\section{The Ponderomotive Force Model}
\label{sec:ponderomotive}

\subsection{Overview}

We now turn to a description of our model of the FIP effect, which invokes the ponderomotive
force due to magneto-hydrodynamic waves as the agent that separates ions from neutrals.
Ponderomotive forces in magnetospheric and space plasmas are reviewed by \citet{lundin06}.
According to them, ``Ponderomotive forces are time-averaged nonlinear forces acting on media in
the presence of oscillating electromagnetic fields. The word \textit{ponderomotive} comes from
the Latin words \textit{pondus (ponderis)} meaning ``\textit{heaviness}'' and \textit{motor}.'' The complex
dynamics of a system acting under Lorentz forces may be considerably simplified by averaging over
the period of oscillations and described instead by ponderomotive forces.
They can be seen to arise from the effects of wave refraction
in an inhomogeneous plasma. In a nonmagnetic plasma, the refractive
index, $\sqrt{\epsilon}$, is given by $\epsilon=1-\omega_p^2/\omega^2$ where $\omega_p$ is the plasma frequency. Waves are refracted
to high refractive index, which means low plasma density. The
increased wave pressure can then expel even more plasma from the low
density region, leading to ducting instabilities. In magnetic plasma, $\epsilon=1-\omega_p^2/\left(\omega ^2-\Omega ^2\right)$ for linearly polarized parallel propagating
transverse waves, where $\Omega$ is the ion
cyclotron frequency. Thus waves with $\omega << \Omega$ refract to high density regions, and
plasma is attracted to regions of high wave energy density, especially when the
energy and momentum of the waves are large. A simple
expression for the ponderomotive force on an ion is derived below.

The refraction of waves to high density regions, and the corresponding attraction of
ions to locations of high wave energy density when the wave pressure dominates over the thermal pressure of the ionized component of the plasma, means that the FIP effect depends crucially
on details of wave propagation through the chromosphere. A high wave energy density in the
corona will lead to a strong FIP effect. Weak coronal waves and strong waves lower down
may lead to an inverse FIP effect, as chromospheric ions are attracted downwards. We will
explore precise conditions under which these various phenomena may occur in more detail below.

We concentrate solely on the fractionation, referring the reader to the reviews of \citet{reale14} for the properties of plasma in closed coronal loops, and to \citet{marsch06}, \citet{cranmer09} and \citet{ofman10} for the physics of the acceleration of the fractionated gas into the fast and slow speed solar wind. In the subsections that follow, we discuss in detail the various components of the model.
Figure \ref{fig:loopmodel} illustrates the basic scenario. We consider a coronal loop meeting the chromosphere
at both footpoints. For the purposes of calculating the Alfv\'en wave propagation, we ignore the
loop curvature. A variety of wave processes can occur at each footpoint, some illustrated at one and some illustrated at the other in the figure for clarity. A steady evaporative flow with speed much less than the local Alfv\'en speed is assumed to take material from the chromosphere to the corona. Such a flow arises from the chromospheric response to coronal heating, as heat is conducted downwards increasing the chromospheric temperature \citep[e.g][]{bray91}. \citet{imada12} calculate evaporative upflows for a variety of coronal heating scenarios, and find upflows of order one to a few km s$^{-1}$ in the chromosphere, similar to previous work \citep{warren02,bray91}. We ignore transient effects, although in a magnetic filament heated
episodically, these could play an important role. For instance, if Alfv\'en waves released by
a coronal heating event reach the chromosphere before the heat conduction front, the chromospheric flow velocity under which the fractionation occurs will be much smaller than these estimates. In the opposite case, of heat conducting down before the Alfv\'en waves arrive, no
fractionation could occur. More exotic cases where the Alfv\'en waves themselves carry the
heat that causes the evaporation, for example in solar flares \citep{haerendel09} are not
considered here.

In the absence of
the ponderomotive force, we assume that the chromosphere is unfractionated, i.e. that sufficient turbulence exists to inhibit any gravitational settling of other forms of diffusion that might
occur. A similar approach was taken by \citet{hansteen97}. We suggest that the reflection and refraction of acoustic waves at discontinuities associated with chromospheric shock waves set
up the conditions necessary for a turbulence. Interactions between oppositely directed
waves lead to fluctuations of smaller and smaller size scales, right down to microscopic dimensions where mixing can occur. Following on from the discussion above, we neglect diffusion
processes. Although the thermal force is expected to develop over a similar region to the
ponderomotive force (the region of strong density and temperature gradients in the upper
chromosphere), the acceleration associated with the thermal force is of order 1\% - 10\% of
the ponderomotive acceleration, and is negligible. The thermal acceleration is also dependent
on $1/A$, where $A$ is the element atomic mass, unlike the ponderomotive acceleration which is
mass independent, and can never produce an inverse fractionation as seen for example in the coronae of M dwarfs.

\epubtkImage{f1}{%
\begin{figure}[htbp]
\centerline{\includegraphics[scale=1.0]{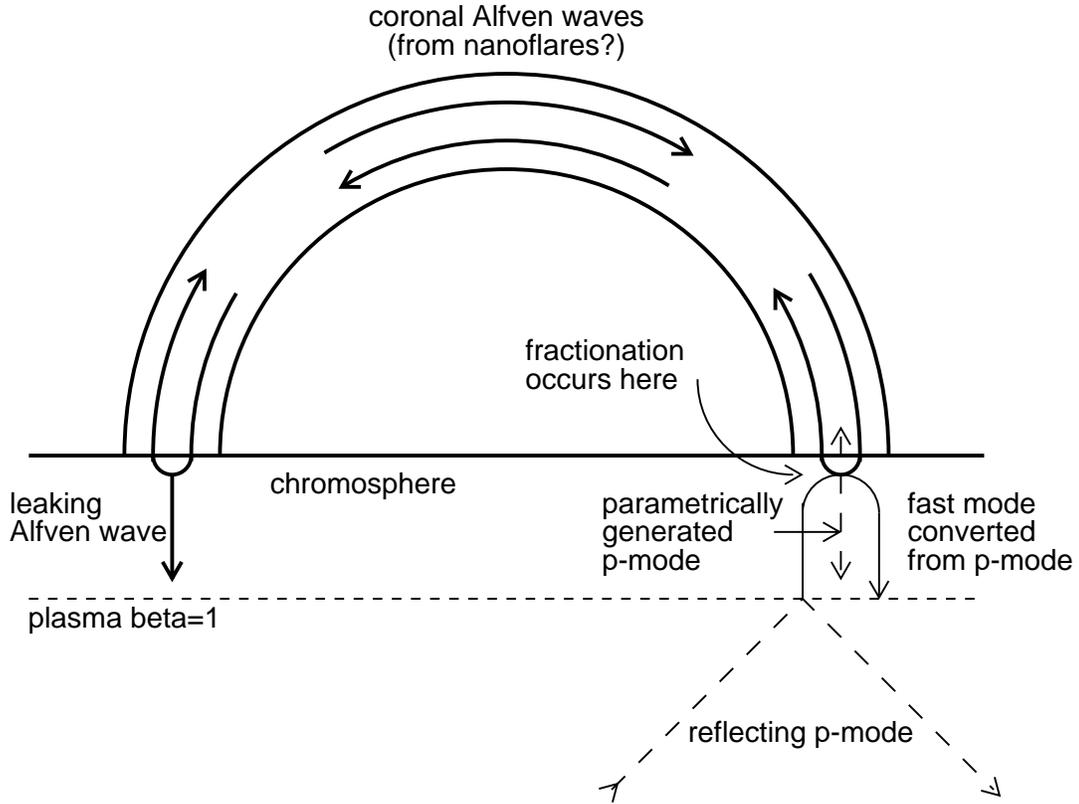}}
\caption {Schematic diagram of model loop and wave processes, adapted from \citep{laming12}, which follows \citet{hollweg84}. All footpoint wave processes may happen at both footpoints, but are
here split between the two for clarity. Alfv\'en waves shown as thick solid lines are assumed to be generated inside the coronal portion of the loop, and to bounce back and forth from the loop
footpoints, with a probability of leaking out and being transmitted deeper into the chromosphere at each bounce (shown on left hand side). Approximately equal amplitudes of waves propagating in each direction result. Reflecting Alfv\'en waves can also
generate slow mode (here labelled as ``p-mode'') waves by a parametric process (shown on the right hand side as thin dashed lines).
Other acoustic waves (``p-modes'' in helioseismology parlance) propagating within the solar envelope can mode convert to fast mode waves upon reaching the chromospheric layer where the sound speed and Alfv\'en speed are equal (approximately where the plasma $\beta =1$). The upgoing fast mode waves (shown as thin solid lines) are refracted back downwards in the chromospheric region where the Alfv\'en speed increases with height.
Fast mode wave may also mode convert to Alfv\'en waves and then propagate up to the loop to be reflected or transmitted, depending on the match between their frequency and the loop resonance.
This provides an alternative means of seeding the loop with propagating Alfv\'en waves. Our non-WKB wave propagation calculations in fact follow an Alfv\'en wave being injected at one loop
footpoint, being reflected or transmitted into the corona, and the successive bounces it undergoes. For waves at the loop resonance, this behaviour (without growth or damping included)
is indistinguishable from waves
generated within the loop itself, necessarily at the loop resonance or its harmonics.
Reproduced by permission of the AAS.\label{fig:loopmodel}}
\end{figure}}

\subsection{The ponderomotive force}

An expression for the ponderomotive force is derived as follows, updated from Appendix~A in \citet{laming09}.
Consider the Lagrangian for a system of $n$ particles with mass $m$ and electromagnetic waves
in a box of unit volume
\begin{equation}
L=\sum_i{1\over 2}m_i\left({\bf v_{th,i}}+\delta {\bf v_i}\right)^2
+\sum_i {q_i\over c}\left({\bf v}_{th,i}+\delta {\bf v}_i\right)\cdot
\delta {\bf A} +
{\epsilon\delta {\bf E}^2-\delta {\bf B}^2\over 8\pi}
\end{equation}
where ${\bf v_{th,i}}$ is the thermal speed and $\delta{\bf v_i}$ is the
oscillatory speed induced by the wave of particle $i$, with mass $m_i$,
and charge $q_i$. Wave electric and magnetic fields are given by $\delta {\bf E}$ and
$\delta {\bf B}$, respectively, $\delta {\bf A}$ is the wave vector potential,
and $c$ is the speed of light. We have chosen the radiation gauge where the electrostatic
potential $\phi =0$. In any case, this is constant in an electrically neutral plasma (in the absence of electrostatic waves).

For Alfv\'en waves, energy is partitioned according to
$\delta B^2/8\pi = \sum_imv_{\mathrm{osc},i}^2/2 +\delta E^2/8\pi$. Also $\delta {\bf B}\cdot{\bf B}=0$ where ${\bf B}$ is the ambient magnetic field, and $\delta {\bf v}_i\cdot \delta {\bf A}=0$, not just on average but for all time. We also take the
time average of ${\bf v}_{th,i}\cdot\delta {\bf v}_i =0$ and ${\bf v}_{th,i}\cdot\delta {\bf A} =0$ to find the Lagrangian
\begin{equation}
L=\sum_i{1\over2}m v_{th,i}^2
+{\left(\epsilon-1\right)\delta E^2\over 8\pi}\rightarrow\sum_i{1\over2}m v_{th,i}^2+\sum
_i{q_i^2\delta E\left(z_i\right)^2\over 2m_i\left(\Omega_i^2-\omega ^2\right)},
\end{equation}
where in the second line we have written $\epsilon -1 = \omega_p^2/\left(\Omega ^2-\omega ^2
\right)= 4\pi nq^2/m\left(\Omega ^2-\omega ^2\right)$ and $\left(\epsilon -1\right)\delta E^2
\rightarrow \Sigma_i 4\pi q_i^2\delta E\left(z_i\right)^2/m_i
\left(\Omega_i^2-\omega ^2\right)$ for linearly polarized parallel propagating transverse waves.
The ``$z$'' Euler--Lagrange equation for particle ``i'' gives
\begin{equation}
{d\over dt}\left(mv_{th,iz}\right)= {q_i^2\over 2m_i}
{d\over dz}{\delta E^2\over\left(\Omega_i^2-\omega ^2\right)},
\end{equation}
evaluating for the component of $v_{th,i}$ orthogonal to $\delta{\bf A}$ and $\delta{\bf v}_i$.
In uniform magnetic field, this is
the same as the expression derived by \citet{landauECM},
and agrees with earlier work \citep[e.g][]{lee83,li93} if
$\delta E^2=\delta E_p^2/2$, where $\delta E_p$ is the peak electric field
in the wave, giving a ponderomotive force
\begin{equation}
F_i={q_i^2\over 4m_i\left(\Omega_i^2-\omega ^2\right)}{d\delta E_p\left(z_i\right)^2\over
dz}.
\end{equation}
When $\omega \ll \Omega_i$, the ponderomotive acceleration is thus {\em
independent of ion mass}, which is one crucial property relevant to
obtaining an almost mass independent fractionation as observed. It
is also independent of ion charge, so long as the ion is charged
(and not neutral). \citet{litwin98} give a similar expression
derived from the ${\bf j}\times {\bf B}$ and other second order terms in the MHD momentum
equation. Away from the low frequency limit, circular polarized waves may give slightly
different forces, with left and right circular polarization being different from each other as well
as linear polarization \citep[e.g.,][]{nekrasov13}.
In nonuniform ${\bf B}$ in the low frequency limit, the ponderomotive force is given by
\begin{equation}
F_i={m_ic^2\over 4}{d\over dz}\left[{\delta E_p\left(z_i\right)^2\over B\left(z_i\right)^2}\right],
\label{eq:pondforce}
\end{equation}
which agrees with the first two terms in Equation~2.6 of \citet{lundin06}.

In fast mode waves, energy is now partitioned according to
$\delta B^2/8\pi +\delta E_{th}= \sum_imv_{\mathrm{osc},i}^2/2 +\delta E^2/8\pi$, where $\delta E_{th}$ is the wave thermal energy arising from compression and rarefaction. For the fast mode wave $\delta {\bf B}\cdot{\bf B}=0$ only on average,
but not instantaneously, though $\delta {\bf v}_i\cdot \delta {\bf A}=0$ remains unchanged. The
thermal energy also time averages to zero, leaving the Lagrangian and ponderomotive force described
by the same expressions as above. However the extra longitudinal pressure associated with obliquely
propagating fast mode waves will reduce the eventual fractionation. See Sections~\ref{subsec:6.5} and \ref{subsec:6.6} below.

With slow mode waves, energy partitioning is similar to that for fast mode waves. Whereas Alfv\'en and fast mode waves always have $\delta {\bf v}\perp {\bf B}$, slow modes have $\delta {\bf v}\Vert {\bf k}$. In parallel propagation the electric field, and hence a ponderomotive force capable of ion-neutral separation, are absent.
Transverse electric field may arise with a slow mode wave in parallel propagation along an
isolated flux tube (a ``sausage'' mode) \citep[e.g.,][]{mikhalyaev05}. We do not consider such
possibilities further.

\subsection{Chromospheric model}
\label{subsec:chromodel}

The ponderomotive force depends on the gradient of the wave transverse electric energy,
$\partial\delta E_{\perp}^2/\partial z$. Since $\delta E_{\perp} =\delta v_{\perp}B/c$,
such a gradient will develop where $\delta v_{\perp}$ varies
in response to varying chromospheric density. Previous \citep[mainly analytic; e.g.][]{hollweg84}
works on wave propagation through the chromosphere have approximated the vertical density structure as exponential, with a scale height of order 200~km. While adequate for studying wave transmission
into the corona, more detail is required for the FIP effect. The chromosphere is heated, presumably by means similar to those thought to heat the corona, and cooled in the region where most FIP fractionation occurs principally by radiation
in H Lyman $\alpha$. In this region where H starts to become ionized, its radiative cooling becomes increasingly
inhibited and the temperature must rise. Correspondingly, the density drops, and the density gradient
so produced is steeper than the typical hydrostatic scale height elsewhere. Consequently, the region of the chromosphere where the FIP fractionation has long been assumed
to occur, i.e. the region where H becomes ionized, is highly likely to be a region with a strong density gradient due to the physics of radiative cooling, and hence a location where the ponderomotive force will be strongest. Figure~\ref{fig:chromodel} shows
this structure in the left panel, taken from the empirical model (C7) of \citet{avrett08}, an
update to the older VALC model \citep{vernazza81}.

\epubtkImage{f8}{%
\begin{figure}[htbp]
\centerline{\includegraphics[scale=0.4]{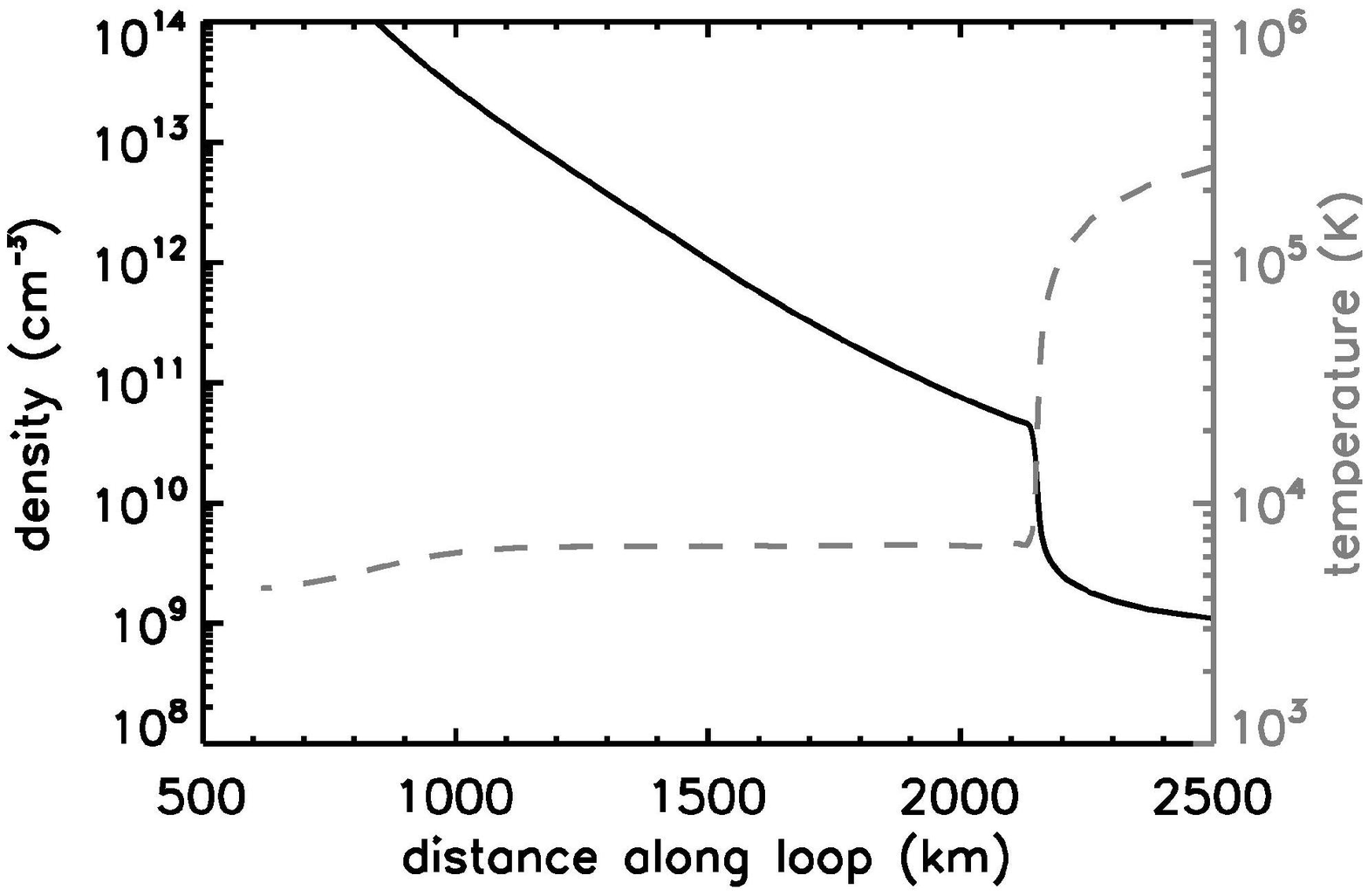}\includegraphics[scale=0.4]{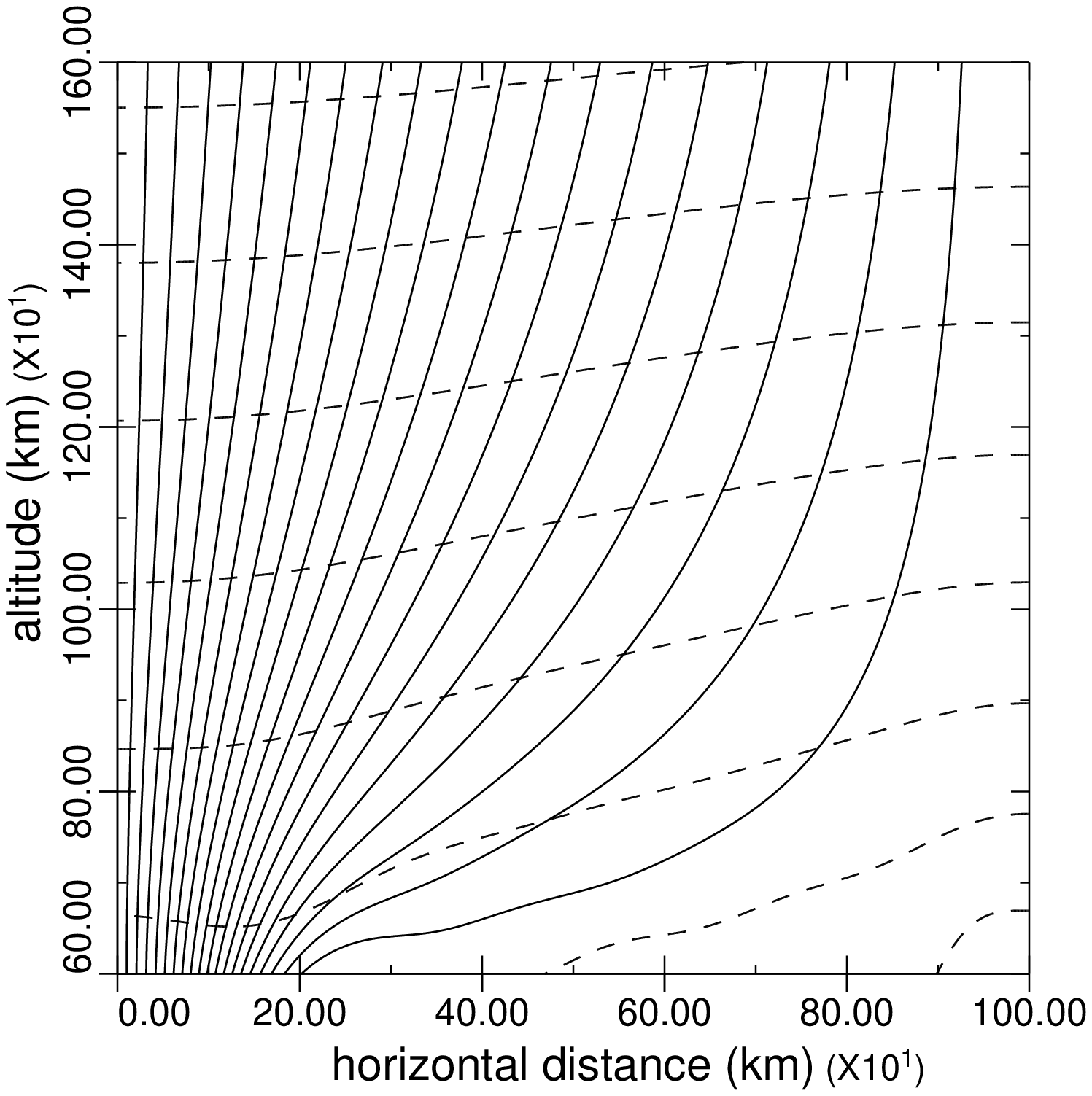}}
\caption{\emph{Left:} Empirical model of the solar chromosphere plotted from data given in \citet{avrett08}. The density is the solid black line, to be read on the left axis, while the temperature is the dashed gray line, to be read on the right axis. The ordinate (``distance along the loop'') is the altitude above the photosphere. The steep density gradient at 2150~km altitude is where strong ponderomotive force can develop. \emph{Right:} Force-free magnetic field structure calculated in \citet{rakowski12} following \citet{athay81}. Altitude is given by the $y$-axis. The $x$-axis give lateral expansion. The dashed lines show contours of the Alfv\'en speed. In this plot the plasma $\beta = 1$ layer is at an altitude of approximately 650~km in the Sun. Reproduced by permission of the AAS. \label{fig:chromodel}}
\end{figure}}

The right panel of Figure~\ref{fig:chromodel} shows the model magnetic field. Altitude here is given by the $y$-axis. The $x$-axis gives lateral expansion
on the same scale. The dashed lines show contours of the Alfv\'en speed. The magnetic field emerges from the photosphere
in tight fibrils. At the atmospheric layer, typically in the low chromosphere, where gas pressure
and magnetic pressure are equal, the field begins to expand and fill the whole volume. In
Figure~\ref{fig:chromodel} (right panel) this corresponds to $y=650$ km approximately. In the real
Sun, this occurs at altitudes typically between 400 and 800~km, depending on the magnetic field
strength. The total magnetic field expansion from $y=600$ km upwards in this model is a factor of 5, similar to that suggested by \citet{gary01}.

Observations of hard X-ray bremstrahlung during flares with RHESSI \citep{lin04} corroborate
many of these features \citep{kontar08,saint-hilaire10}. An average scale density scale height of
130\,--\,140~km is derived. \citet{kontar08} infer chromospheric flux tube expanding in radius by a
factor of 3 at an altitude between 900~km and 1200~km. This would imply magnetic field decreasing
by a factor of 9, albeit at a higher height than suggested above.

\citet{avrett08} also give an empirical electron density coming from the ionization balance for H, from which we calculate the ionization balance for all other elements of interest. Collisional
processes (ionization, radiative and dielectronic recombination) are all taken from the tabulations
provided by \citet{mazzotta98}. Subsequent refinements to the ionization and dielectronic recombination rates as implemented by \citet{bryans09} are not crucial here, since our concern is
only with neutrals and singly ionized species, but the suppression of dielectronic recombination
by finite density effects \citep{nikolic13} is important and is now included as an update to
previous calculations \citep{laming09,laming12}. We also include charge exchange ionization using
rates given by \citet{kingdon96}, updated and amended as detailed in \citet{laming12}. This treats the capture by a free proton of an electron from an initially neutral atom. It is most important for O, and helps lock the O ionization balance to that of H, but can be of significance for other low FIP elements as well. In the
\citet{avrett08} model, H begins to
become ionized at heights between 1000\,--\,1500~km, reaching about 50\% ionization at 2000~km. For
comparison, \citet{saint-hilaire10} infer a unit step change in ionization level at a height of
$1.3\pm 0.2$ Mm.

We evaluate photoionization rates using incident spectra based on
\citet{vernazza78} with the extensions and modifications outlined in
\citet{laming04}. In most cases, the ``quiet region'' spectrum
is used. An additional contribution from trapped Lyman $\alpha$ in regions where this line is
optically thick is also included, with approximate flux
$\int n_{\mathrm{H}}\,dlAC_{ex}/\left(
A+\tau C_{de-ex}\right)$ where $\int n_{\mathrm{H}}\,dl$ is the column density of H atoms, $\tau =\sigma\int n_{\mathrm{H}}\,dl$ is the opacity in terms of the absorption cross section $\sigma$ at line centre, $A$ is the upper
level decay rate, and $C_{ex}$ and $C_{de-ex}$ are the collisional excitation and de-excitation rates for the upper level. We take photoionization cross sections from the compilation of \cite{verner96}. Our ionization balance so computed is necessarily a steady state model, but is based on the empirical electron density found by \citet{avrett08}, which is higher than that which
we would compute based on their inferred chromospheric temperature.
\citet{carlsson02} study the effects of dynamic ionization of H, and find that the passage of shock
waves through the chromosphere does indeed increase the electron density over that coming from
simple photoionization-recombination equilibrium for H. Individual photoionization and recombination rates however are too slow for the ionization balance to respond to individual shock waves. The
elevated electron density instead represents a time averaged response to the dynamics, and is
itself quasi-steady state. \citet{wedemeyer11} consider the Ca$^+$ to Ca$^{2+}$ ionization balance.
It is more variable than that between H and H$^+$, but is of less concern to us, since the
ponderomotive force experienced by an ion is independent of ion charge, so long as the wave
frequency is much lower than the ion gyrofrequency.

In isolation, the action of the ponderomotive force would produce a local abundance variation
close to the region where it is strongest, around chromospheric altitudes of 2150~km. A region where
low FIPs are depleted by the upwards acceleration would sit below this altitude, and above it a region of FIP enhancement. A coronal
abundance anomaly would not result without either a flow through the chromosphere, or diffusion (in
a static case). The typical timescales associated with such processes would be $H_D/v\sim
2\times 10^7/10^3\sim 2\times 10^4$ s where
$H_D$ is the density scale height and $v$ the flow velocity lower down in the chromosphere, or $H_D^2/D\sim 4\times 10^{14}/6\times 10^9\sim 6\times 10^4$ s where $D$ is the diffusion coefficient for low FIP ions in a neutral H atmosphere. Both timescales
evaluate to several hours to days, similar to that observed \citep{widing01}, as do similar timescales evaluated for the coronal loop. The chromospheric response may be altered by the mixing
induced by turbulence associated with shocks in the lower chromosphere \citep[e.g.,][]{reardon08},
and so the coronal diffusion is likely to be
the controlling timescale. We do not consider such processes further, restricting
ourselves to a steady state chromosphere and upward flow speed, and emphasizing again that collisional coupling in the chromosphere
is crucial to the coronal abundance anomaly, since it is the source for the extra low FIP ions.
Once collisionless (in the solar wind) no further bulk fractionation can occur. A local FIP enhancement must be accompanied by another local depletion, since the collisional coupling to the
lower solar atmosphere has been lost.

\subsection{The Alfv\'en wave transport equations}
\label{subsec:6.4}
Alfv\'en wave propagation is often treated in the Wentzel-Kramers-Brillouin (WKB) approximation,
where it is assumed that the wavelength is much shorter than the typical length scale of
inhomogeneities, and hence that the effects of reflection and refraction can be neglected. This
is emphatically not the case in the solar chromosphere. Since
the ponderomotive force depends on the spatial variation of the Alfv\'en wave electric field in the
inhomogeneous plasma of the chromosphere, a direct integration of the transport equations is required. We evaluate this spatial variation paraphrasing and updating the treatment in Appendix~B of \citet{laming09}.
We start from the linearized MHD force and induction equations,
\begin{equation}
\rho{\partial\delta {\bf v}\over\partial t}+ \rho\left({\bf
u}\cdot\nabla\right)\delta {\bf v}+\left(\delta {\bf
v}\cdot\nabla\right)\rho{\bf u}
={\left(\nabla\times\delta{\bf
B}\right)\times{\bf B}\over 4\pi}={\left({\bf
B}\cdot\nabla\right)\delta {\bf B}-\left(\nabla\delta{\bf
B}\right)\cdot {\bf B}\over 4\pi},
\label{eq:momentum}
\end{equation}
and
\begin{eqnarray}
{\partial\delta {\bf B}\over\partial t}&=&\nabla\times\left(\delta
{\bf v}\times {\bf B}\right) +\nabla\times\left({\bf u}\times\delta
{\bf B}\right)\nonumber\\ &=& -{\bf B}\nabla\cdot\delta {\bf v} -\left(\delta {\bf v}\cdot\nabla\right){\bf B}+
\left({\bf B}\cdot\nabla\right)\delta {\bf
v}-\delta {\bf B}\nabla\cdot {\bf u}-\left({\bf
u}\cdot\nabla\right)\delta {\bf B} +\left(\delta {\bf B}\cdot\nabla\right){\bf u}\,,
\label{eq:maxwell}
\end{eqnarray}
where ${\bf u}$ and ${\bf B}$ are the unperturbed velocity and
magnetic field, $\delta {\bf v}$ and $\delta {\bf B}$ are the
perturbations, and $\rho$ is the density.
Equation~(\ref{eq:momentum}) is rewritten using  $\left(\nabla\delta{\bf
B}\right)\cdot {\bf B}=\nabla\left({\bf B}\cdot\delta {\bf B}\right)-\left(\nabla {\bf
B}\right)\cdot \delta{\bf B}$ to yield
\begin{equation}
{\partial\delta {\bf v}\over\partial t}+\left({\bf
u}\cdot\nabla\right)\delta {\bf v}={\bf V}_A\cdot\nabla\left(\delta
{\bf B}\over\sqrt{4\pi\rho}\right)+{\delta {\bf
B}\over\sqrt{4\pi\rho}}{{\bf V}_A\cdot\nabla\rho\over 2\rho}+
{\left(\nabla {\bf B}\right)\cdot\delta {\bf B}\over
4\pi\rho}-{\left(\delta {\bf v}\cdot\nabla\right)\left(\rho{\bf
u}\right)\over\rho}\,,
\end{equation}
where ${\bf V}_A={\bf B}/\sqrt{4\pi\rho}$ is the Alfv\'en velocity, and we have assumed ${\bf B}\cdot\delta {\bf B}=0$ for Alfv\'en waves and parallel propagating fast modes.

Specializing to plane Alfv\'en waves, we write
$\left(\nabla {\bf B}\right)\cdot\delta {\bf
B}=\left(\partial B_x/\partial x\right)\delta {\bf
B}=-\left(\partial B_z/\partial z\right)\delta {\bf B}/2$ since
$\nabla\cdot {\bf B}=0$ (assuming $\partial B_x/\partial x=
\partial B_y/\partial y$ in cylindrical symmetry,
with $\delta {\bf B}=\delta B{\bf \hat{x}}$ where ${\bf \hat{x}}$ is a
unit vector along the x-axis). Similarly, $\left(\delta {\bf v}\cdot\nabla\right)\left(\rho {\bf
u}\right)=\delta v_x\partial\left(\rho u_x\right)/\partial x=-\delta v_x\partial\left(\rho u_z\right)/\partial z/2$ since $\nabla\cdot\left(\rho {\bf u}\right)=0$, and using $\partial\left(\rho u_z/B_z\right)/\partial z =0$ gives
\begin{equation}
{\partial\delta {\bf v}\over\partial t}+\left({\bf
u}\cdot\nabla\right)\delta {\bf v}={\bf V}_A\cdot\nabla\left(\delta
{\bf B}\over\sqrt{4\pi\rho}\right)+{\delta {\bf
B}\over\sqrt{4\pi\rho}}{V_A\over 2H_D}-{\delta {\bf
B}\over\sqrt{4\pi\rho}}{V_A\over 2H_B}+\delta {\bf v}{u\over 2H_B}\,.
\label{eq:10}
\end{equation}
Here $1/H_B=\partial\ln B_z/\partial z$, $1/H_D=\partial\ln\rho
/\partial z$, and below $1/H_A=\partial\ln V_A/\partial z$.

Similar
manipulations using $\nabla\cdot {\bf u}=-{\bf u}\cdot\nabla\rho /\rho = -u/H_D$ and
assuming $\nabla\cdot\delta {\bf v}=0$ give the induction equation in the form
\begin{equation}
{\partial\over\partial t}\left(\delta {\bf
B}\over\sqrt{4\pi\rho}\right)+\left({\bf u}\cdot\nabla\right){\delta
{\bf B}\over\sqrt{4\pi\rho}}=\left({\bf V}_A\cdot\nabla\right)\delta
{\bf v}+{\delta {\bf B}\over\sqrt{4\pi\rho}}{u\over 2H_D}+\delta
{\bf v}{V_A\over 2H_B}-{\delta {\bf B}\over\sqrt{4\pi\rho}}{u\over 2H_B}\,.
\label{eq:11}
\end{equation}
Taking Equation~(\ref{eq:10}) plus or minus Equation~(\ref{eq:11}) and rearranging
gives the final result,
\begin{equation} {\partial I_{\pm}\over\partial t}+\left(u\mp
V_A\right){\partial I_{\pm}\over\partial z}= \left(u\pm
V_A\right)\left({I_{\pm}\over 4H_D}+{I_{\mp}\over 2H_A}\right),
\label{eq:12}
\end{equation}
where $I_{\pm}=\delta {\bf v}\pm \delta {\bf B}/\sqrt{4\pi\rho}$,
representing waves propagating in the $\mp$ z-directions. We have focussed on plane polarized
Alfv\'en waves. Clearly circularly polarized Alfv\'en waves will obey the same transport
equations, as do torsional waves in cylindrical coordinates. These different wave polarizations
may still produce subtly different fractionations, due to their different couplings to other
wave modes. The case of parametric generation of slow mode waves is discussed below.

Such transport equations have been presented in several different forms by various authors.
\citet{cranmer05} review these in their Appendix B and demonstrate their equivalence. We have
assumed a cylindrically symmetrical magnetic field above. One could easily generalize this
treatment to incorporate Alfv\'en wave transport on arbitrary (i.e., observed or extrapolated)
magnetic field lines for a more realistic description of observations. We have also restricted
treatment to field aligned wave propagation, rendering the waves incompressible. For plane and
circularly polarized waves this amounts to assuming that the wave follows the field line on
which it propagates. For torsional waves on a flux tube, it requires that the flux tube magnetic
field have no $B_{\phi}$ component. In the case that this is not so, the torsional wave is
inevitably of mixed Alfv\'en/fast mode polarization and has ${\bf B}\cdot\delta {\bf B}\ne 0$.
In this case Equations~15 are augmented by extra terms $-\nabla\left({\bf B}\cdot\delta
{\bf B}\right)/4\pi\rho \mp {\bf V_A}\nabla\cdot\delta {\bf v}$ on the right hand side.

In integrating Equations~(\ref{eq:12}), we put $\partial I_{\pm}/\partial t = i\omega I_{\pm}$ to derive
four coupled equations for the real and imaginary parts of $I_{\pm}$.
Equations~(\ref{eq:12}) are integrated from a starting point in the left hand
side chromosphere where Alfv\'en waves
leak down into the chromosphere, back through the corona to the
right hand side where waves are fed up from
below. In this way the reflection and transmission of Alfv\'en waves at
the loop footpoints and elsewhere is naturally reconstructed.
The velocity and magnetic field perturbations are
calculated from
\begin{eqnarray}
\delta v&=&{I_++I_-\over 2}\cr
{\delta B\over\sqrt{4\pi\rho}}&=&{I_+-I_-\over 2}\,.\label{eq:13}
\end{eqnarray}
The wave energy density and positive and negative going energy fluxes are
\begin{eqnarray}
U&=&{\rho\delta v^2\over 2} +{\delta B ^2\over 8\pi}={\rho\over 4}
\left(I_+^2 + I_-^2\right)\cr
F_+&=&{\rho\over 4}I_+^2V_A\cr
F_-&=&{\rho\over 4}I_-^2V_A\,,\label{eq:14}
\end{eqnarray}
and the wave peak electric field appearing in equation 4 is
\begin{equation}
\delta E_p^2={B^2\over 2c^2}\left(I_+^2+I_-^2\right).\label{eq:15}
\end{equation}

\subsection{Fractionation}
\label{subsec:6.5}

We recap and expand the treatment given in \citet{laming04}. Our model starts with
a fully mixed chromosphere, upon which pondermotive forces due to
Alfv\'en wave reflection and transmission
act to provide the fractionation. The low solar chromosphere is of
much higher density than the upper layers where the FIP fractionation
will occur in our models. Consequently the lower boundary condition
of completely mixed photospheric composition material gives an
essentially infinite particle ``reservoir'' to supply the extra
fractionated elements. To calculate the fractionation, we follow in part the approach and notation of \citet{schwadron99}.
Consider first the motion of ions and neutrals of element $s$
in a background flow of protons and hydrogen
with speed $u$. We neglect the ambipolar force, which is generally
much less than gravity. We also neglect an inertial
term $\partial /\partial z\left(\rho_su_s^2/2\right)$ since the flow
speed will turn out to be much lower than the particle thermal speeds.
Then the momentum equations for ions and neutrals are
\begin{eqnarray}
{\partial P_{si}\over\partial z}&=&-\rho_{si}g-\rho_{si}\nu_{si}
\left(u_{si}-u\right) \label{eq:17}\\
{\partial P_{sn}\over\partial z}&=&-\rho_{sn}g-\rho_{Sn}\nu_{sn}
\left(u_{sn}-u\right)\label{eq:16},
\end{eqnarray}
where $P_{si}$ and $P_{sn}$ are the partial pressures of ions and neutrals
of element $s$, $\rho_{si}$ and $\rho_{sn}$ are the corresponding densities,
$\nu_{si}$ and $\nu_{sn}$ the collision rates with ambient gas (assumed
hydrogen and protons), $u_{si}$ and $u_{sn}$ the flow speeds, and $u$ the
hydrogen flow speed imposed on the loop. Detailed expressions for $\nu_{si}$ and
$\nu_{sn}$ are given in \citet{laming04}. The partial pressures are given by
$P_{s,i,n}=\left(k_{\mathrm{B}}T/m_s+v_{\mathrm{turb}}^2+v_{\mathrm{wave}}^2\right)\rho_{s,i,n}/2$, where the
various terms on the right hand side represent the particle thermal speed, the microturbulent
velocity coming from the chromospheric model, and the particle motion in longitudinal
(i.e., acoustic) waves, either induced by the Alfv\'en waves themselves, or propagating up
from the photosphere. These are discussed in more detail in the next subsection.
In true collisionless plasma, neutrals would not respond to the wave.
However the solar chromosphere is sufficiently collisional that
neutrals move with the ions in the wave motion
\citep[e.g.,][]{vranjes08} for wave frequencies well below the charge
exchange rate that couples neutrals and ions, and so neutrals
require the same form for their partial pressure as the ions.
We also neglected inertial terms above. Inclusion of such terms would lead to
\begin{equation}
P_{si,n}=\left(k_{\mathrm{B}}T/m_s +v_{\mathrm{turb}}^2+v_{\mathrm{wave}}^2+u_s^2\right)\rho_{si,n}/2 =v_s^2\rho_{si,n}/2
\label{eq:pressure}
\end{equation}
assuming $u_{si}-u_{sn} \ll u_s\sim u$.

The momentum equations
are combined by forming $\nu_{sn}\partial P_{si}/\partial z + \nu_{si}\partial P_{sn}/\partial z$ to yield
\begin{equation}
{\partial P_s\over\partial z}=-\rho_sg
-\nu_{\mathrm{eff}}\rho_s\left(u_s-u\right)+{\partial\xi_s\over\partial z}
{\rho_sv_s^2\over 2}{\nu_{si}-\nu_{sn}\over \left(1-\xi_s\right)
\nu_{si} +\xi_s\nu_{sn}}\,,
\end{equation}
with $\nu_{\mathrm{eff}}=\nu_{si}\nu_{sn}/\left(\xi_s\nu_{sn} +\left(1-\xi_s\right)
\nu_{si}\right)$  and $\xi_s$ being the ionization fraction of element $s$.
For
\begin{equation}
u_s=u-g/\nu_{\mathrm{eff}}\left(
1-\mu /m_s\right) +\partial\xi_s/\partial z
\left(v_s^2/2\right)\left(\nu_{si}-\nu_{sn}\right)/\nu_{si}/\nu_{sn}\,,\label{eq:19}
\end{equation}
where $\mu$ is the mean molecular weight, all elements are lifted by the
background flow to the same scale height given by $k_{\mathrm{B}}T/\mu g$.
This obviously requires $u\nu_{\mathrm{eff}} > g$, assuming the term in
$\partial\xi_s/\partial z$ negligible. Generally $u-g/\nu_{\mathrm{eff}}\left(
1-\mu /m_s\right)\sim 10^3$ cm s$^{-1}$ or more, while ${\partial\xi_s\over\partial z}\sim 10^{-8}$ cm$^{-1}$ and ${\rho_sv_s^2\over 2}\left(\nu_{si}-\nu_{sn}\right)/\nu_{si}/\nu_{sn}\sim 10^9$ cm$^2$s$^{-1}$.
For $u\nu_s \ll g$, we get gravitationally stratified solutions with
$\rho_s\propto\exp\left(-m_sgz/k_{\mathrm{B}}T\right)$.
With $g=2.74\times 10^4$ cm s$^{-2}$, and $\nu_s\sim 10^2 - 10^3$ the
former case is valid for the Sun for flow speeds in the chromosphere
greater than a relatively modest 10\,--\,100 cm s$^{-1}$.
\citet{laming12} considers the upward flow speed
due to chromospheric evaporation, and finds speeds of at least $10^3$ cm s$^{-1}$.

A gravitational settling timescale may be estimated from equations \ref{eq:17} and \ref{eq:16}. We write
$\partial P_s/\partial z\simeq c_S^2\partial\rho _s/\partial z\simeq \left(c_S^2/u_{sett}\right)
\partial\rho _s/\partial t$ where the sound speed is $c_S$ and the settling speed $u_{sett}\sim g/\nu _{eff}$. Integrating with respect to $t$ gives $t\sim c_S^2\nu _{eff}/g^2\sim 10^4\left(n_H/10^{10}\right)/A$ s, where $A$ is the atomic mass of element $s$, assuming that $\nu _{eff}$ is dominated by collisions with
neutral H. The settling time increases with the density as one moves deeper into the chromosphere, and with the presence of turbulence with amplitude added in quadrature to $c_S$, and can easily be on the order of days or weeks or more. Higher in the chromosphere, $\nu _{eff}$ will be dominated by collisions with protons, and a longer timescale would result as the chromospheric density approaches $10^{10}$ cm$^{-3}$.

The solar chromosphere is doubtless a more dynamic environment than
represented by Equations~(\ref{eq:16})\,--\,(\ref{eq:19}). For our purposes the net result
of this dynamic behavior is assumed to completely mix up the plasma to give
uniform elemental composition with height, which is obtained in our
model with the above choice for $u_s$. Other choices may be possible
which would provide chemical fractionation in the unperturbed
chromosphere, and one could choose $u_s$ to provide the required FIP
effect. However the physics behind such a specification for $u_s$ in
most cases remains obscure, and is probably unrealistic, leading to
an unsatisfactory explanation for the FIP effect.

We now include a ponderomotive force, $\rho_{si}a$ on the ions
in the momentum equations;
\begin{eqnarray}
{\partial P_{si}\over\partial z}&=&-\rho_{si}g-\rho_{si}\nu_{si}
\left(u_{si}-u\right) +\rho_{si}a \\
{\partial P_{sn}\over\partial z}&=&-\rho_{sn}g-\rho_{sn}\nu_{sn}
\left(u_{sn}-u\right).
\end{eqnarray}
Again writing $P_s=\rho _sc_S^2$,
taking $u_s$ as specified above and integrating with respect to $z$ we find
\begin{equation}
{\rho_s\left(z_u\right)\over\rho_s\left(z_l\right)}=
\exp\left\{2\int_{z_l}^{z_u}
{\xi_sa\nu_{\mathrm{eff}}\over \nu_{si}v_s^2}\,dz\right\},
\label{eq:22}
\end{equation}
where the constant of integration has been chosen to keep $\rho_s\left(z_u\right)
=\rho_s\left(z_l\right)$ when $a=0$.
A quantitative assessment of coronal
element abundances anomalies requires an evaluation of Equation~(\ref{eq:22}) with a
realistic model chromosphere in the region of Alfv\'en wave reflection, with $a$ coming
from Equation~(\ref{eq:pondforce}).

As will be seen below, most the the FIP fractionation develops over a small range in $z$,
where the ponderomotive acceleration $a$ is strong. The timescale to establish this local abundance anomaly is short, basically $\left(z_u-z_l\right)/u \sim  \left(10^6-10^7\right)/10^4\sim 10^3$ s. By
itself, the ponderomotive force would only produce a deficit of low FIP ions just below this region, and a surplus just above. To produce the fractionation seen in coronal loops and the solar wind, thermal transport or diffusion processes must continually supply further low FIP ions from below to erase the deficit, and communicate the resulting fractionation above to still higher levels
of the solar atmosphere. Thus while the fractionation can be produced locally rather quickly, and
is therefore immune from processes that would otherwise disrupt fractionation produced solely by diffusive processes, the timescale to change the composition of a coronal loop is still on the
order of days, as observed \citep{widing01} due to the necessity of thermal transport of the fractionation.

\subsection{Compressional chromospheric waves}
\label{subsec:6.6}

\subsubsection{Introduction}

We have introduced an extra longitudinal pressure associated with the Alfv\'en waves
proportional to $v_{\mathrm{wave}}^2$ (see Equation~(\ref{eq:pressure})), which has the effect of causing
some saturation of the FIP fractionation. Here we give some physical motivations for this
extra term. It has three possible sources. The first is the inevitable
generation of slow mode waves by the Alfv\'en driver. Physically, the periodic
variation in magnetic pressure of the Alfv\'en wave drives longitudinal compressional
waves. These generated acoustic waves can act back on the Alfv\'en driver, as the
compressional wave introduces a periodic variation in the Alfv\'en speed, which
generates new Alfv\'en waves. The second will be obliquely propagating fast mode waves, which are
necessarily compressive. Torsional Alfv\'en wave on a twisted magnetic flux tube are inevitably
mixed with the kink mode, and hence are compressive. The third relates to acoustic waves
propagating up to the chromosphere from the convection zone.

\subsubsection{Parametric generation by Alfv\'en waves}

Following \citet{laming12} we illustrate the generation of slow mode or acoustic
waves by the ponderomotive force associated with plane Alfv\'en waves with a simple 1D
calculation. The linearized momentum equation keeping
terms to all orders in perturbed quantities is (all symbols have their usual meanings as defined above),
\begin{equation}
\left(\rho +\delta\rho\right){\partial\delta v_z\over\partial t}+
\left(\rho +\delta\rho\right)\delta v_z{\partial\delta v_z\over\partial z}=
\left(\rho +\delta\rho\right){\partial\over\partial z}{\delta B^2\over 8\pi\left(\rho +\delta\rho\right)}-{\partial\delta P\over\partial z} -g\delta\rho \,,
\label{eq:24}
\end{equation}
where
\begin{eqnarray}
\delta\rho &=-\rho\nabla\cdot{\bf \delta {\bf r}}-\delta r_z{\partial\rho\over\partial z} =-\rho\delta {\bf r}\left(ik_S+{1\over H_D}\right)\cr
\delta P &=-\gamma P\nabla\cdot{\bf \delta {\bf r}}-\delta r_z{\partial P\over\partial z}
=-P\delta {\bf r}\left(ik_S\gamma +{1\over H_P}\right),
\label{eq:25}
\end{eqnarray}
for Eulerian displacement ${\bf \delta r}\propto\exp i\left(\omega _St+k_Sz\right)$,
where $H_P=P/\left(\partial P/\partial z\right)$ and
$H_D=\rho /\left(\partial\rho /\partial z\right)$
(signed) pressure and density scale heights, respectively. The first term on the
right hand side of Equation~(\ref{eq:24})
represents the instantaneous ponderomotive force. In cases
where the WKB approximation applies, $\delta B\propto\rho ^{1/4}$, and this expression is
equivalent to the more usual form $-\partial\left(\delta B^2/8\pi\right)/\partial z$.
Substituting for $\delta\rho$ and $\delta P$ from Equations~(\ref{eq:25}),
keeping terms as high as second order in small quantities, Equation~(\ref{eq:24}) becomes
\begin{equation}
-i{\omega_S\over H_D}\delta v_z^2+\left(-\omega_S^2 +k_S^2c_S^2-i{k_Sc_S^2\over H_P}-
{c_S^2\over\gamma H_P^2}-i{k_Sc_S^2\over\gamma H_P}-ik_Sg-{g\over H_D}\right)\delta v_z
-i\omega_S{\partial\over\partial z}{\delta B^2\over 8\pi\rho}=0.
\end{equation}
This is considerably simplified in isothermal conditions, $\gamma =1$, $H_P=H_D=-c_S^2/g$, so that
\begin{equation}
-i{\omega_S\over H_D}\delta v_z^2+\left(-\omega_S^2 +k_S^2c_S^2+ik_Sg\right)\delta v_z
-i\omega_S{\partial\over\partial z}{\delta B^2\over 8\pi\rho}=0.
\end{equation}
We put $\Im k_S=-g/2c_S^2$, and $\sqrt{\left(\Re k_S\right)^2+g^2/4c_S^4}
=2\Re k_A/n$, $\omega =2\omega_A/n$, where $k_A$ and $\omega_A$ are the wavevector and angular frequency of the Alfv\'en wave with $n=1,2,3,..$
(anticipating the result below). We find
\begin{equation}
\delta v_z^2-\delta v_z iH_D\omega_S\left(1-{c_S^2\over V_A^2}\right)+
H_D{\partial\over\partial z}{\delta B^2\over 8\pi\rho}=0
\end{equation}
with solution
\begin{equation}
\delta v_z={-i\over 2}\left[\sqrt{\delta v_A^2+H_D^2\omega_S^2
\left(1-{c_S^2\over V_A^2}\right)^2}-H_D\omega_S\left(1-{c_S^2\over V_A^2}\right)\right]
\end{equation}
where we have put ${\partial\over\partial z}\left(\delta B^2/ 8\pi\rho\right)=
\left(\delta B^2/4\pi\rho\right)/4H_D=\delta v_A^2/4H_D$ using the
WKB result for Alfv\'en waves in a density gradient, and assuming $1/H_D \gg \Re k_A$.
When $c_S^2\sim V_A^2$ or $H_D\rightarrow 0$, $\delta v_z\simeq-i\delta v_A/2$. In
the opposite limit $\delta v_z\simeq -i\delta v_A^2/4H_D\omega_S
\left(1-c_S^2/V_A^2\right)$. In these two cases $\omega_S=
\omega_A$ or $\omega_S=2\omega_A$, respectively. In fact, acoustic waves can be
generated with $\omega_S=2\omega_A/n$, with higher $n$ becoming more intense as the
nonlinearity increases \citep{landauCM}. This is shown explicitly by \citet{laming12}, who extends the treatment above to include higher powers of $\delta v_z$.
\citet{vasheghani11} treat the case of slow
mode wave generation by a torsional Alfv\'en wave. This is subtly different to the case of
a plane Alfv\'en wave considered here, and the FIP fractionation resulting from such a model
will be investigated in future work.

When the Alfv\'en wave becomes evanescent in a sufficiently steep density gradient, $k_A\rightarrow i\kappa _A$ and
\begin{equation}
\delta v_z={-i\over 2}\left[\sqrt{\delta v_A^2+H_D^2\omega_S^2
\left(1+{\kappa _A^2c_S^2\over\omega _A^2}\right)^2}-H_D\omega _S\left(1+{\kappa _A^2c_S^2\over\omega _A^2}\right)\right],
\end{equation}
resulting in reduced $\delta v_z$ compared to the case with propagating Alfv\'en waves. This will be apparent in results to be presented in Section \ref{sec:results}.

\subsubsection{Fast-mode waves from above}

The treatment of \citet{vasheghani11} also considers a second source of longitudinal pressure.
Torsional waves on a flux tube with magnetic twist are no longer pure Alfv\'en waves, but have an
admixture of fast mode or kink polarization. Since the magnetic perturbation (in the $\phi$ direction in cylindrical polar coordinates) now has a component along the background magnetic field
(the twist gives it a $\phi$ component also), the wave becomes compressive. This is also true
more generally of any fast mode wave in oblique propagation, and in \citet{laming09} it was shown
that such longitudinal pressure would lead to a natural saturation of the FIP effect at high
wave amplitudes. In the case where transverse and longitudinal pressures are equal, the FIP
effect saturates at the commonly observed value of about 4. However such compressive waves are
also subject to increased damping, and are therefore less likely to develop to large amplitudes.
The existence or otherwise of twisted magnetic loops in the corona is also controversial.
Accordingly, we do not consider such possibilities further in this review.

\subsubsection{\textit{P}-modes from below}

The third contribution to the longitudinal pressure in the chromosphere comes from the continued propagation upwards of slow modes waves or shocks developing from turbulence lower down in the
atmosphere. We include a longitudinal pressure with amplitude 6.25\kms motivated by the
simulations in \citet{heggland11}. The velocity amplitude of these waves would be expected to grow
with altitude, due to WKB effects, although some reflection back downwards may also occur, and should damp mainly due to the effect of radiation. The principal coolant in the photosphere is
H$^-$, with cooling timescale $\tau \simeq \left(1.2\times 10^{11}/n_e\right)\left(5000\mathrm{\ K}/T\right)^2$ minutes \citep{ayres96}. Cooling of a similar magnitude will be present in the low chromosphere, although different processes may also contribute. Although the plasma density varies, throughout large
sections of the chromospheric model, the electron density and temperature are roughly constant and of order $10^{11}$ cm$^{-3}$ and 5000~K, respectively, leading to a constant cooling time of order
1 minute. Thus for 3 minute or 5 minute waves, the product $\omega\tau\sim 1$, and the wave
decay length would be comparable to the wavelength \citep{mihalasFRH}, and similar to the WKB
growth. In fact, the simulations of \citet{heggland11} show approximately constant slow mode
wave amplitude with height for a variety of models, as do the observations of \citet[albeit at a lower wave amplitude]{beck13}, so we accordingly adopt a constant slow mode wave amplitude in our models.

While slow mode waves reduce the fractionation by increasing $v_s^2$ in the denominator of the
integrand in Equation~(\ref{eq:22}), when these steepen into shock waves, we expect
fractionation to cease altogether due to the extra turbulence. Accordingly
we also neglect fractionation at chromospheric altitudes where the
slow mode wave amplitudes added in quadrature come to a value greater than the local sound speed.
The rationale for this is that in such conditions, a shock will form, and the mixing produced by turbulence generated in such conditions would preclude further fractionation. Reflection and
refraction of sound waves at the discontinuity will produce oppositely directed packets of
sound waves which can initiate a cascade to smaller, and ultimately microscopic, spatial scales,
assumed to induce microscopic mixing. This has the effect
in most models of limiting FIP fractionation to the upper regions of the chromosphere.

\newpage

\section{Results and Interpretation}
\label{sec:results}

\subsection{Closed loop}

As a first illustration of the action of the ponderomotive force in causing FIP fractionation,
we consider the case of a closed magnetic loop. The model loop is taken to be 100\,000~km long, and
plasma in the loop has a density scale height of 125\,000~km, extrapolated upwards from the
chromospheric model. The coronal temperature is constant at the maximum temperature of
the chromospheric temperature ($\sim 1.6\times 10^6$ K) but this value is not significant since we
are uninterested in the ionization balance in this region and the Alfv\'en wave propagation
only depends on the density. The coronal magnetic field is 20 G,
increasing by a factor of 5 in the photosphere, as in Figure~\ref{fig:chromodel} (right panel).
The chromospheric model is taken from \citet{avrett08}, and the plasma $\beta =1$ layer is at 615~km altitude above
the photosphere. We simply treat the parallel propagation of undamped shear Alfv\'en waves. With
reference to Figure~\ref{fig:loopmodel}, the solution of Equations~(\ref{eq:12}) is started
with a downgoing wave at the left-hand chromosphere, and then integrated back to the chromosphere at the other end of the loop. Choosing a wave angular frequency of 0.07 s$^{-1}$ means the wave is on resonance with the coronal loop, and
in our calculation, 98\% of the upgoing wave energy from the right-hand chromosphere is transmitted to the corona. Hence the chromospheric wave patterns in both chromospheres are very similar, with the
roles of upgoing and downgoing waves interchanged (with 100\% transmission, they would be identical). The ponderomotive force is insensitive to the direction of wave motion, and so is the
same at each footpoint.

\epubtkImage{loop}{%
\begin{figure}[htbp]
\centerline{\includegraphics[scale=0.9]{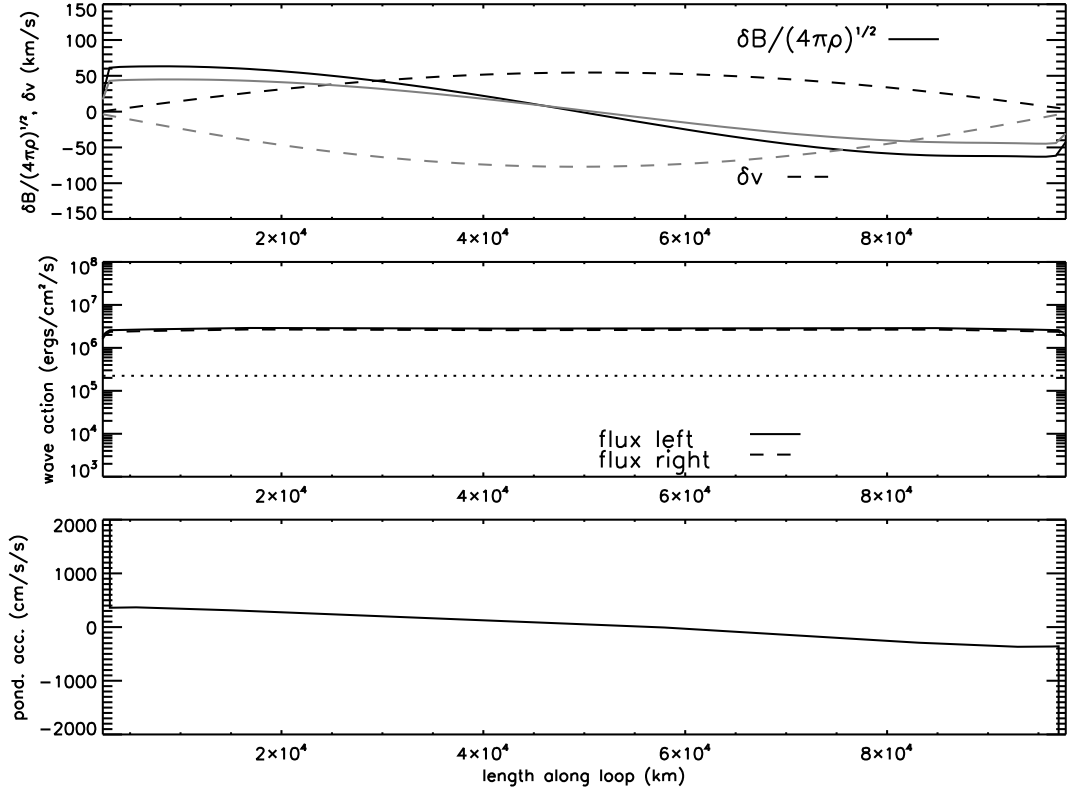}}
\caption{Solution for the coronal portion of the loop. The top panel
shows $\delta {\bf v}$ and $\delta {\bf B}/\sqrt{4\pi\rho}$ (Equations~(\ref{eq:13})), real components in black, imaginary components in grey. The middle panel shows the wave energy fluxes,
left and right going, and their difference, divided by the magnetic flux density. Normalized thus,
this difference should be constant in the absence of wave growth or damping, and provide a check
on energy conservation in the integration. The bottom panel shows the ponderomotive acceleration, $a$.\label{fig:loop}}
\end{figure}}

Figure~\ref{fig:loop} shows the solution for the coronal portion of the loop. The top panel
shows $\delta {\bf v}$ and $\delta {\bf B}/\sqrt{4\pi\rho}$ (Equations~(\ref{eq:13})), real components in black, imaginary components in grey. The middle panel shows the wave energy fluxes,
left and right going, and their difference, divided by the magnetic flux density. Normalized thus,
this difference should be constant in the absence of wave growth or damping, and provides a check
on energy conservation in the integration. The bottom panel shows the ponderomotive acceleration, $a$, calculated from Equations~(\ref{eq:pondforce}) and (\ref{eq:15}). Throughout the corona (2500~km $< z <$ 97500~km), $a \ll 2.7\times 10^4$ cm s$^{-2}$, the solar gravitational acceleration. In the chromospheric footpoints, $a$ is much larger.

\epubtkImage{chromo_loop}{%
\begin{figure}[htbp]
\centerline{\includegraphics[width=\textwidth]{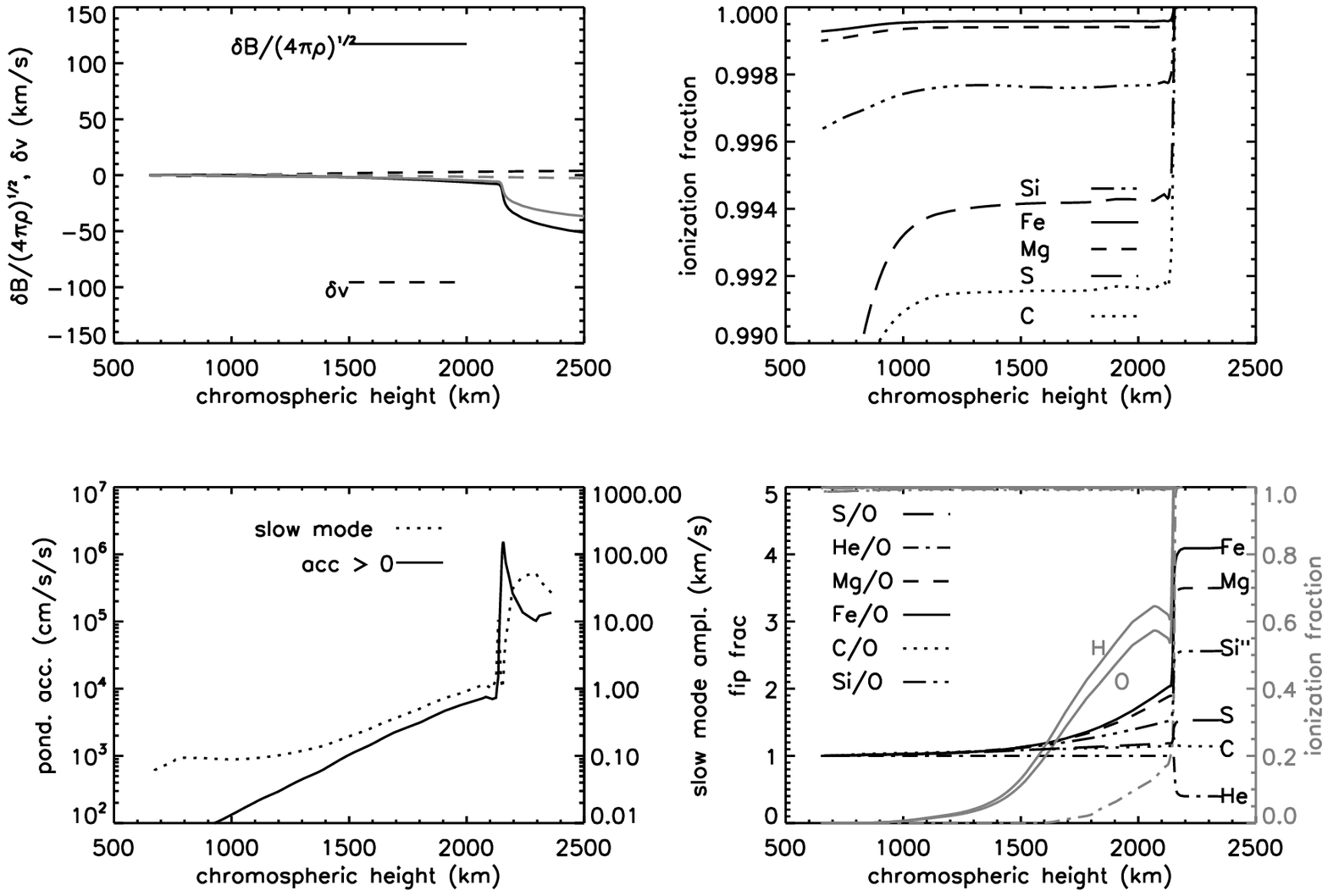}}
\caption{The top and bottom panels on the left hand side show the same variables as the
top and bottom panels in Figure~\ref{fig:loop} (the wave energy fluxes are not shown). The panels on the right hand side show the chromospheric ionization balance and
FIP fractionation for selected elements. The bottom panel shows FIP fractionation for the
abundance ratios S/O, He/O, Mg/O, Fe/O and Si/O, with the linestyles indicated as black lines, to
be read on the left axis. One can see Fe/O and Mg/O fractionating to a value close to 4,
Si/O to 2.6. S/O to 1.6 and He/O to 0.4. The ionization balance is shown as grey lines with the same linestyles for Fe, Mg, Si, S, and He. He stays neutral the longest, and
matches calculations in \citet{golding14}. Also shown are H and O as solid lines, labelled respectively. The O ionization balance follows that for H due to the strong coupling by charge exchange. The ionization balances for Fe, Mg, Si, and S are hard to distinguish on this plot,
so the top right panel shows these in more detail, with the charge fraction scale running from
0.99 to 1.00.\label{fig:chromo_loop}}
\end{figure}}

Figure~\ref{fig:chromo_loop} illustrates important features of the solution in the chromospheric
portion of the loop. The top and bottom panels on the left hand side show the same variables as the
top and bottom panels in Figure~\ref{fig:loop} (the wave energy fluxes are not shown). The
ponderomotive acceleration shows a large ``spike'' at an altitude of 2150~km, coinciding with the
steep chromospheric density gradient shown in Figure~\ref{fig:chromodel} (left panel). Also
shown on the bottom left panel is the amplitude of slow mode waves arising from parametric generation by the Alfv\'en driver, shown as a dotted line. This is added in quadrature to the
amplitude of slow mode waves propagating upwards from the photospheric, assumed here to be a constant 6.25\kms, following the discussion in Section~\ref{subsec:chromodel}. This
value is also consistent with the macroscopic velocity field inferred in the solar chromosphere by
\citet[][see their Figure~3ab]{vieytes05}.
The parametric slow mode amplitude goes through a sharp minimum where the chromospheric density
gradient is steepest at about 2150 km altitude. \citet{rakowski12} show that this occurs where the Alfv\'en wave becomes
evanescent. The panels on the right hand side show the chromospheric ionization balance and
FIP fractionation for selected elements. The bottom panel shows FIP fractionation for the
abundance ratios S/O, He/O, Mg/O, Fe/O and Si/O, with the linestyles indicated as black lines, to
be read on the left axis. One can see Fe/O and Mg/O fractionating to a value close to 4,
Si/O to 2.6. S/O to 1.6 and He/O to 0.4. The ionization balance is shown as grey lines with the same linestyles for Fe, Mg, Si, S, C, and He.  Also shown are charge states fractions for H and O as labelled. The ionization balances for Fe, Mg, Si, C, and S are hard to distinguish on this plot,
so the top right panel shows these in more detail, with the charge fraction scale running from
0.99 to 1.00.

Fe and Mg have similar charge state fractions for all altitudes of relevance, and hence fractionate
to about the same degree. Si, although low FIP and 99.8\% ionized in the region 1500~km to 2500~km, fractionates noticeably less. Around 1000~km altitude, Si is more highly ionized due to the increased flux in Ly $\alpha$, which is now trapped by the increased opacity in the line. In the
model presented here, no fractionation occurs at this height. S is also has a high charge state
fraction, at 99.4\%, but its ionization potential is higher than the energy of Ly $\alpha$, and so it is not affected by the trapped radiation. This dependence on the ionization fraction comes
from the form of $\nu _{eff}=\nu _{si}\nu _{sn}/\left(\xi _s\nu _{sn}+\left(1-\xi _s\right)\nu _{si}\right)$ in the integrand of equation \ref{eq:22} where $\nu _{si} >> \nu _{sn}$, \citep{laming04}, which is the case when the background gas is mainly ionized H. Only when $\xi _s\simeq 1$ to a high degree of precision does the second term in the denominator become negligible compared to the first, allowing strong fractionation to occur. Lower down in the chromosphere, where neutral H dominates, $\nu _{sn}\sim \nu _{si}$, and this precise
dependence on element charge state is lost.

%\begin{landscape}
\begin{table}[htbp]
\caption[FIP Fractionations in Closed Magnetic Field]{FIP Fractionations in Closed Magnetic Field (see text for details).}
\label{table:closedfield}
\centering
{\small
\begin{tabular}{l ccc cccccccc}
\toprule
ratio & \multicolumn{3}{c}{models}&\multicolumn{7}{c}{observations} \\
~ & 75& 95& 110& a& b& c& d& e& f& g\\
~ & \multicolumn{3}{c}{(\kms)}& \\
\midrule
He/O & 0.56& 0.40& 0.22&  0.68\,--\,0.60& 0.54\,--\,0.98& ~ & ~ & ~ & ~& 0.29\\
C/O & 1.14& 1.15& 1.03&  1.36\,--\,1.41& 1.23\,--\,1.55& ~ & ~ & ~ &~ & 0.76 \\
N/O & 0.85& 0.73& 0.55&  0.72\,--\,1.32& 0.59\,--\,0.68& ~ & ~ & ~ &~ &  1.02 \\
Ne/O& 0.71& 0.55& 0.36&  0.38\,--\,0.75& 0.79\,--\,1.15& ~ & ~ & ~ & ~ & 0.75\\
Na/O& 2.57& 3.95 & 5.83&  & 2.51\,--\,3.31& 1.8${+2\atop -1}$& 7.8${+13\atop -5}$& ~ &$3.43\pm 1.7$\\
Mg/O& 2.37& 3.52&  4.95& 2.58\,--\,2.61& 1.95\,--\,3.55&  2.7$\pm 0.3$&2.8${+2.3\atop -1.3}$& ~ &~ & 2.71\\
Al/O& 2.20& 3.16&  4.28& ~ & 2.29\,--\,3.02& 5.6${+3.3\atop -2.1}$&3.6${+1.7\atop -1.2}$& & \\
Si/O& 1.92& 2.60&  3.26& 2.49\,--\,3.11& 2.14\,--\,3.26& ~ & 4.9${+2.9\atop -1.8}$& ~ & 2.5$\pm 1.0$ & 2.37 \\
P/O & 1.52& 1.83& 2.01& ~ & ~ & 5${+11\atop -3.4}$\\
S/O & 1.37& 1.56&  1.61& 1.62\,--\,1.92& 1.20\,--\,2.09& 2.1$\pm 0.2$&2.2$\pm 0.2$&  $1.7\pm
0.3$& $1.00{+0.48\atop -0.32}$&  1.00\\
Cl/O& 1.04& 0.97& 0.84& ~ & ~ & ~ & ~ & ~ & 1.8$\pm 1.8$\\
Ar/O& 0.93& 0.83&  0.65& ~ & ~ & ~ & ~ & $1.1\pm 0.1$& 1.00$\pm 0.15$& 0.73\\
K/O & 2.71& 4.29& 6.58&  & ~ & 4.7${+7.0\atop -2.8}$& 1.8${+0.4\atop -0.6}$&
  $3.5\pm 0.9$& 5.5${+9\atop -3.7}$\\
Ca/O& 2.71& 4.29&  6.58& ~ &2.09\,--\,3.88 & 2.7$\pm 0.25$&3.5${+4.3\atop -1.9}$&
 & 3.0\,--\,9.7& 2.86\\
Ti/O& 2.70& 4.28& 6.57\\
Cr/O& 2.66& 4.18& 6.36& ~ & 2.40\,--\,3.47& & & & \\
Fe/O& 2.66& 4.18&  6.35& 2.28\,--\,2.90& 1.95\,--\,3.55& ~ &7.0${+1.4\atop -1.2}$& ~ &  & 2.27 \\
Ni/O& 2.39& 3.57&  5.11& ~ & ~ & ~ & ~ & \\
Kr/O& 0.99& 0.89&  0.73& ~ & ~ & ~ & ~ & ~ \\
\bottomrule
\end{tabular}}
\end{table}
%\end{landscape}

Table~\ref{table:closedfield} gives model fractionations for three different assumed coronal
Alfv\'en wave amplitudes, 75, 95, and 110\kms, which are initiated with amplitudes 0.6, 0.75, and 0.9\kms low in chromosphere A. Figures \ref{fig:loop} and \ref{fig:chromo_loop} correspond to the middle entry, 95\kms. The model results
are compared with observations taken from the literature; (a) \citet{zurbuchen02}, given relative to O, (b) \citet{bochsler07a}, relative to O, (c) \citet{giammanco07, giammanco08}, relative to H, (d) \citet{bryans09}, given
relative to the mean of O, Ne and Ar
and, (e) \citet{phillips03}, relative to H, (f)
Ar, \citet{sylwester10a}; K, \citet{sylwester10p}; Cl, \citet{sylwester11}; S, \citet{sylwester12};
Si, \citet{sylwester13}; and Na \citet{phillips10}, all relative to H and given relative to the
photospheric abundance of \citet{asplund09}, and (g) SEP measurements from
\citet{reames14}, given relative to O. \citet{zurbuchen02}, \citet{bochsler07a} and
\citet{giammanco07,giammanco08} give abundances derived from \textit{in situ} measurements in the
slow solar wind. They generally agree best with model with coronal Alfv\'en wave amplitudes of
75-95\kms. \citet{bryans09} give abundances derived from a deep SOHO/SUMER \citep{wilhelm95,wilhelm97} spectrum
of the quiet solar corona, and are more consistent with a higher Alfv\'en wave
amplitude, 110\kms. The remaining references give abundance measured in solar flares
with RESIK \citep{sylwester05}. All models assume a constant amplitude with height for slow
mode waves propagating up from the photosphere of 6.25\kms. Fractionations are
somewhat sensitive to this, in that increased slow wave wave amplitude decreases the degree of
fractionation. However for the case considered here, that of a coronal Alfv\'en wave on resonance
with the coronal loop, most of the fractionation is restricted to a small region close to the
top of the chromosphere where the slow mode waves generated parametrically by the Alfv\'en wave
driver are also significant. \citet{rakowski12} extended such considerations to Alfv\'en waves away from the loop resonance. A similar FIP effect is produced, but the
fractionation occurs over a wider layer of the chromosphere, as Alfv\'en waves penetrate
deeper. Some subtle differences occur. S is more strongly fractionated in such a case, and He is depleted much less, both as a result of the difference in Alfv\'en wave behaviour. Slow mode waves are also more strongly generated deeper in the chromosphere, and in cases such as this, their generation has a stronger effect on the fractionation. \citet{laming04} also considered the effect of different chromsopheric models from
\citet{vernazza81}. If VALC may be considered as a forerunner to the \citet{avrett08} model used here, and it does indeed give similar FIP fractionation, the result of
\citet{laming04} in that stronger fractionations are found within cells at dark points
(VALA, VALB) and weaker fractionation at network segments (VALD, VALE). However these
chromospheric models should probably be considered obsolete at the current time of writing.

\subsection{Open field}

We next consider the FIP effect in open magnetic field. The chromospheric model is extrapolated
upwards assuming a density scale height of 170,000 km, to give a density profile comparable to
that observed and modeled elsewhere \citep{laming04a}. The temperature is set at a maximum of
$10^6$K, but is similarly inconsequential as in the closed field case. Figure~\ref{fig:ch} shows the same
panels as Figure~\ref{fig:loop}, but extending to an altitude of 500\,000~km. Five Alfv\'en waves
are included, designed to match the spectrum given in Figure~3 of \citet{cranmer07}, with amplitude
in the transition region from their Figure~9. Waves of
angular frequency 0.010, 0.031, 0.062, 0.093, and 0.124 s$^{-1}$ are included with amplitudes
at 500\,000~km of 12.5, 150, 75, 50, and 12.5\kms, respectively. The integration of Equations~(\ref{eq:12}) begins at 500\,000~km with outgoing waves only following \citet{cranmer07},
and is taken back to the chromosphere. No account is taken of the
motion of the coronal hole plasma. For the range of altitudes considered, the fast solar
wind speed is always much less than the Alfv\'en speed, and is probably still insignificant
compared to the uncertainty in the Alfv\'en speed.

\epubtkImage{ch}{%
\begin{figure}[htbp]
\centerline{\includegraphics[scale=0.9]{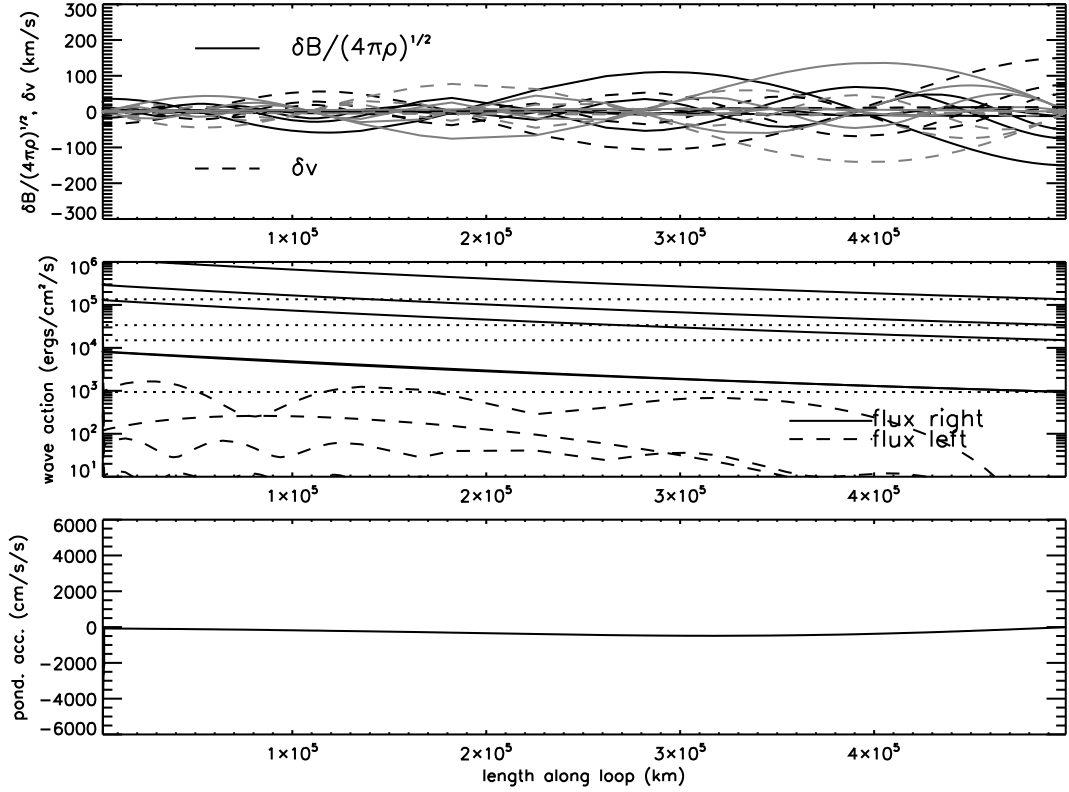}}
\caption{Solution for the coronal hole. The top panel
shows $\delta {\bf v}$ and $\delta {\bf B}/\sqrt{4\pi\rho}$ (Equations~(\ref{eq:13})), real components in black, imaginary components in grey. The middle panel shows the wave energy fluxes,
left and right going, and their difference, divided by the magnetic flux density. Normalized thus,
this difference should be constant in the absence of wave growth or damping, and provide a check
on energy conservation in the integration. The bottom panel shows the ponderomotive acceleration, $a$.\label{fig:ch}}
\end{figure}}

\epubtkImage{chromo_ch}{%
\begin{figure}[htbp]
\centerline{\includegraphics[width=\textwidth]{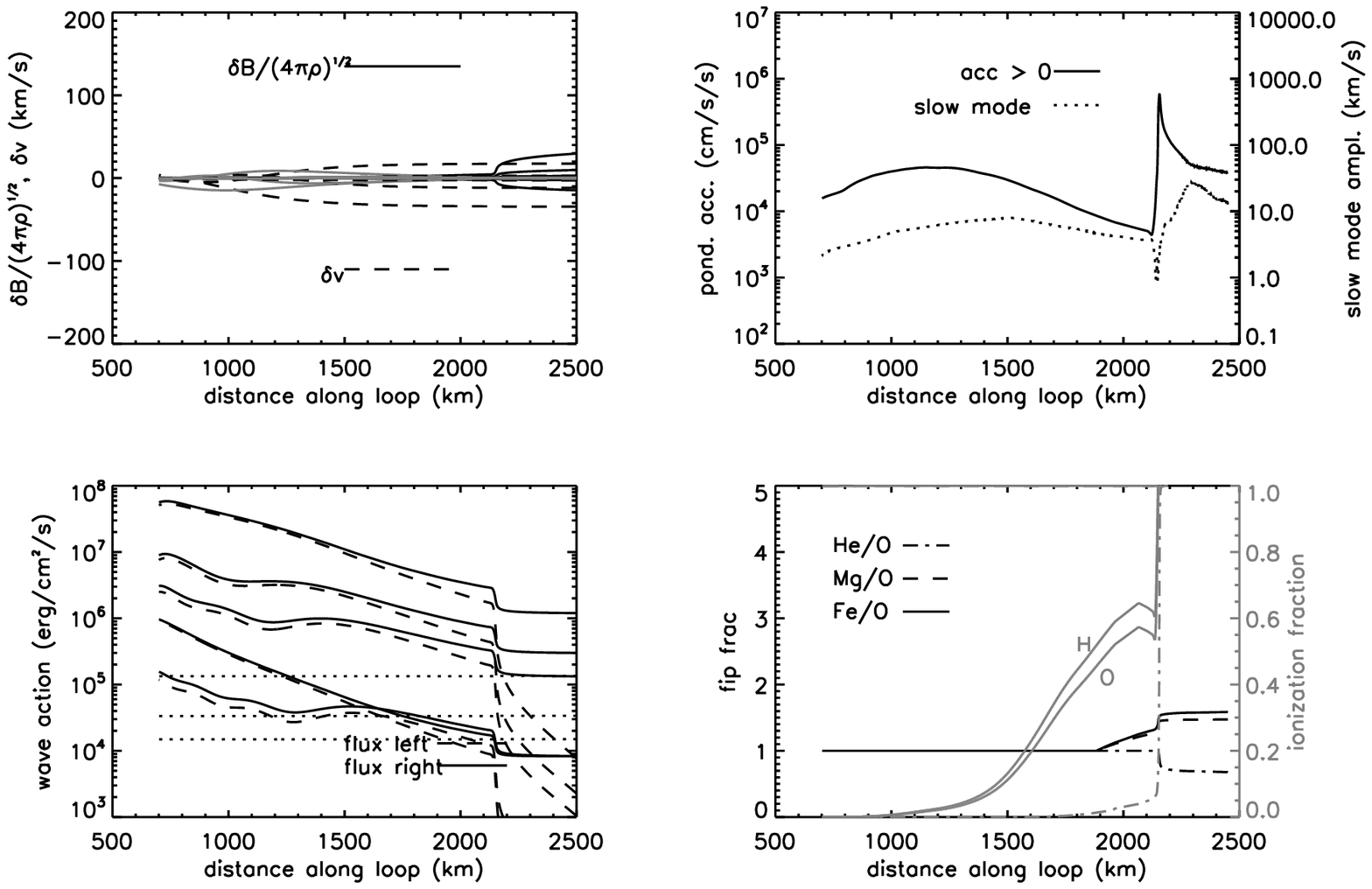}}
\caption{The top and bottom panels on the left hand side and the top panel on the right hand side show the same variables as in Figure~\ref{fig:ch}. Additionally, the amplitude of slow
mode wave generated parametrically by the Alfv\'en waves is also shown in the top right panel.
Compared to the closed loop case with a wave on resonance, the ponderomotive acceleration has
a stronger contribution lower down in the chromosphere.
The bottom right hand side panel shows the chromospheric ionization balance and
FIP fractionation for selected elements. The bottom panel shows FIP fractionation for the
abundance ratios He/O, Mg/O, and Fe/O, with the linestyles indicated as black lines, to
be read on the left axis. The degree of fractionation is reduced from the closed field case. The ionization balance is shown as grey lines with the same linestyles for Fe, Mg, Si, S, and He, and is very similar to the previous, except that in the reduced radiation field of the coronal hole, He stays neutral longer.\label{fig:chromo_ch}}
\end{figure}}

Figure \ref{fig:chromo_ch} shows a similar set of panels to Figure~\ref{fig:chromo_loop} above. Top left shows $\delta {\bf v}$ and $\delta {\bf B}/\sqrt{4\pi\rho }$ for the five waves in the
chromosphere, with the real and imaginary parts distinguished as before. Bottom left now shows
the chromospheric wave energy fluxes. Although we have made no assumption about where the waves
originate, in the case of the coronal hole the usual interpretation is that Alfv\'en waves
start out as kink mode waves on photospheric flux tubes, and propagate up into the coronal hole.
This bottom left panel shows that most of the upgoing wave energy flux is reflected back down again, at the steep chromospheric density gradient. Of order 10\% of the wave energy makes it out of the
chromosphere, and more reflection occurs higher up (not included in our model). Higher frequency
waves suffer less reflection than the low frequency waves.

The top right panel shows the ponderomotive acceleration. Superficially it is very similar to the
previous case, with a ``spike'' at the altitude where the chromospheric density gradient is steepest. However, the strong wave reflection leads to a stronger component of the ponderomotive
acceleration at lower altitudes, around 1000~km above the photosphere, and a reduced acceleration
associated with the ``spike''. In principle, stronger fractionation can occur low down
in the chromosphere because here the hydrogen is mainly neutral, whereas around the ``spike''
it is becoming ionized. This difference in the background H makes a big difference to the
value of $\nu_{\mathrm{eff}}$ in Equation~(\ref{eq:22}), and also to the resulting fractionation.
The degree to which this actually happens is determined by the amplitude of slow mode waves
appearing in $v_s^2$ in the denominator of the integrand in Equation~(\ref{eq:22}). The model
shown in Figure~\ref{fig:chromo_ch} assumes the same slow mode wave amplitude as in the closed loop case, a constant value of 6.25\kms, which when added in quadrature with the parametrically
generated slow mode wave amplitude eliminates the fractionation low down. Thus the fractionation
is restricted to the top of the chromosphere, and generally agrees well with coronal hole
observations. Treating the Alfv\'en and acoustic waves in a more complete fashion (but only considering Fe/O), \citet{cranmer07} reach a similar result. The amplitude of acoustic waves
developing low in the chromosphere in their model is higher than in ours, of order 10\kms from their Figure~9, though varying with altitude.

\begin{table}[htbp]
\caption[FIP Fractionations in Open Magnetic Field]{FIP Fractionations in Open Magnetic Field (see text for details).}
\label{table:openfield}
\centering
{\small
\begin{tabular}{l ccc ccc}
\toprule
ratio & \multicolumn{3}{c}{models}&\multicolumn{3}{c}{observations} \\
~ & 6.15& 6.25& 6.5& a& b& c\\
~ & \multicolumn{3}{c}{(\kms)}& \\
\midrule
He/O & 0.68& 0.68& 0.68&  0.59\,--\,0.63& 0.55\,--\,0.69&  0.45\,--\,0.55\\
C/O & 3.69& 1.00& 0.99&   1.41\,--\,1.68& 1.41\,--\,1.68&  0.9\,--\,1.1\\
N/O & 0.89& 0.89& 0.89&   1.07\,--\,1.32& 1.07\,--\,1.32& \\
Ne/O& 0.81& 0.81& 0.81&   0.44\,--\,0.52& 0.44\,--\,0.52& 0.3\,--\,0.4\\
Na/O& 7.12& 1.56 & 1.43&           & 1.45\,--\,1.91& \\
Mg/O& 6.74& 1.48&  1.37&  1.61\,--\,1.85& 1.29\,--\,2.82&  0.95\,--\,2.45\\
Al/O& 6.46& 1.41&  1.33& ~ & 1.51\,--\,2.00& \\
Si/O& 5.97& 1.31&  1.26&  1.86\,--\,2.26& 1.29\,--\,2.34& 0.9\,--\,1.8 \\
P/O & 5.22& 1.18& 1.15\\
S/O & 4.80& 1.13&  1.11&  1.46\,--\,1.60& 1.17\,--\,1.86& \\
Cl/O& 1.01& 1.00& 1.00\\
Ar/O& 0.95& 0.95&  0.95& \\
K/O & 7.76& 1.62& 1.48&  \\
Ca/O& 7.77& 1.62&  1.48& ~ & 1.38\,--\,1.82& \\
Ti/O& 7.83& 1.62& 1.48\\
Cr/O& 7.78& 1.60& 1.47& ~ & 1.81\,--\,2.63& \\
Fe/O& 7.80& 1.60&  1.47&  1.45\,--\,1.80& 1.51\,--\,2.29& 0.65\,--\,1.35 \\
Ni/O& 7.25& 1.50&  1.39& \\
Kr/O& 0.98& 0.98&  0.98& \\
\bottomrule
\end{tabular}}
\end{table}

The effect of varying the slow mode wave amplitude is illustrated in Table \ref{table:openfield},
where models are given for amplitudes of 6.15, 6.25 and 6.5\kms. Above about 6.25\kms, the FIP fractionation slowly decreases with increasing slow mode waves, assuming the
Alfv\'en waves are kept constant. Below this value, the FIP fractionation increases dramatically.
Observational ratios are taken from, (a) \citet{zurbuchen02}, given relative to O, (b) \citet{bochsler07a}, relative to O, and (c) \citet{ko06}, relative to H.
FIP fractionations at the level of those in the first model column in Table~\ref{table:openfield},
or higher, have been reported \citep{widing92,young97}, and are always observed \emph{in open field
structures}. Other authors \citep{doschek00,delzanna03a,delzanna03b} have not found large FIP effects in polar plumes. This last reference actually challenges the finding of \citet{widing92},
\citep[but not that of][]{young97}, arguing that an isothermal plasma at $\log T=5.9$ would
produce the same Ne VI/Mg VI intensity ratio with photospheric abundances. Hence the reality, and certainly the ubiquity of strong FIP effects in polar plumes is questionable. Even so, the possibility of strong FIP effects in these structure is supported by our model, as Alfv\'en waves can have considerable amplitudes low in the chromosphere, and the sensitivity of our open field FIP model to the assumed slow mode wave amplitude is perhaps realistic.

\subsection{The helium abundance}

The ponderomotive force was originally invoked to explain just the FIP effect \citep{laming04}, i.e., the enhancement in coronal abundance of the low FIP ions. Once more accurate treatments involving
the non-WKB analysis of Alfv\'en wave transport were implemented \citep{laming09,laming12}, it
was noticed that as well as providing a satisfactory explanation of the FIP effect, the model also
predicted a depletion of the He abundance, (and to a lesser extent Ne also) with respect to O
in the corona and
solar wind. As noted in Section~\ref{sec:solar_FIP}, in the slow solar wind especially, the
He/H ratio is reduced from its photospheric value and becomes more variable in slower speed slow
solar wind. The He abundance is also depleted in the fast solar wind, but to a lesser extent, and
is much less variable.

\citet{rakowski12} investigated models for the depletion of He in more detail. They showed that
similar behavior is seen in the ratio He/O measured by ACE/SWICS (see Figure~\ref{fig:He} left panel); a depletion of He to about 0.4\,--\,0.8 of the solar photospheric value, with greater depletion seen the slower the solar wind
speed, although the solar cycle dependence is not apparent in the time interval 1998\,--\,2011.
\citet{rakowski12} constructed models for a variety of loop lengths and corona magnetic fields,
all using the \citet{avrett08} model chromosphere, and studied the fractionation produced as a
function of the assumed Alfv\'en wave frequency with respect to the loop resonance frequency.
For each case, the coronal wave amplitude was chosen to give a fractionation Fe/O by a factor
of 4, as observed in the slow speed solar wind. As we have seen above, for waves that resonate
with the coronal loop, most of this fractionation occurs at the top of the chromosphere, whereas
off resonance, (or in open field regions) fractionation lower down in the chromosphere is possible.

\epubtkImage{Hef4}{%
\begin{figure}[htbp]
\centerline{
\includegraphics[width=0.5\textwidth]{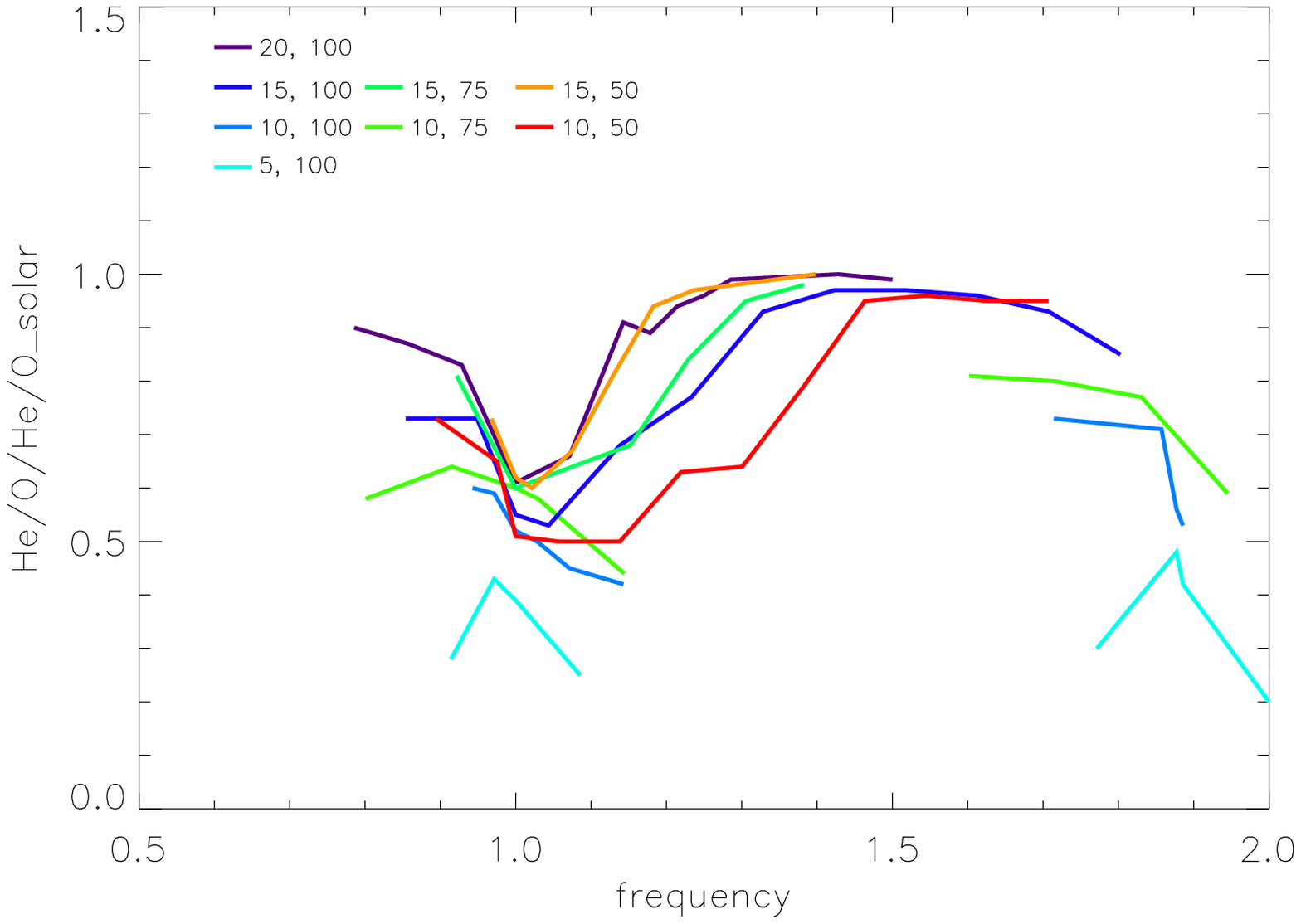}
\includegraphics[width=0.5\textwidth]{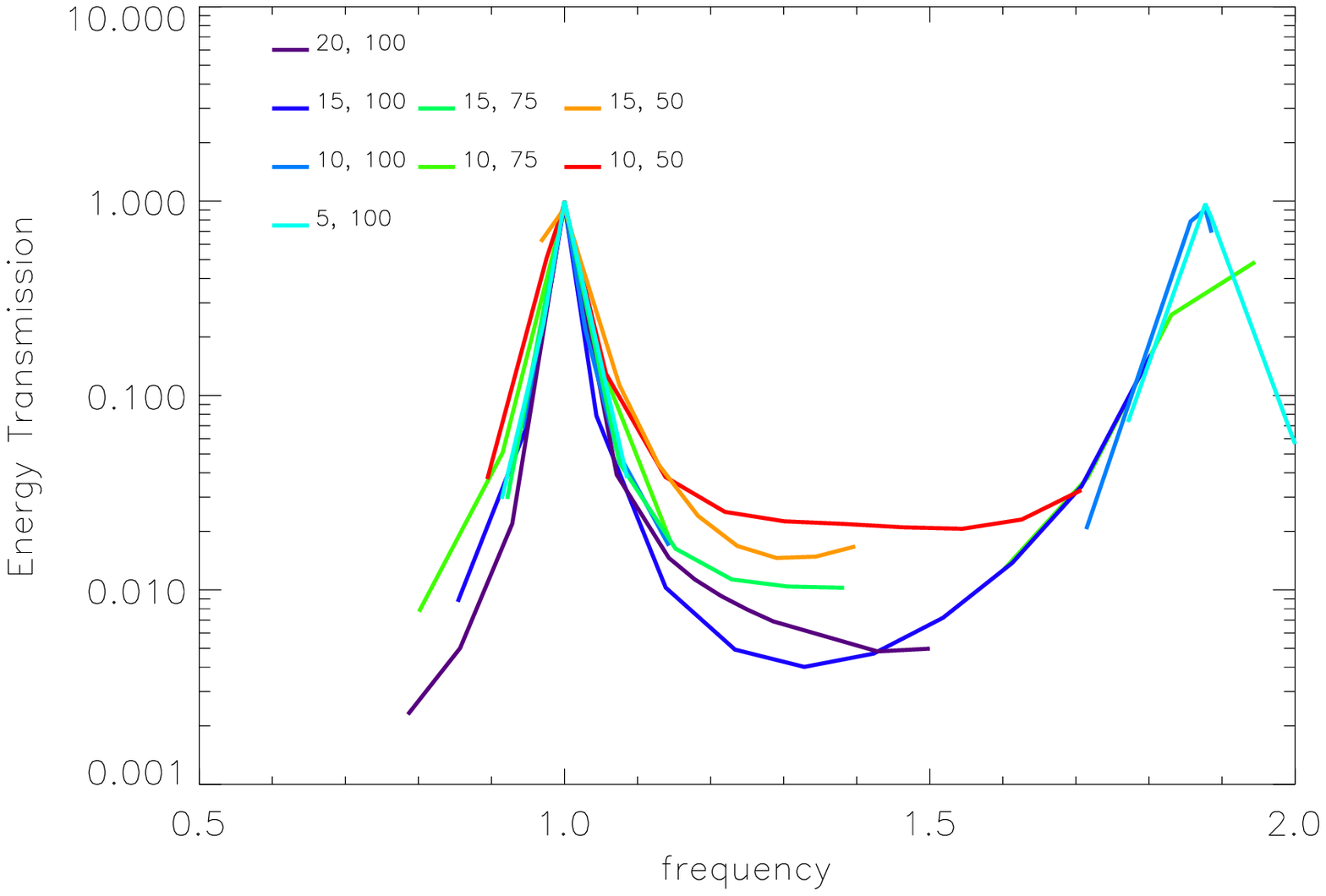}
}
\caption{\emph{Left:} He/O relative to the solar photospheric ratio as a function of the wave frequency, in units of the loop resonance frequency, for different loop lengths and magnetic fields. A trend of increasing He depletion with decreasing magnetic field strength, and possibly also increasing loop length, can be seen. \emph{Right:} Wave energy transmission coefficients for the
same loop models as a function of wave frequency. Shorter wavelengths (lower magnetic field, shorter loops) have broader resonances and are more easily transmitted the chromosphere-corona boundary. From \citet{rakowski12}. Reproduced by permission of the AAS.\label{fig:He}}
\end{figure}}

Results are shown in Figure~\ref{fig:He} left panel. Helium is seen to be strongly depleted
with respect to oxygen when the wave frequency coincides with the loop resonance. In these
cases, the ponderomotive acceleration is restricted to the top of the chromosphere. This means
that O can be accelerated into the corona once it becomes ionized, while He, the last element to remain neutral, is left behind. Elsewhere, where waves are not on resonance, the ponderomotive
acceleration develops over a greater range of chromospheric altitudes. Correspondingly
the acceleration in the region where O is ionized is smaller, since we are restricting ourselves
to cases where Fe/O fractionates to a factor of 4. Hence O is not accelerated relative to He to the same extent, and the He/O ratio is unaffected, while otherwise the usual FIP effect still develops,
albeit with subtle changes in the fractionation pattern.

Closer inspection of the left panel of Figure~\ref{fig:He} also reveals greater depletion of
He at the resonance with weaker magnetic fields. Again, the ponderomotive acceleration is becoming
more concentrated towards the top of the chromosphere, but for a different reason. As the
magnetic field weakens, the chromospheric layer where the plasma $\beta =1$, the equipartition
layer, moves to higher altitudes. At this layer, a myriad of wave phenomena occur; reflection,
transmission and mode conversion, and we do not attempt to model this except to say that FIP
fractionation must occur above this layer. Hence as this layer moves upwards, ponderomotive
acceleration is more restricted to the upper chromospheric layers, and He/O becomes more
depleted. This is most easily seen in Figure~\ref{fig:He} for frequencies just above the
resonant frequency.

In principle, the He/O depletion should also depend on loop length. Longer wavelength waves,
resonating with longer loops, are more effectively reflected at density gradients. Thus we would
expect more He/O depletion to be associated with longer coronal loops, and this seems to be
borne out by the three loop lengths with magnetic fields of 15G. The three cases with 10G coronal fields are less clear, mainly because for many wave frequencies, and Fe/O enhancement of 4 could
not be achieved, and so these points are not plotted. The increased penetration of shorter wavelength waves into a density gradient before reflection is illustrated in the right panel of
Figure~\ref{fig:He}. Wave energy transmission coefficients for the
same loop models are plotted as a function of wave frequency. Waves with shorter wavelengths (lower magnetic field, shorter loops) have broader resonances and are more easily transmitted across the  chromosphere-corona boundary.

The fact that lower values of He/O are to be found for longer loops and weaker magnetic fields
may have some bearing on the origin of the slow speed solar wind, and more specifically the
origin of its different speed components. Some authors \citep[e.g.,][]{cranmer07} argue that
both fast and slow speed solar wind originate in open magnetic flux tubes, and that the difference
in fractionation arise from subtle differences in the wave propagation on such structures. We
would argue here that different slow mode wave amplitudes lower down in the chromosphere are
the most likely variable. But in this case, the He depletion in fast and slow should be similar.
Observationally this is clearly not the case, and our inference that the He/O depletion is strongly
dependent on the Alfv\'en wave frequency with respect to the loop resonance, being strongest when
on resonance suggests that the slow solar wind plasma must originate in closed magnetic loops
where it becomes fractionated, and that the Alfv\'en wave causing the fractionation must be
generated in the loops themselves, in order that the loop resonant frequency is selected.

\subsection{Significance of coronal Alfv\'en waves}

The Alfv\'en wave levels suggested above are similar to those observed in solar flares
\citep[e.g.][]{alexander98}, but higher than those usually considered and observed in the solar corona \citep[cf.][]{peter01,depontieu07,peter10,mcintosh11}. The predicted slow mode wave
amplitudes actually match quite well with the non-thermal broadening observed close to loop footpoints in an active region by \citet{baker13}, and correlated with the locations of FIP fractionation (see Fig. \ref{fig:baker}). The Alfv\'en wave values can be quite
reasonable if the MHD fluctuation is confined to a small part
of the coronal flux tube cross section. In this case a two component line profile should be
expected, as in \citet{peter01}, with a narrow component with nonthermal line broadening with peak amplitude of $\sim 25$ km s$^{-1}$, and a broad component
corresponding to Alfv\'en waves with peak amplitude of $\sim 100$ km s$^{-1}$. Depending on the
filling factor of the strongly oscillatory plasma, this second component may or may not
be readily detectable.
Filamentary models of coronal and flare heating have been invoked to
explain flare lightcurves \citep{warren05,warren06} and stellar
coronal emission measure distributions \citep{cargill06}.
Figure~\ref{fig:nanoflare} shows in the left panel the results of a 3D compressible MHD
simulation including parallel heat conduction and radiation, of a coronal loop subject to
forcing at its footpoints by photospheric motions \citep{dahlburg12}. The plane depicted is
across the midpoint of the loop, and one can easily see localized temperature hotspots developing.
Physically, the magnetic field is becoming stressed and periodically releasing the stored
magnetic energy in small scale reconnection event. The right panel show observations with
the Hi-C instrument \citep{testa13} of the footpoint region of a coronal loop. Filamented, and
variable emission in Fe XII can be see, interpreted as the thermal conductive response to
filamented and sporadic heating events higher up in the coronal portion of the loop. The left panel
is scaled to 4000~km on a side, and so represents a similar spatial scale to that observed by
\citet{testa13}. The Fe XII emission peaks at a temperature of about $1.6\times 10^6$ K, just above
the top of the temperature scale in the simulation. The filamentation appears on an angular
scale of arseconds or smaller, and thus is only resolvable by Hi-C. Scales of this size would
have been below the angular resolution of prior imaging spectrometers or imagers, and consequently any Alfv\'en waves associated with such structures would have been difficult to
detect.

\epubtkImage{3temp}{%
\begin{figure}[htbp]
\centerline{
\includegraphics[scale=0.85]{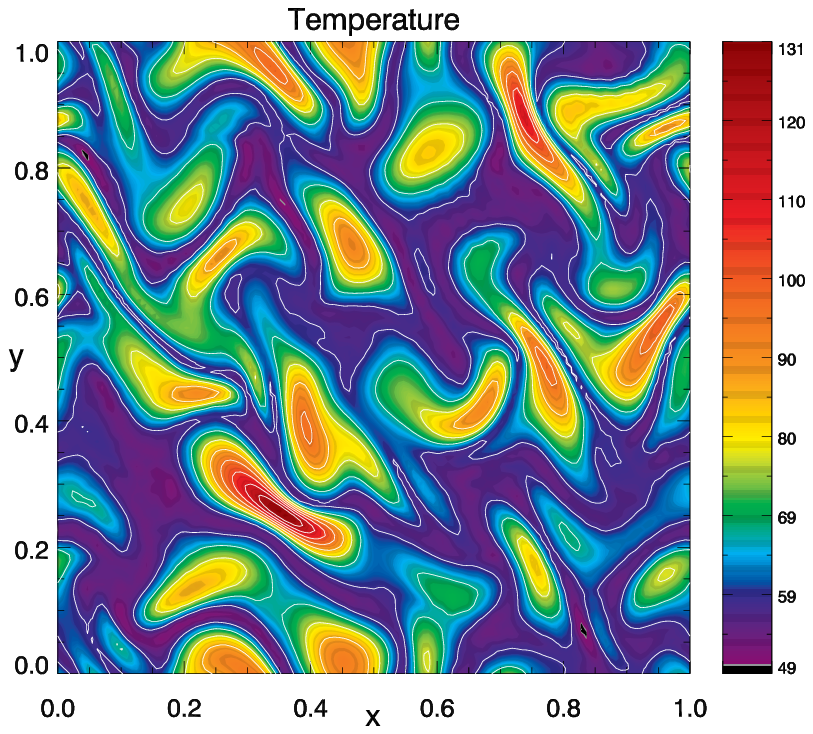}
\includegraphics[scale=0.375]{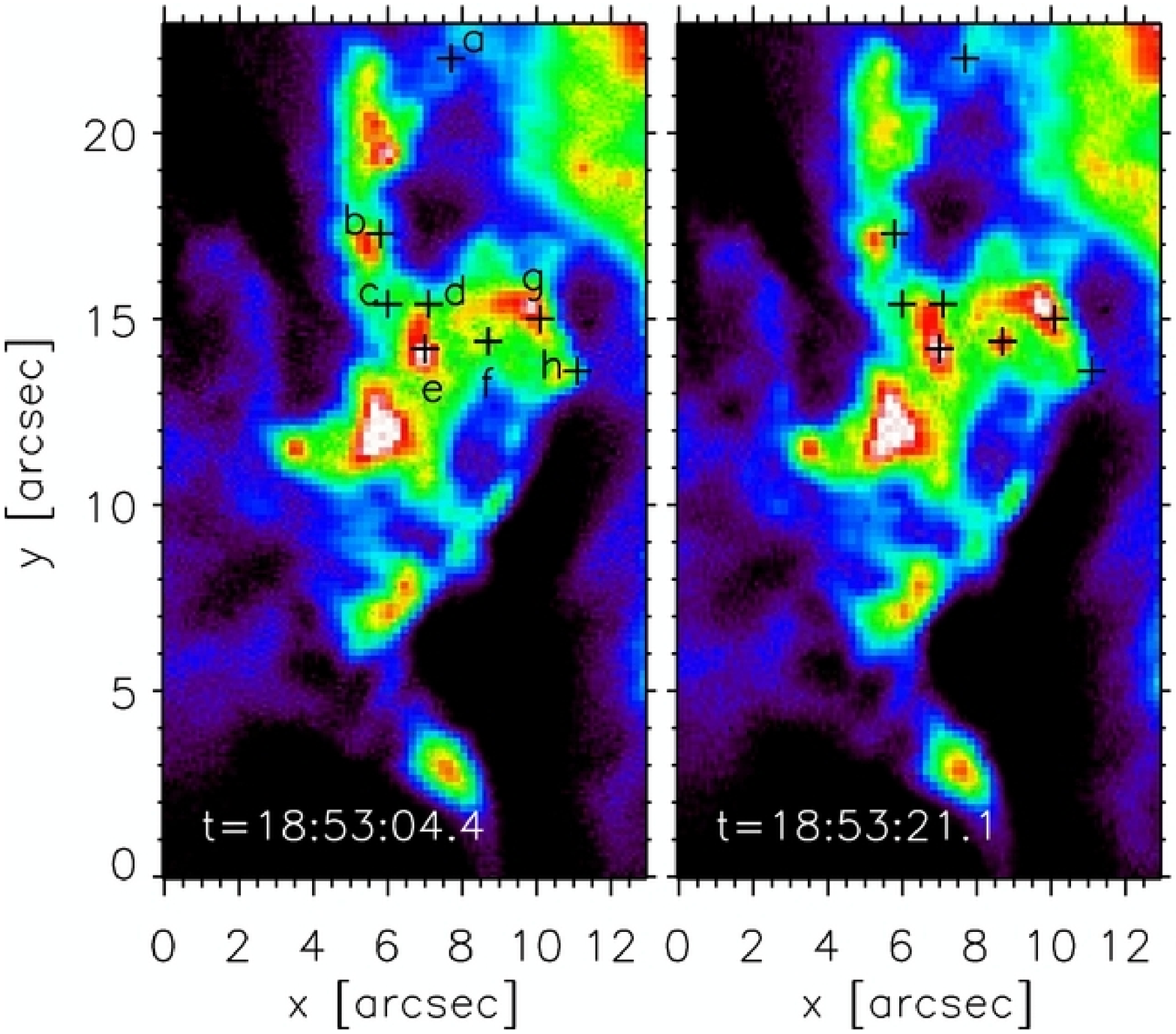}
}
\caption{\emph{Left:} Results of 3D MHD simulation from \citet{dahlburg12}. Temperature contours in the loop mid-plane are shown, illustrating the filamentary nature of the loop heating. The length scale is 4000~km on a side. The temperature scale runs from $4.9\times 10^5$~K to
$1.3\times 10^6$~K.
Reproduced by permission of the ESO. \emph{Right:} Observations of active region moss from \citet{testa13}. Filamentary and variable emission is seen, most likely as a response to filamentary heating in the coronal portion of the loop, communicated to lower altitudes by heat conduction. At
1 arcsec $\sim$~750~km, this structure is of comparable size to the simulation. Reproduced by permission of the AAS.\label{fig:nanoflare}}
\end{figure}}

The filamentation is usually taken to derive from nanoflare heating (see below),
though it could also refer to models of coronal heating by Alfv\'en
resonance. In a flux tube with cross-B density gradient scale length
$l$, a kink mode oscillation of the flux tube becomes Alfv\'enic at a
resonant surface where the wave frequency $\omega = 2L/V_A$, where $L$
is the loop length and $V_A$ is the Alfv\'en speed \citep[see,
 e.g.,][]{ruderman02}.  The width of this resonant layer is $\delta
\sim \left(l\nu /\omega\right)^{1/3}$, where $\nu = aV_A/R$ is the
kinematic viscosity, $a$ is the loop radius and $R$ is the Reynolds
number. The Alfv\'enic velocity fluctuations are larger than those of
the kink mode by a factor $l/\delta =\left(l^2R/aL\right)^{1/3}\sim
0.1 R^{1/3}$ under typical conditions.
Such wave
motions observed in the solar corona have been identified as kink (a
subset of fast mode) mode waves \citep[e.g.,][]{nakariakov99,wang04}. Alfv\'en
waves which generate the ponderomotive force in the chromosphere have been much harder to detect \citep{vandoorsse08}, though see
\citet{erdelyi07} and \citet{tomczyk07}.
The global kink modes above presumably derive from acoustic motions in
the chromosphere driving loop footpoints initiating the loop
oscillation, that then subsequently decay into Alfv\'en waves at the
resonant surface. Turbulence associated with heating on
particular field lines will only produce fractionation if
chromospheric upflows are restricted to these field lines. However,
if the upflow is the result of chromospheric evaporation, this is
precisely what we should expect.

The nanoflare paradigm suggests that the dominant
loop footpoint motions are of much lower frequency, and do not
excite oscillations in the loop but act so as to build up magnetic
stresses in the corona. These stresses periodically release
themselves, in what has become known as a ``nanoflare'', as a
current sheet develops \citep{p88}. \citet{rappazzo07,rappazzo08} discuss the
buildup of magnetic stresses within the framework of turbulence
phenomenology, where it appears that velocity perturbations similar
to the $\sim$~30\kms observed
should be expected. \citet{longcope09} conjecture that
in impulsive reconnection in post-flare loops, only about 10\% of
the liberated magnetic energy is converted directly into heat, the rest
reappearing as kinetic energy that ultimately drives turbulence. In the case of reconnection of field lines at angle $\theta$, equating magnetic energy destroyed
to kinetic energy gained, $B^2\sin ^2\left(\theta /2\right)/8\pi = \rho v^2/2$, suggests $v=v_A\sin\theta /2$.
For $B=20$ G and $\rho = 1.67\times 10^{-24}\times 10^9$ g cm$^{-3}$, $v_A=1400$\kms,
and $v=100$\kms implies $\theta = 8^{\circ}$, slightly larger than the expectation of
\citet{rappazzo13}. Of course some magnetic energy may go
directly to heat, but it is plausible that such wave generation explains why
surveys to find localized hot plasma as evidence of nanoflare reconnection have generally
been unsuccessful \citep[e.g.,][]{warren11}. Instead, energy goes from magnetic field to
waves, and is thus gradually dissipated as heat throughout the corona, and not quickly and
locally as might have been expected.

\citet{sturrock99} gives a pedagogic
review of the mechanisms by which various wave modes may be excited by
reconnection. The reconnected
field line is generally distorted, and this can either propagate away
from the reconnection site as an Alfv\'en wave, or emit
magnetoacoustic waves traveling perpendicularly to the magnetic field
direction. \citet{isobe08} model small scale reconnection in the
chromosphere. The emerging magnetic flux reconnects with the
previously open field, to produce a reconnection jet accompanied by an
upward propagating Alfv\'en wave, which appears to be of appropriate
frequency (0.01~Hz) and amplitude (20\kms in the transition
region) to give rise to some fractionation.  Observational evidence of
low-lying reconnection producing jets and transverse waves of similar
frequencies and amplitudes has been reported
\citep{nishizuka08,he09,vasheghani09}. \citet{kigure10} explicitly
consider the generation of Alfv\'en waves by magnetic reconnection,
and find that a significant fraction of the magnetic energy released
(several tens of \%, depending on geometry and plasma $\beta$) can be
carried off by Alfv\'en or magnetoacoustic (fast or slow mode) waves,
with Alfv\'en waves dominating for $\beta <1$. \citet{liu11a} and \citet{liu11b} discuss the role
of temperature anisotropies and wave generation by the firehose instability
in the outflow.

\Citet{vanball11} challenge some of these ideas and offer a more traditional view of an
Alfv\'en wave heated corona, where Alfv\'en waves are introduced at loop footpoints from below and are either transmitted or reflected, following \citet{hollweg84}. They argue that ``nanoflare'' heating, as in Parker's concept, cannot supply sufficient heat to the corona. However, the simulations of \citet{dahlburg12}, \citep[see also][]{gudiksen04,bingert11},
discussed above seem to contradict this statement. A million degree corona can indeed be heated and maintained by random motions of loop footpoints with timescale longer than the loop resonance.
Further, \citet{rakowski12} show that extended Alfv\'en wave propagation throughout the chromosphere, as in \citet{vanball11}, is unlikely to correctly predict the depletion in the coronal abundance of helium.

\epubtkImage{cally1}{%
  \begin{figure}[htb]
    \centerline{
      \includegraphics[height=7cm]{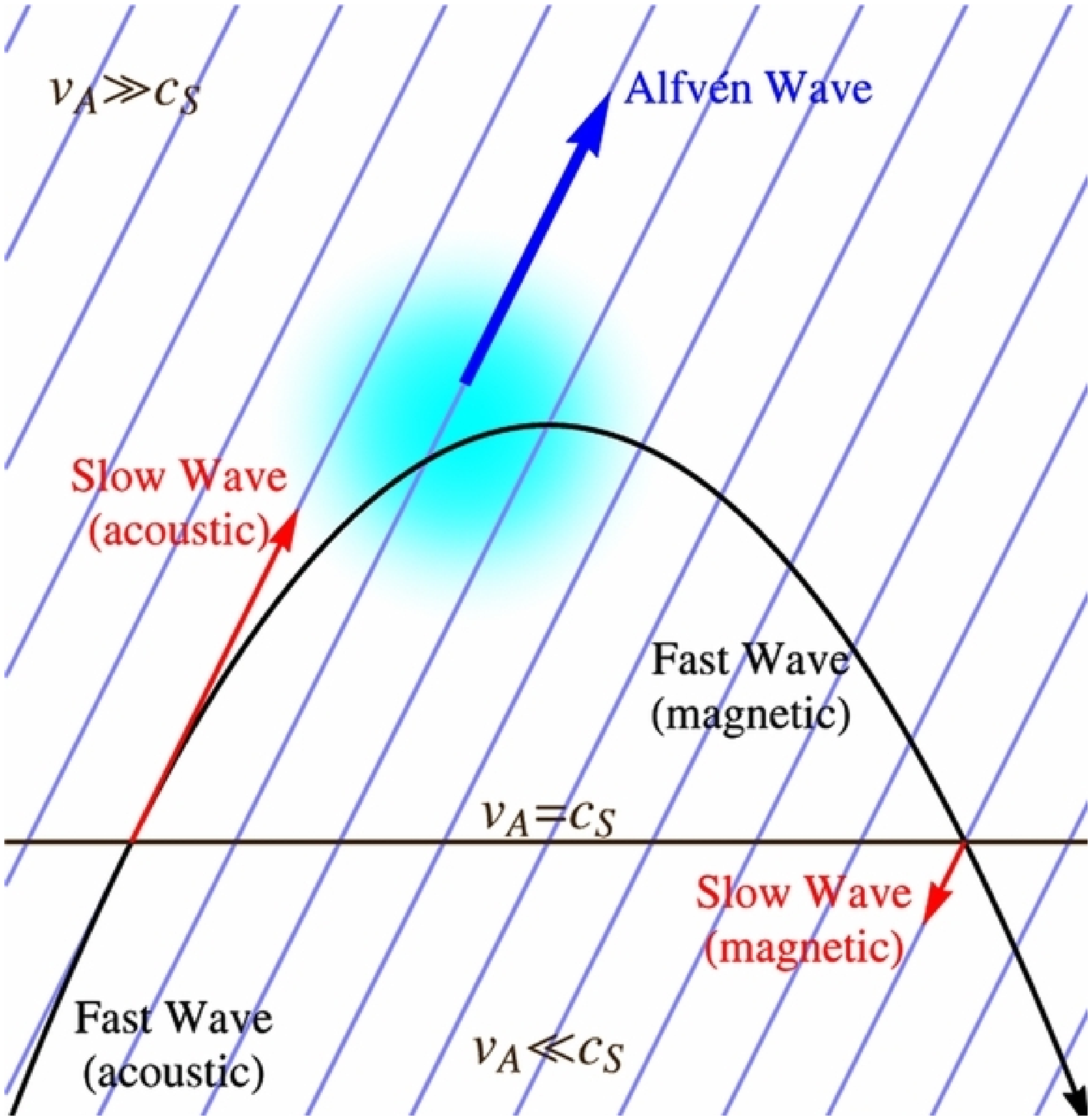}\qquad
      \includegraphics[height=7cm]{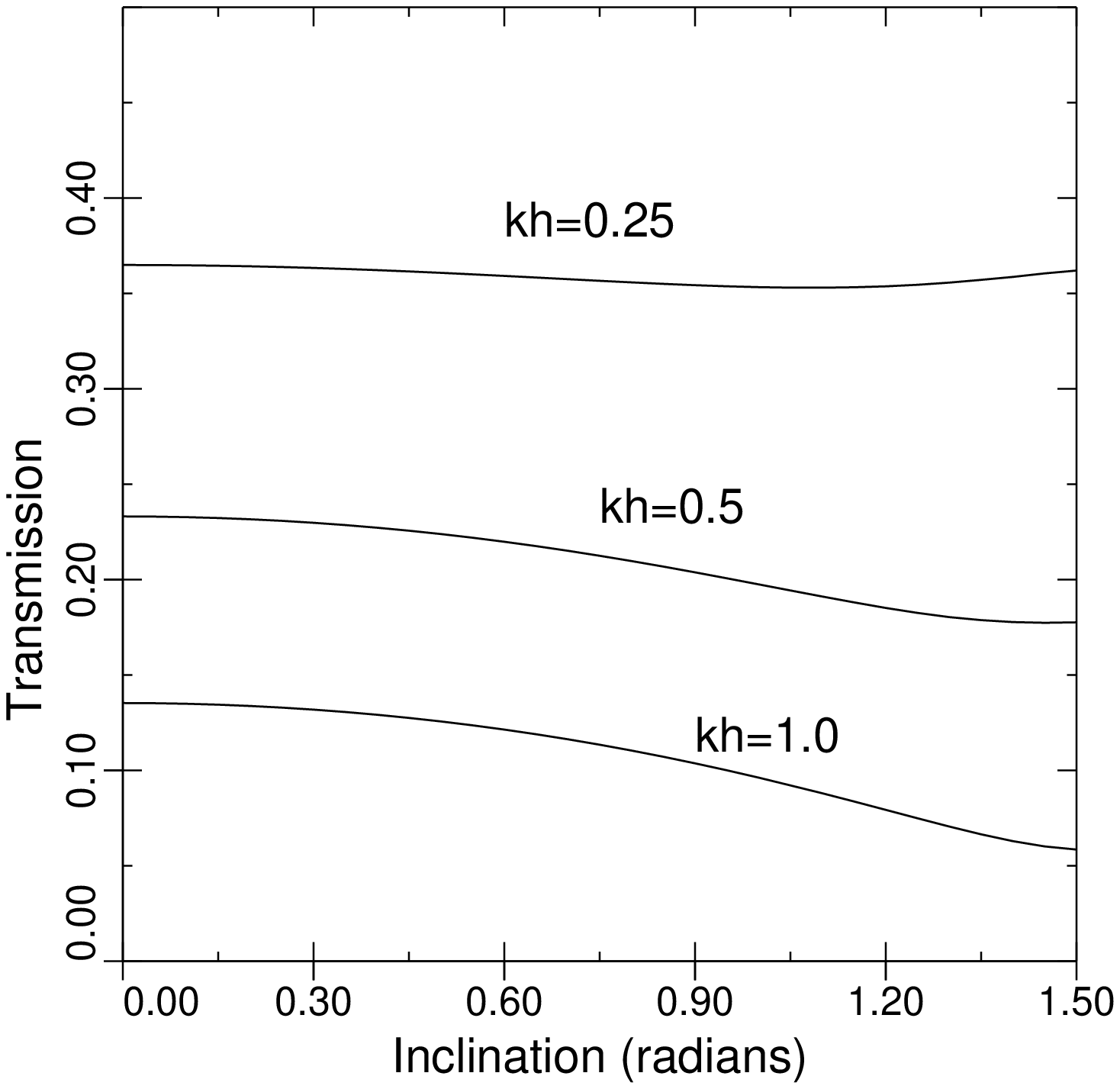}
    }
\caption{\emph{Left:} Schematic diagram from \citet{khomenko12} of acoustic to magnetosonic wave mode conversion at the plasma $\beta =1$ layer of the chromosphere. An upcoming acoustic (fast mode) wave where $v_A \gg c_S$ ($\beta \gg 1$) can be transmitted into the overlying region where $v_A \ll c_S$ ($\beta \ll 1$) as an acoustic (slow mode) wave, or mode converted to a magnetoacoustic (fast mode) wave. The magnetoacoustic wave refracts in the region where $v_A$ is increasing with height
and eventually rejoins the lower $\beta \gg 1$ plasma. The transmitted slow mode can continue propagating until it reaches an acoustic cut-off. When the plane in which the fast mode refracts
is distinct from the plane of inclination of the magnetic field (indicated here by blue lines), it may also mode convert to an Alfv\'en wave, which is not considered further in this paper. Reproduced by permission of the AAS. \emph{Right:} Transmission coefficients for isotropic acoustic waves (in the upgoing hemisphere) as a function of the inclination of the magnetic field, for different values of $kH_D$, calculating by integrating a generalization of Equation~(\ref{eq:transmission}) in the region $0\le\theta\le\pi //2$ and $0\le\phi\le 2\pi$. \label{fig:modeconv}}
\end{figure}}

\subsection{Modeling the inverse FIP effect}

One of the striking features of the Wood--Linsky relation illustrated
in Figure~\ref{fig:woodlinsky} is how the coronal abundance anomaly
smoothly changes from solar-like FIP bias at spectral types G to early
K, to inverse FIP in M dwarfs. This suggests that a model for
solar-like FIP fractionation should, with suitably chosen parameters, be capable of predicting an inverse FIP effect.

The model described above and in \citet{laming04,laming09,laming12} for the FIP effect suggests
that Alfv\'en waves of amplitude approximately 100\kms are generated in the coronal loop
at the resonant frequency, and remain trapped in the loop ``resonant cavity''. Upon
reflection from chromospheric footpoints, the waves develop a ponderomotive force in the
steep density gradients there, and this force, acting on chromospheric ions (but not neutrals),
preferentially accelerates these ions up into the corona. \citet{laming12} studies the
fractionations produced by waves on and off resonance, and \citet{rakowski12} extend
this to different loop lengths and magnetic fields, concentrating mainly on the
fractionation of He with respect to O. These works only consider coronal Alfv\'en waves,
with chromospheric acoustic waves included following \citet{heggland11} and \citet{cranmer07}
as terms in the denominator of the integrand in Equation~(\ref{eq:22}). When these upcoming
chromospheric acoustic waves are allowed to mode convert, at the layer where sound and
Alfv\'en speeds are equal, to what in the magnetically dominated upper chromospheric
become fast mode waves, inverse FIP fractionation can result. This arises because the fast
mode waves undergo reflection back downwards as the Alfv\'en speed increases, giving rise
to a downwards directed ponderomotive force than can compete with that due to the coronal
Alfv\'en waves. Inverse FIP requires $\partial\delta E^2/\partial z < 0$. If waves propagate from chromosphere to corona (or from corona to chromosphere) without reflection, then
$\partial\delta E^2/\partial z > 0$ always. This is because $\delta E = \delta v B$ (where $B$ is ambient field, assumed uniform here) and $\delta v$ increases as the density decreases.
If waves propagating up from beneath are reflected back down again, $\delta v$ and $\delta E$ must decrease with height, and $\partial\delta E^2/\partial z < 0$, giving rise to a downward
ponderomotive force.

The process of mode conversion is shown schematically in the left panel of Figure~\ref{fig:modeconv}, taken from
\citet{khomenko12}. An upcoming acoustic (fast mode) wave where $v_A \ll c_S$ ($\beta \gg 1$) can be transmitted into the overlying region where $v_A \gg c_S$ ($\beta \ll 1$) as an acoustic (slow mode) wave, or mode converted to a magnetoacoustic (fast mode) wave. The magnetoacoustic wave refracts in the region where $v_A$ is increasing with height
and eventually rejoins the lower $\beta \gg 1$ plasma. The transmitted slow mode can continue propagating until it reaches an acoustic cut-off. When the plane in which the fast mode refracts
is distinct from the plane of inclination of the magnetic field (indicated here by blue lines), it may also further mode convert to an Alfv\'en wave, which is not considered further in this paper.
The acoustic waves have an energy transmission coefficient of \citep{cally08}
\begin{eqnarray}
T&=&\exp\left(-\pi \left|{\bf k}\right|\sec\theta\left(1-\left(\sin\alpha\cos\phi\sin\theta
+\cos\alpha\cos\theta\right)^2\right)\over \left[d\left(V_A^2/c_S^2\right)/dz\right]_{\beta =1}\right)\cr
&=&\exp\left(-\pi \left|{\bf k}\right|H_D\sec\theta\left(1-\left(\sin\alpha\cos\phi\sin\theta
+\cos\alpha\cos\theta\right)^2\right)\right)\cr
&&\rightarrow \exp\left(-\pi\left|{\bf k}\right|H_D\sec\theta\sin ^2\theta\right),
\label{eq:transmission}
\end{eqnarray}
where $\alpha$ and $\theta$ are the angles to the vertical made by the magnetic field and the
wavevector respectively,
and $\phi$ is the polar angle between them. The last line gives the more familiar
result in vertical magnetic field. The right panel of Figure~\ref{fig:modeconv} shows the
acoustic to acoustic energy transmission coefficient for initially isotropic waves (in the
upward going hemisphere) as a function of magnetic field inclination to the vertical (the
angle $\alpha$ above), for values of $\left|{\bf k}\right|H_D=0.25$, 0.5, and 1. For solar
oscillations with period 300 seconds, $kH_D =\sqrt{\omega ^2H_D^2/c_S^2 -1/4}=0.25$ where
we have taken $H_D=168$~km from Figure~\ref{fig:chromodel} and $c_S=kT/m_p\simeq 6.3$\kms.
In later type stars with deeper convection zones, we might expect $kH_D$ to increase, since
$\omega\sim g/c_S$ increases with increasing $g$ and decreasing $c_S$ in the colder photospheres
\citep[see, e.g.,][]{bruntt10,kjeldsen11},
while $H_D$ in the low chromosphere remains approximately constant, as far as can be determined from model chromospheres available in the literature \citep[e.g.,][]{fuhrmeister05,houdebine97,vieytes05}.
These factors would increase the degree of mode conversion in later type stars. Solar $p$-modes are also known to decrease in intensity and increase in width from solar minimum
to maximum \citep{chaplin00,simoniello10}. This suggests that the trapping of acoustic modes within the solar envelope becomes less effective at solar maximum, and allows energy to leak out
\citep{pinter01}. Sunspots have long been known to be sinks of $p$-mode energy \citep[see][and
references therein]{braun95}, most likely through mode conversion or resonant absorption to
Alfv\'en or fast mode waves. $P$-modes are also known to decrease in intensity for later spectral
types, modeled by \citet{kjeldsen11} as due to extra ``leakage''.

A further reason for the transition from FIP to inverse could be that in stars on the left hand side of Figure~\ref{fig:woodlinsky}, upcoming acoustic waves
encounter an acoustic cut-off before reaching the $\beta =1$ layer, and never have a chance to mode
convert. In the stronger magnetic fields of later type stars, the $\beta =1$ layer is deeper in
the chromosphere, and upcoming acoustic waves will reach this first, before any acoustic cut off,
and mode convert to fast mode waves. The fast mode waves are immune to the acoustic cut off, but
ultimately refract back downwards.

We treat the fast mode waves as approximately isotropic in the upward moving hemisphere,
following \citet{wood12}. Then
the fraction reflected at chromospheric height $z$ is
\begin{equation}
f_R\left(z\right)\simeq\sqrt{1-{c_S^2\left(z_{\beta =1}\right)\over V_A^2\left(z\right)+c_S^2\left(z\right)}}
\end{equation}
where $z_{\beta =1}$ is the chromospheric height where mode conversion occurs. The ponderomotive
acceleration due to fast modes waves is then
\begin{equation}
a={c^2\over 4}{\partial\over\partial z}\left(\delta E^2\over B^2\right)={\delta v^2\over 2}\left(1-f_R\right){1\over\delta v}{\partial\delta v\over\partial z}-{\delta v^2\over 4}
{\partial f_R\over\partial z}\,.
\end{equation}
The two terms represent an upwards contribution arising as the fast mode waves increase
in amplitude as they propagate through lower density plasma, and the downwards contribution
arising from fast mode wave reflection. Evaluating
\begin{equation} {\partial f_R\over\partial z} ={c_S^2\left(z_{\beta =1}\right)V_A\over\left(V_A^2+c_S^2\right)^2f_R}{\partial V_A\over\partial z}= {c_S^2\left(z_{\beta =1}\right)V_A^2\over\left(V_A^2+c_S^2\right)^2f_R} \left({1\over H_B}-{1\over 2H_D}\right)
\end{equation}
and assuming from the WKB approximation
\begin{equation}
{1\over \delta v}{\partial \delta v\over\partial z}={-1\over 2H_B}-{1\over 4H_D}\,,
\end{equation}
where $H_D$ and $H_B$ are the signed density and magnetic field scale heights, we find from equation 5
\begin{equation}
a={\delta v^2\over f_R}\left\{\left(f_R-1\right)\left(-{1\over 8H_D}-{1\over 4H_B}\right)
+{c_S^2\left(z_{\beta =1}\right)\over\left( V_A^2+c_S^2\right)^2}\left(-{c_S^2\over 8H_D} -{c_S^2\over 4H_B}-{V_A^2\over 2H_B}\right)\right\}.
\label{eq:fastmodepond}
\end{equation}
Remembering that both $H_D$ and $H_B$ are negative, and $f_R < 1$, the first term in curly
brackets is negative, giving rise to Inverse FIP effect, and the second term is positive, giving
the more usual FIP effect. In conditions where $V_A \gg c_S$, an overall downwards pointed ponderomotive acceleration requires $\left|H_D\right| < \left|H_B\right|/6$ in this simple model.

Additional reflection of fast mode wave from, e.g., density fluctuations (not included in this model) would relax the requirement. So too would fast mode waves spreading out laterally from a horizontally localized source. Even so, Equation \ref{eq:fastmodepond} implies that inverse FIP effect is more likely to found
in stars with minimal magnetic field expansion through the chromosphere, which fits with its
observation in M~dwarfs. While the magnetic fields measured in these stars are similar to those
in the Sun, the filling factor is higher \citep[e.g.,][]{donati09,reiners09}, allowing less volume for
expansion with increasing altitude.

\epubtkImage{fig5}
\begin{figure}[htbp]
\centerline{\includegraphics[width=\textwidth]{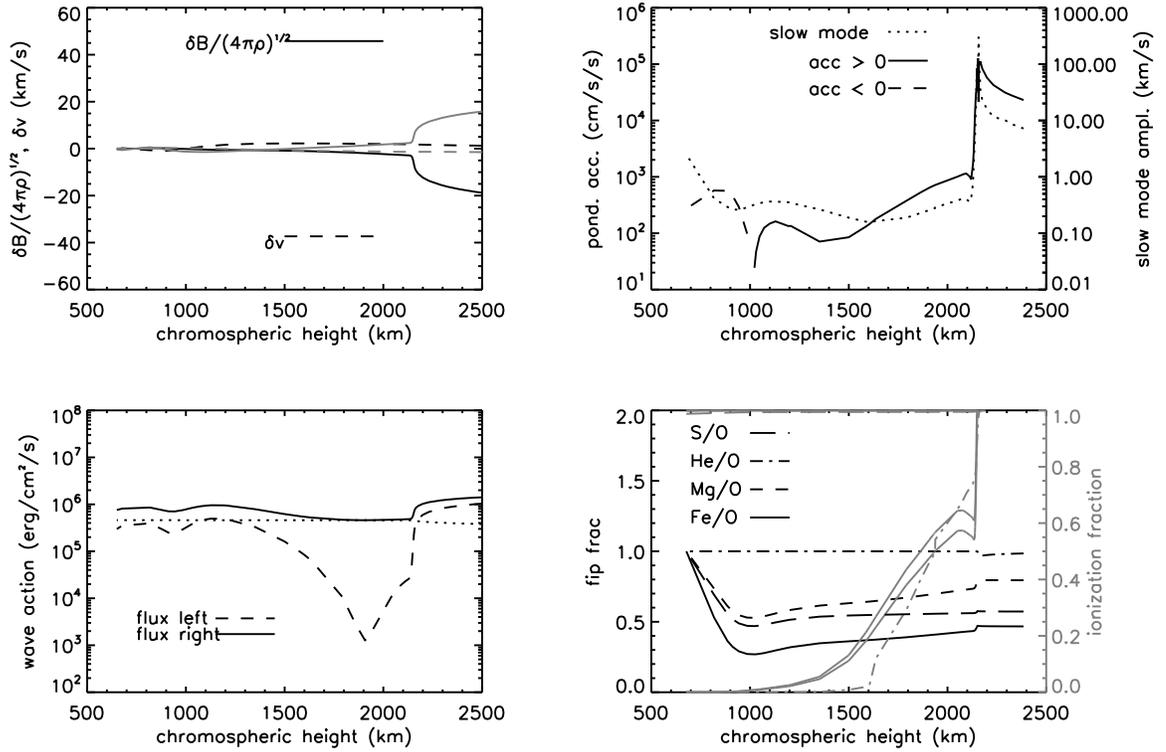}}
\caption{Illustration of a model demonstrating the origin of the Inverse FIP Effect. The top
left panel shows the variables $\delta v$ and $\delta B/\sqrt{4\pi\rho}$ for the coronal
Alfv\'en wave. The bottom left panel shows the upgoing (dashed curve) and downgoing (solid curve)
wave energy fluxes. The dotted line shows their difference. The top right panel shows the ponderomotive acceleration. The positive contribution in the upper chromosphere
(solid curve) comes from the coronal Alfv\'en waves. The negative contribution lower down (dashed curve) comes from the total internal reflection of fast mode waves. The positive fractionation induced by the Alfv\'en waves is suppressed by the choice of evaporative flow speed, which is higher in the lower density regions of the chromosphere due to continuity. Suppression
of fractionation low in the chromosphere by slow mode waves is reduced in this example, because the slow modes mode convert to fast modes.} The dotted curve gives the amplitude of acoustic waves through the chromosphere. The bottom right panel gives the FIP fractionations for the ratios S/O, He/O, Mg/O and Fe/O. He/O remains unchanged, but all others display an inverse FIP effect. The choice of an ``active region'' spectrum from \citet{vernazza78} has increased the ionization fraction of He, compared to the earlier examples. \label{fig:fig5}
\end{figure}

Figure~\ref{fig:fig5} illustrates a calculation designed to give an inverse FIP effect.
The model is similar to that
shown in Figure~3 of \citet{laming12}. A loop of length 100\,000~km, with a 80~G magnetic field is considered, with a resonant Alfv\'en wave of frequency 0.28 rad s$^{-1}$. The magnetic field is compressed by a factor 0.5 through the chromosphere (previously it was 0.2), and a fast mode wave amplitude
of 10\kms is included at the $\beta = 1$ layer, which is allowed to propagate and refract
as described above. The model chromosphere derives from \citet{avrett08}, though future work
should implement a model stellar chromosphere. The top left panel shows the variation of
the perturbations $\delta v$ and $\delta B/\sqrt{4\pi\rho}$ associated with the coronal Alfv\'en wave, and the bottom right panel shows the Alfv\'en wave energy fluxes, both upward and downwards
directed. The dotted line in the bottom right panel shows the difference in wave energy fluxes.
The top
right panel shows the ponderomotive acceleration. The positive contribution in the upper chromosphere
(solid curve) comes from the coronal Alfv\'en waves. The negative contribution lower down (dashed curve) comes from the total internal reflection of fast mode waves. The dotted curve gives the amplitude of acoustic waves through the chromosphere, modeled as outlined above. The bottom right panel gives the FIP fractionations for the ratios S/O, He/O, Mg/O and Fe/O. He/O remains unchanged, but all others display an inverse FIP effect, coming from the fast mode waves. The positive FIP
that would arise from the Alfv\'en waves higher in the chromosphere is suppressed by the upward flow speed through the chromosphere, taken here to be $10^6$ cm s$^{-1}$ at a chromospheric density of
$10^{10}$ cm$^{-3}$, and of course smaller lower down in the chromosphere due to the increased density. This upflow is now included in the Alfv\'en wave transport equation 11. Upflows of at least this speed are typical of the gradual phase of solar flares \citep[e.g.,][]{czaykowska99}. High in the chromosphere, the He ionization fraction is
increased relative to the earlier examples by the choice of an ``active region'' spectrum from\citet{vernazza78} with which to illuminate the chromosphere. This will reduce any
depletion of He that might otherwise occur in conditions giving rise to positive FIP
effect, and might be relevant to the relatively high He abundances observed in flares
and CMEs \citep[e.g.][]{feldman05,wimmer06}.

The transition from FIP effect to inverse FIP effect with increasing fast mode wave amplitude
at the $\beta =1$ layer is illustrated in Table~\ref{table:inverseFIP}. Models with fast modewave amplitudes of 5, 10, and 15\kms are compared with abundances in M dwarfs taken from
\citet{liefke08}, who compare observed coronal abundances with solar photospheric abundances of
\citet{asplund05c}. The coronal abundances for EV Lac given by \citet{laming09} are in better
agreement with the models, especially for Ne/O and S/O, but even so, it is clear that the broad
systematics of the inverse FIP effect are well reproduced by the models in Table~\ref{table:inverseFIP}. Results for further minor ions observed in $\sigma$ Gem and HR 1099 are taken from \citet{huenemoerder13a,huenemoerder13b}. The Alfv\'en wave amplitude
has been reduced from that in Figure~\ref{fig:loop} by a factor of two. \citet{wood13}
argue that this might be appropriate in the strong magnetic fields of
late type stellar coronae: \citet{drake06} studied the efficiency of electron acceleration in reconnection with
the ambient plasma $\beta$ (the electron plasma $\beta_e = 8\pi n_ek_{\mathrm{B}}T_e/B^2$, to be more precise). They
found maximum energy input to electrons at $\beta_e = 0$, with reduced electron heating at higher
$\beta_e$, or lower magnetic field. We suggest that at the left hand side of the
Wood--Linsky relation,
coronal reconnection primarily generates Alfv\'en waves that end up causing positive FIP
fractionation when they reflect from the chromosphere. As one moves to the right, to later
spectral type, coronal reconnection puts more energy into electrons, and less into waves.
Consequently, the positive FIP effect diminishes.

While inevitably a highly simplified model of chromospheric wave processes, we take the potential
for the ponderomotive force to explain the Inverse FIP effect at later spectral types, as well as
the more usual FIP effect in the Sun, as a significant point in its favor. The Inverse FIP
effect is also observed in many other more active stars, and we expect a similar explanation to
hold there. However, the wave origin will be more complicated when additional processes due to
stellar rotation, tidal interaction (e.g., RS CVn binaries) and accretion (e.g., T Tauri stars) are included.

\begin{landscape}
\begin{table}[htbp]
\caption{FIP and Inverse FIP Effects}
\label{table:inverseFIP}
\centering
{\small
\begin{tabular}{l ccc cccccc c cc}
\toprule
Ion &  \multicolumn{3}{c}{Ampl. (km s$^{-1}$)} &\multicolumn{6}{c}{from \citet{liefke08}}
& from \citet{laming09}& \multicolumn{2}{c}{from \citet{huenemoerder13a}}\\
~ &5  &7.5 & 10 & EQ Peg A& YY Gem& AU Mic& EV Lac& AD Leo& Prox. Cen& EV lac& $\sigma$ Gem& HR 1099\\
\midrule
N/O & 0.99& 0.99& 0.99& 1.00& 1.75& 2.08& 1.52& 1.67& 2.94& & 2.88& 1.58\\
Ne/O& 0.98& 0.98& 0.98& 2.7&  2.1&  2.31& 1.71& 2.33& 2.06& 0.96& 2.45& 2.45\\
Na/O& 1.53& 0.87& 0.39&    &     &      &     &     &     &     & 0.78& 1.07\\
Mg/O& 1.49& 0.83& 0.36& 0.52& 0.34& 0.28& 0.38& 0.49& 0.72& 0.23& 0.38& 0.32\\
Al/O& 1.44& 0.75& 0.30&    &     &      &     &     &     &     & 0.63& 0.54\\
Si/O& 1.37& 0.71& 0.28& 1.21& 0.87& 0.51& 0.85& 1.20& 1.80& 0.52& 0.44& 0.36\\
S/O&  1.14& 0.57& 0.22& 1.52& 1.27&     &     &     &     &0.86& 0.46& 0.41\\
Ar/O& 0.98& 0.98& 0.98&     &     &     &     &     &     &    & 1.95& 1.48\\
K/O & 1.57& 0.64& 0.18&     &     &     &     &     &     &    & 6.46& $< 4.07$\\
Ca/O& 1.57& 0.63& 0.17&     &     &     &     &     &     &    & 1.20& 0.93\\
Fe/O& 1.55& 0.47& 0.09& 0.45& 0.29& 0.26& 0.33& 0.52& 0.46& 0.24& 0.38& 0.32\\
\bottomrule
\end{tabular}}
\end{table}
\end{landscape}

\subsection{Saturation}
\label{sec:saturation}
We conclude this section with a simple discussion of what might limit the absolute
magnitude of the FIP fractionation. Alfv\'en waves of arbitrarily high amplitude will
eventually erase the density gradients that give rise to the ponderomotive force. We
estimate the ponderomotive acceleration at which this will occur, and hence a limit
on how high this acceleration can go, as follows.

The density in a gravitationally stratified atmosphere with a ponderomotive acceleration is
\begin{equation}
\rho = {P\over v_s^2}\exp\left(-gz/v_s^2 +\int\xi a\nu _{eff}/\nu _s/v_s^2dz\right)
\end{equation}
where all symbols are defined above in subsection \ref{subsec:6.5}, and here the element
$s$ of interest is H. Then
\begin{equation}
{1\over H_D}={\partial\ln\rho\over\partial z}=-2{\partial\ln v_s\over\partial z}-{g\over v_s^2}
+{2gz\over v_s^2}{\partial\ln v_s\over\partial z}+\xi a{\nu _{eff}\over\nu _sv_s^2}.
\end{equation}
We substitute $a= -\delta v^2/8H_D$ to find
\begin{equation}
{1\over H_D}\left(1+{\xi\over 8}{\nu _{eff}\over \nu _s}{\delta v^2\over v_s^2}\right)=
-2{\partial\ln v_s\over\partial z}\left(1-{gz\over v_s^2}\right)-{g\over v_s^2} ={1\over H_{D0}}
\end{equation}
where $H_{D0}$ is the density scale length when $a=0$, i.e. in the absence of the ponderomotive
acceleration. Consequently
\begin{equation}
a={a_0\over 1+{\xi\over 8}{\nu _{eff}\over \nu _s}{\delta v^2\over v_s^2}}
=\left({1\over a_0} -{H_{D0}\over v_s^2}{1\over 1+\left(1/\xi -1\right)\nu _{si}/\nu _{sn}}\right)^{-1}
\end{equation}
where $a_0=-\delta v^2/8H_{D0}$, the ponderomotive acceleration in the unmodified density gradient. Hence as $a_0\rightarrow\infty$, $a\rightarrow
-\left(1+\left(1/\xi -1\right)\nu _{si}/\nu\right)v_s^2/H_{D0}$ which is in general comparable
to the ponderomotive accelerations invoked in this paper to explain the FIP effect, if we
take $\xi =1$, with $v_s\sim 10^6$ cm s$^{-1}$ and $H_{D0}\sim 10^6$ cm. A small departure
from full ionization ($\xi < 1$ increases this estimate quite quickly, because
$\nu _{si} >> \nu _{sn}$. In this case we would have to appeal to properties of the Alfv\'en waves themselves to explain the relative constancy of the FIP effect. If the waves are restricted to filamentary sections of a coronal loop, and are not monolithic oscillations of the whole
loop itself, then regions of velocity shear will exist between oscillating and non-oscillating
regions of the loop. This velocity may be expected to excite further wave motions, for example
drift waves, which could provide another means of saturating the ponderomotive acceleration. Further developments along these lines are beyond the scope of this review.

\newpage

%===================================================================================
%===================================================================================

\section{Conclusions \& Future Work}
\label{sec:conclusions}

In this review we have attempted to show how the interpretation of abundance enhancements or
depletions arising due to the action of the ponderomotive force due to various MHD waves offers
the potential of a rather complete description of the phenomenon. It may also be hoped that
advances along these lines, together with the detection and observation of waves in the solar
atmosphere will be a profitable route towards solving the problem of coronal heating. Here we
highlight some of the weak spots in the model, and areas where extra effort might provide a
significant advance.

\begin{description}
\item[1. Coronal Waves:] Concentrating first on the solar FIP effect,
our models above give a good account of the observed
abundance anomaly with a coronal Alfv\'en wave amplitude of around 100\kms. This is a
\emph{peak} wave amplitude, and should therefore be expected to give rise to a nonthermal
line broadening around 70\kms, measured at half maximum line intensity. This is higher
than usually observed, but is not necessarily  a problem if the line width is ``diluted'' by
the filamentary structure across the loop. One important
question is then to understand quantitatively where this velocity amplitude originates. We have sketched out some possibilities above, nanoflare associated reconnection and Alfv\'en resonance, but this is an area where numerical simulation \citep[e.g.,][]{dahlburg12} should be expected to yield new insights. The amplitude of waves generated will probably depend on the values of the
coronal resistivity and viscosity, and thus have implications for theories of coronal heating. Different mechanisms of coronal heating might also be expected to produce Alfv\'en waves of different polarizations. The examples given in this review have treated shear Alfv\'en waves, such as might be expected to derive from reconnection. Alfv\'en waves resulting from resonant absorption would be torsional waves, and those generated by a reconnection outflow by a firehose instability would exhibit some circular polarization.
The Alfv\'en wave transport equations and fractionation (Sections~\ref{subsec:6.4}
and \ref{subsec:6.5}) do not change with
polarization, but the coupling to other wave modes (Section~\ref{subsec:6.6}) does.
Pure parallel propagating circularly polarized waves do not couple to slow modes at all, while the case of torsional waves has been considered by \citet{vasheghani11}. On twisted
flux tubes, pure torsional Alfv\'en waves do not exist, and some degree of mixing with the compressional kink mode is inevitable. It remains to be seen whether these changes in the
wave physics will result in detectable changes in FIP fractionation, and if so, whether this represents a new avenue of approach to the problem of coronal heating.

\item[2. Chromospheric Waves:]
The solar FIP effect is subtly different depending on the altitude in the chromosphere
where the fractionation occurs. High in the chromosphere, where H is becoming ionized, Fe and Mg fractionate more than Si, and S behaves more like a high FIP element. He can also be depleted relative to O. In the lower chromosphere where H is neutral, the low FIPs fractionate to essentially the same degree, and S behaves more like a low FIP element. He/O remains unchanged. Such fractionation can occur with off-resonant Alfv\'en waves, e.g., in an open field region, or with upcoming fast mode waves in sufficiently diverging magnetic field. An important goal for spectroscopic and in-situ observations should be to obtain data of sufficient quality to distinguish
between these two possibilities. Possibly the best extant spectroscopic analysis, that of
quiet solar corona by \citet{bryans09} (given as column ``d'' in Table~\ref{table:closedfield}) strongly favors fractionation at the top of the chromosphere.

Measurements of the S abundance made {\it in situ} in the slow speed solar wind often show significantly higher values than those obtained from spectroscopy \citep[e.g.][]{giammanco07,reisenfeld07}. The comparison between these two forms of measurement
is very clear in Figure 1. of \citet{schmelz12}. Another element predicted to behave similarly
\citep{rakowski12} is C. Unfortunately, spectroscopic measurements of the C abundance in the solar corona are very difficult, because at coronal temperatures, C is typically fully ionized and emits no lines.

Fractionation low down in the chromosphere can occur with Alfv\'en waves well away from
the loop resonant frequency, such that they are not trapped in the coronal loop, or with
upward propagating fast mode waves in region where the magnetic field expands, giving rise
to positive FIP fractionation from Equation~(\ref{eq:fastmodepond}). He is well known to be present with high abundance in CMEs \citep{wimmer06}, consistent with fractionation low in the chromosphere, but S, the other element crucial in this regard is typically not measured. Whether fractionation occurs low down or high up depends not only on the coronal
Alfv\'en wave, but also on wave physics in the chromosphere. The mode conversion acoustic waves to fast mode waves at the $\beta =1$ layer makes a connection with the fields of
helioseismology and asteroseismology. There is much observational and theoretical work ongoing
on these aspect of chromospheric wave propagation that is relevant to the FIP effect. One key
component to this will be chromospheric vector magnetic fields. Currently chromospheric and
coronal magnetic fields are extrapolated from photospheric magnetic field observations.
Uncertain knowledge of the magnetic field expansion through the chromosphere makes it difficult
to know where mode conversion occurs, and whether upcoming fast mode waves should give FIP or inverse FIP effects.

Considerable interest has arisen recently in the idea that the solar corona and wind might be supplied by Type II spicules accelerated directly from the chromosphere \citep[e.g.,][]{depontieu11,martinez11}, rather than plasma being evaporated into a coronal loop before being released into the solar wind. We comment here that in such a case, it is difficult to see how a FIP effect could arise. With reference to Equation~(\ref{eq:22}), the flow velocity associated with the spicule motion is likely to
increase $v_s^2$ in the denominator of the integrand to an extent such that the integral
tends to zero, and $\rho_s\left(z_u\right)=\rho_s\left(z_l\right)$, yielding no fractionation.
In simulations, \citet{martinez11} find Type II spicule material extending to temperatures close to $10^6$ K, but no higher, consistent with the temperature range over which \citet{laming95} observed photospheric abundances (see Fig. \ref{fig:LDW}). Discontinuities in several other observables have been located between 500,000 K and $10^6$K in the solar atmosphere \citep{feldman83,feldman87}, and further considerations of element abundances \citep{feldman98a} reinforce the conclusion of \citet{laming95}. Other authors have also
questioned the connection between coronal mass supply and Type II spicules on the basis of
spectral line profiles \citep{patsourakos14,klimchuk14}. \citet{goodman14} argues on energetics
grounds that Type II spicules may power the quiet solar corona and coronal holes, but not
active regions.

Type II spicules are most likely the cause of the absence of the FIP effect at temperatures below about $10^6$ K. However in the case of Procyon, with its lower surface
gravity, Type~II spicules are possibly the most likely reason why its corona appears to have photospheric composition, unlike the other F dwarfs of similar spectral type in Figure~\ref{fig:woodlinsky}. It is interesting to note that \citet{drake95b}, without knowing the cause, recognized that the FIP effect appeared in the full disk solar spectrum at about the temperature where the supergranulation disappeared from solar images, and speculated that if a similar transition happened on Procyon, it would occur at higher temperature.

\item[3. Solar Observations:] Detailed spectroscopy and \textit{in situ} measurements should aim to distinguish between the two regimes of FIP fractionation mentioned above, i.e., low or high in the chromosphere. The aim should be to try and measure several element  abundances simultaneously, not just the evaluation of one or two abundance ratios. This will ultimately require spectrometers with bandpasses specially designed for the purposes. The Hinode/EIS instrument, for example, was designed in the 1990s (well before the ponderomotive force model of the FIP effect was published) to maximize the coverage of lines from Fe ions, and so has rather few lines from high FIP ions within its bandpass, S being the most prominent. However, Hinode/EIS does offer imaging spectroscopy sensitive to the FIP effect, as demonstrated by \citet{baker13}. The observation of FIP enhanced regions correlated with high non-thermal broadening at loop footpoints is an important advance. The filamentary coronal heating, and correspondingly filamentary FIP fractionation could be potentially observable as a variation in FIP effect \emph{across} the cross section of a coronal loop. Increased efforts to measure absolute abundances (i.e. relative to H) would also yield insights into other mechanisms working to modify element abundances, and be advantageous in comparisons of abundances from remotely sensed spectroscopic observations with those measured {\it in situ} in the solar wind.

\item[4. Stellar Observations:] The quality of stellar observations is possibly less likely to improve in the future than those of solar observations, since missions in X-ray astronomy are fewer and further spaced in time than solar missions. Imaging spectroscopy is of course not possible, so the same comments about detailed spectroscopy to measure as many element abundances simultaneously apply. The main advance here will come from
    measuring abundances in a wider range of stellar targets. Observing more F dwarfs would flesh out the bottom left hand corner of Figure~\ref{fig:woodlinsky} and really establish a saturation of the FIP effect at about the level observed in the Sun, and no higher. This would then be interpreted as the FIP effect established solely by coronal Alfv\'en waves, with no contribution from upcoming fast modes. Further observation with Chandra of $\tau$ Bootis A would establish whether contamination from its M dwarf companion is responsible for its anomalous inverse FIP effect, or whether this is intrinsic to the F dwarf, and presumably due to the close-in Jupiter mass planet.

    Observations of inverse FIP effect exist in many more active and exotic stellar targets than those plotted in Figure~\ref{fig:woodlinsky}. We have somewhat neglected discussion of these, for the simple reason that in trying to reach theoretical understanding of where the Inverse FIP effect comes from, it is obviously advantageous to start with simple objects first, before moving onto more complicated ones. Strong
    Inverse FIP seen in RSCVn binaries with tidal interaction would suggest that the planet hosted by $\tau$ Boo A might produce the same effect. However, $\epsilon$ Eridani also hosts a Jupiter mass planet, apparently with no unusual coronal abundance effects. While on the subject of stellar ``complexity'', we note the poor state of knowledge of M dwarf chromospheres and photospheric abundances. This is an intrinsic issue, due to the complicated stellar spectra with many molecular bands and otherwise unidentified lines. There appears to be no ``magic bullet'' in sight, other than detailed careful work on obtaining and modeling such data. Further knowledge of asteroseismology and associated wave physics may be expected to come from the Kepler mission \citep{koch10}.

\end{description}

%==============================================================================
\newpage
\section*{Acknowledgements}
\label{sec:acknowledgements}

This work has been supported by NASA grants from the Astrophysics Theory Program, the
Heliophysics Supporting Research Program, and by basic research funds of the Office of Naval
Research. I am grateful to the editors of \textit{Living Reviews in Solar Physics} for the initial
invitation to write this review, and for their patience with me as various other commitments threatened to derail progress. I acknowledge permission from Deb Baker, Paul Cally, Russ Dahlburg, Jeremy Drake, Justin Kasper, Elena Khomenko, Cara Rakowski, Paola Testa and Brian Wood to reproduce figures from their published work in this review. I am also grateful to Deb Baker, Jeremy Drake, and Brian Wood for their comments on an early draft of this paper, and to John Raymond and other referees for helpful reviews of a more final version.

\newpage

%==============================================================================

% To use the bibtex bibliography in 'refs.bib' do:
% 'latex article'
% 'bibtex article'
% 'latex article'
% 'latex article'

\bibliography{refs}

\end{document}